\newif\ifCM 
\newif\ifSF 
\CMfalse
\SFfalse
\ifCM
\documentclass[AMA,STIX1COL]{WileyNJD-v2}
\else
\documentclass[11pt,a4paper]{article}
\fi
\usepackage{float}
\ifCM
\usepackage[caption=false]{subfig}
\else
\usepackage{subfig}
\usepackage[ocgcolorlinks,allcolors={blue}]{hyperref}
\usepackage[numbers,sort&compress]{natbib}
\usepackage{appendix}
\usepackage{amsmath}
\usepackage{amssymb}
\fi
\usepackage[utf8]{inputenc}
\usepackage[T1]{fontenc}
\usepackage{graphicx}
\usepackage{xcolor}
\usepackage{multirow}
\usepackage{xspace}
\usepackage{ifthen}
\usepackage{ifmtarg}
\usepackage{calc}
\usepackage{algpseudocode}
\usepackage{algorithm}
\usepackage{algorithmicx}
\usepackage[mathlines]{lineno}
\usepackage{placeins}

\usepackage{hho_package}
\ifCM
\else
\fi
\ifSF
\else
\graphicspath{{Images/}}
\fi
\makeatletter
\newcommand*{\red}[1]{{\color{red}#1}} 
\ifCM

\else
\newtheorem{theorem}{Theorem}
\newtheorem{lemma}[theorem]{Lemma}
\newtheorem{hypothesis}[theorem]{Hypothesis}

\newtheorem{remark}[theorem]{Remark}

\DeclareMathSymbol{\bigtimes}{1}{mathx}{"91}
\fi
\makeatother
\ifCM

\articletype{Article Type}%

\received{}
\revised{}
\accepted{}
        
\fi
\begin{document}
\ifCM

\title{A Hybrid High-Order method for finite elastoplastic deformations within a logarithmic strain framework}

\author[1,2]{Mickaël Abbas}
\author[3,4]{Alexandre Ern}
\author[1,2,3,4]{Nicolas Pignet*}   

\authormark{MICKAEL ABBAS \textsc{et al}}

\address[1]{\orgname{EDF R\&D ERMES}, \orgaddress{7 Boulevard Gaspard Monge, 91120 Palaiseau, \country{France}}}
\address[2]{ \orgname{IMSIA}, \orgaddress{UMR EDF/CNRS/CEA/ENSTA 9219,  \country{France}}}
\address[3]{ \orgname{Université Paris-Est, CERMICS (ENPC)}, \orgaddress{6-8 avenue Blaise Pascal, 77455 Marne la Vall\'ee cedex 2, \country{France}}}
\address[4]{ \orgname{INRIA}, \orgaddress{75589 Paris, \country{France}}}

\corres{*Nicolas Pignet, 7 Boulevard Gaspard Monge, 91120 Palaiseau, France. \email{nicolas.pignet@edf.fr}}

\presentaddress{7 Boulevard Gaspard Monge, 91120 Palaiseau, France}

\abstract[Summary]{We devise and evaluate numerically a Hybrid High-Order (HHO) method for finite plasticity within a logarithmic strain framework. The HHO method uses as discrete unknowns piecewise polynomials of order $k\ge1$ on the mesh skeleton, together with cell-based polynomials that can be eliminated locally by static condensation.  The HHO method leads to a primal formulation, supports polyhedral meshes with non-matching interfaces, is free of volumetric locking, the integration of the behavior law is performed only at cell-based quadrature nodes, and the tangent matrix in Newton's method is symmetric. Moreover,  the principle of virtual work is satisfied locally with equilibrated tractions. Various two- and three-dimensional benchmarks are presented, as well as comparison against known solutions with an industrial software using conforming and mixed finite elements.}


\maketitle
\else
\author{Mickaël Abbas$^1,2$,  Alexandre Ern$^{3,4}$ and Nicolas Pignet$^{1,2,3,4}$}
\title{A Hybrid High-Order method for finite elastoplastic deformations within a logarithmic strain framework}
\maketitle
\footnotetext[1]{EDF R\&D ERMES, 7 Boulevard Gaspard Monge, 91120 Palaiseau,
  France}
\footnotetext[2]{ IMSIA, UMR EDF/CNRS/CEA/ENSTA 9219,  France}
\footnotetext[3]{Université Paris-Est, CERMICS (ENPC), 6-8 avenue Blaise Pascal, 77455 Marne-la-Vall\'ee cedex 2, France}
\footnotetext[4]{INRIA, 75589 Paris, France}
\begin{abstract}
We devise and evaluate numerically a Hybrid High-Order (HHO) method for finite plasticity within a logarithmic strain framework. The HHO method uses as discrete unknowns piecewise polynomials of order $k\ge1$ on the mesh skeleton, together with cell-based polynomials that can be eliminated locally by static condensation.  The HHO method leads to a primal formulation, supports polyhedral meshes with non-matching interfaces, is free of volumetric locking, the integration of the behavior law is performed only at cell-based quadrature nodes, and the tangent matrix in Newton's method is symmetric. Moreover,  the principle of virtual work is satisfied locally with equilibrated tractions. Various two- and three-dimensional benchmarks are presented, as well as comparisons against known solutions obtained with an industrial software using conforming and mixed finite elements.
\end{abstract}
\\
\fi
%
%
\section{Introduction}
 Modelling plasticity, particularly in the regime of finite deformations, is of a major importance in industrial applications since this is one of the main nonlinearites that can be encountered in nonlinear solid mechanics. Moreover, finite elastoplastic deformations have a major influence on the life time of a mechanical structure. 
The present contribution is an extension to the finite strain regime of the Hybrid High-Order (HHO) method for incremental associative plasticity with small deformations  \cite{AbErPi2018a}. This extension hinges on a logarithmic strain framework \cite{Miehe2002}  for anisotropic finite elastoplasticity. This framework provides a natural extension of small elastoplastic deformations to finite elastoplastic deformations by means of purely geometric transformations. Indeed, the weak form of the plasticity problem is derived from the minimization of an energy functional based on an incremental pseudo-energy density.

The present work aims at addressing the following important issues. Firstly, the incompressibility of the plastic deformations generally leads to volumetric locking when employing a continuous Galerkin (cG) approximation based on low-order $H^1$-conforming finite elements. In these methods, only the displacement field is approximated globally, whereas the variables associated with the plastic behavior are computed locally in each mesh cell (typically at  quadrature nodes). A way to circumvent the volumetric locking is to use high-order $H^1$-conforming finite elements or NURBS for small \cite{Elguedj2014} and finite \cite{Elguedj2008} elastoplastic deformations. Therein, the displacement is still the only field which is approximated globally. However, the resulting discrete problem is more costly to solve because of the larger support of the basis functions. Another possible way to prevent volumetric locking is to introduce additional global unknowns as in the Enhanced Assumed Strain (EAS) methods \cite{Simo1992a} and in mixed methods  \cite{AlAkhrass2014, Saracibar2006, SouzaNeto2005, Simo1985} on simplicial or hexahedral meshes (the variables associated with the plastic behavior are still computed locally). However, the introduction of additional globally coupled unknowns generally increases the cost of building and solving the discrete problem. Moreover, devising mixed methods on polyhedral meshes with non-matching interfaces is a delicate question. On the positive side, cG methods as well as EAS and mixed methods require to perform the integration of the behavior law only at the quadrature nodes in the mesh cells. Another class of methods free of volumetric locking are discontinuous Galerkin (dG) methods. We mention in particular \cite{Eyck2006, Eyck2008, Eyck2008a, Noels2006} for hyperelasticity. Interior penalty dG methods have been developed for classical plasticity with small \cite{Hansbo2010, Liu2010} and finite \cite{Liu2013} deformations, and for gradient plasticity with small  \cite{Djoko2007, Djoko2007a} and finite \cite{McBride2009} deformations. However, dG methods from the literature generally require to perform the integration of the behavior law also at additional quadrature nodes located at the mesh faces. Moreover, if the plasticity problem is solved using a Newton's method, which is often the case, the tangent matrix from the dG formulation is generally non-symmetric owing to the nonlinear nature of the consistency term. Thus, the solving cost can increase significantly, particularly with iterative solvers (the memory requirements can become important for direct solvers). We also mention the lowest-order Virtual Element Method (VEM) for inelastic problems with small deformations \cite{BeiraodaVeiga2015}  (and its  two-dimensional higher-order extension \cite{Artioli2017}), whereas the case of finite deformations is treated in \cite{Wriggers2017a, Hudobivnik2018}, still in the lowest-order case. We also mention the recent study of low-order hybrid dG methods with conforming traces \cite{Wulfinghoff2017} and the hybridizable weakly conforming Galerkin method with nonconforming traces in the context of linear \cite{Kramer2016} and nonlinear \cite{Bayat2018} solid mechanics. Moreover,  finite volume methods for plasticity problems have been developped for small deformations \cite{Taylor2003} and for large deformations \cite{Barton2010, Cardiff2014}.

In the present work, we devise and evaluate numerically a HHO method for finite plasticity within a logarithmic strain framework. HHO methods have been introduced a few years ago for diffusion problems \cite{DiPEL:2014} and for linear elasticity problems \cite{DiPEr:2015}. Since then, the development of HHO methods has received a vigorous interest. Examples include  in solids mechanics Biot's problem \cite{BoBoD:2016}, nonlinear elasticity \cite{BoDPS:2017} and associative plasticity \cite{AbErPi2018a} with small deformations, and hyperelasticity with finite deformations \cite{AbErPi2018}, and in fluid mechanics, the incompressible Stokes equations \cite{DPELS:16}, the steady incompressible Navier--Stokes equations \cite{DiPKr:2017}, and viscoplatic flows with yield stress \cite{Cascavita2018}. The discrete unknowns in HHO methods in computational mechanics are face-based vector-valued polynomials of arbitrary order $k \geq 1$ on the mesh skeleton. Cell-based vector-valued polynomials are also introduced for the stability and approximation properties of the method. These cell-based vector-valued polynomials are eliminated locally by using the well-known static condensation technique (based on a local Schur complement).

The devising of HHO methods hinges on two key ideas: \textup{(i)} a local reconstruction operator acting on the face and cell unknowns that builds a tensor-valued polynomial representing the displacement gradient in the polynomial space $\Pkd(T; \Rdd)$, where $T$ is a generic mesh cell and $d$ is the space dimension \cite{BoDPS:2017, AbErPi2018}; \textup{(ii)} a local stabilization operator that weakly enforces on each mesh face the consistency between the local face unknowns and the trace of the cell unknowns \cite{DiPEL:2014, DiPEr:2015}. HHO methods offer several advantages: \textup{(i)} general meshes (including fairly general polyhedral mesh cells and non-matching interfaces) are supported; \textup{(ii)} a local formulation using equilibrated fluxes is available; \textup{(iii)} computational benefits owing to the static condensation of the cell unknowns and the higher-order convergence rates, and \textup{(iv)} the construction is dimension-independent. Moreover, an open-source implementation of HHO methods, the \texttt{DiSk++} library, is available using generic programming tools \cite{CicDE:2018}. In computational mechanics, other salient features of HHO methods are: \textup{(i)} a displacement-based formulation avoiding the need to introduce additional globally coupled unknowns; \textup{(ii)} absence of volumetric locking; \textup{(iii)}  the integration of the behavior law only at the cell quadrature nodes; and \textup{(iv)} the tangent matrix arising in the Newton's method is symmetric. Furthermore, HHO methods have been bridged \cite{CoDPE:2016} to Hybridizable Discontinuous Galerkin (HDG) methods \cite{CoGoL:09} and to nonconforming Virtual Element Methods (ncVEM) \cite{Ayuso2016}. The essential difference with HDG methods is that the HHO stabilization is different so as to deliver $O(h^{k+1})$ energy-error estimates for linear model problems with smooth solutions on general meshes, where $h$ is the mesh-size. Concerning ncVEM, the devising viewpoint is different (ncVEM considers the computable projection of virtual functions instead of a reconstruction operator), whereas the stabilization achieves similar convergence rates as HHO but is written differently. We also notice that, to our knowledge, HDG methods have not yet been devised for finite elastoplasticity problems (in contrast to hyperelasticity problems \cite{KaLeC:2015, Nguyen2012a}). Owing to the close links between HHO and HDG methods, this work can thus be seen as the first HDG-like method for plasticity problems in finite deformations.

This paper is organized as follows: in Section~2, we present the plasticity model within a logarithmic strain framework and the weak formulation of the governing equations. In Section~3, we devise the HHO method and highlight some of its theoretical aspects. In Section~4, we investigate numerically the HHO method on two- and three-dimensional benchmarks from the literature, and we compare our results to analytical solutions whenever available and to numerical results obtained using established cG and mixed methods implemented in the open-source industrial software \CA \cite{CodeAster}. 
\section{Plasticity model}\label{sec::model}
In what follows, we write $v$ or $V$ for scalar-valued fields, $\vv$ or $\vecteur{V}$ for vector-valued fields, $\matrice{v}$ or $\matrice{V}$ for second-order tensor-valued fields, and $\tenseur{4}{V}$ for fourth-order tensor-valued fields. Contrary to the hyperelastic model, the elastoplastic model is based on the assumption that the deformations are no longer reversible.
\subsection{Kinematics}
Let $\mathcal{B}_{0}$ be an elastoplastic material body that  occupies the domain $\Omega_0$ in the reference configuration.  Here, $\Omega_0 \subset \Rd$, $d \in \{2,3\}$, is a bounded connected polyhedral domain with Lipschitz boundary $\Gamma := \partial \Omega_0$ decomposed in the two relatively open subsets $\Bn$ and $\Bd$, where a Neumann and a Dirichlet condition is enforced respectively, and such that $\overline{\Bn} \cup \overline{\Bd} = \Gamma$, $\Bn \cap \Bd = \emptyset $, and $\Bd$ has positive Hausdorff-measure (so as to prevent rigid-body motions).  Due to the deformation, a point $\Xo \in \Omega_0$ is mapped to a point $\xc(t) = \Xo + \vu(\Xo,t)$ in the equilibrium configuration, where $\vu : \Omega_0 \times I \rightarrow \Rd$ is the displacement mapping. The deformation gradient $\Fdef(\vu) = \matrice{I} +\gradX \, \vu$ takes values in $\Rdd_{+}$ which is the set of $\Rdd$-matrices with positive determinant. In what follows, the gradient and divergence operators are taken with respect to the coordinate $\Xo$ of the reference configuration (we use the subscript $X$ to indicate it). 

We use the logarithmic strain framework \cite{Miehe2002} developed for anisotropic finite elastoplasticity. Hence, it allows us to define the logarithmic strain tensor $\Elog \in \Msym$ as 
\begin{equation}
\Elog := \frac{1}{2} \ln(\Fdef^T \Fdef).
\end{equation}
This measure of the deformations $\Elog$ is objective. Moreover, if  the eigenvectors of $\Elog$ do not change with time (the eigenvalues may change in time), then $\dot{\Elog} = \dot{\matrice{U}}\matrice{U}^{-1}$, where $\matrice{U} \in \Msym $ is the right stretch tensor from the polar decomposition $\Fdef = \matrice{R} \matrice{U}$.  The plastic deformations are measured by means of the plastic logarithmic strain tensor $\pElog \in \Msym$. We assume the following additive decomposition of the logarithmic strain tensor $\Elog$:
\begin{equation}
\eElog := \Elog - \pElog,
\end{equation}
where $\eElog \in \Msym$ is the elastic logarithmic strain tensor. Finally, the plastic strains are assumed to be incompressible, i.e. 
\begin{equation}
\trace \pElog = 0.
\end{equation}
\subsection{Constitutive logarithmic strain model}\label{ss:helmhotz}
 In what follows, we place ourselves within the framework of generalized standard materials initially introduced in \cite{Halphen1975} and further developed in \cite{Lemaitre1994}. Moreover,  the plasticity model is assumed to be strain-hardening (or perfect) and rate-independent, i.e., the time and the speed of the deformations have no influence on the strains. For this reason, only the incremental plasticity problem is considered. The local material state is described by the logarithmic strain tensor $\Elog \in \Msym$, the plastic logarithmic strain tensor $\pElog \in \Msym$, and a finite collection of internal variables $\IV:= ( \alpha_1, \cdots, \alpha_m) \in \Reel^m$, which contain typically at least the equivalent plastic strain $p$, see Sect.~\ref{ss:nl_hardening} for a simple example or \cite{Lemaitre1994}  for more detailed examples. For simplicity, we denote $\IVG:=( \pElog, \IV ) \in \IVGS$ the generalized internal variables, where the space of the generalized internal variables is
\begin{equation} \label{eq:def_space_X}
\IVGS := \left\lbrace \IVG=(\pElog, \IV) \in  \Msym \times \Reel^{m}  \: | \: \trace(\pElog)= 0 \right\rbrace.
\end{equation} 
Moreover, we assume that there exists a Helmholtz free energy $\psi: \Msym \times \Reel^m \rightarrow \Reel$ acting on the pair $(\eElog, \IV)$ and satisfying the following hypothesis.
\begin{hypothesis}[Helmholtz free energy]\label{hyp_free_ener}
$\psi$ can be decomposed additively into an elastic and a plastic part as follows:
\begin{equation}
 \psi(\eElog, \IV) = \psi^e( \eElog) + \psi^p( \IV).
\end{equation} 
where the function $\psi^p$ is strictly convex and the function $\psi^e$ is polyconvex. 
\end{hypothesis}

Following the second principle of thermodynamics, the logarithmic stress tensor $\stressT \in \Msym$ and the thermodynamic forces $\IF$ are derived from $\psi$ as follows:
\begin{equation}\label{eq_IF}
\stressT = \partial_{ \eElog} \psi^e(\eElog)  \quad \textrm{and} \quad
\IF =   \partial_{ \IV} \psi^p(\IV). 
\end{equation}
The criterion to determine whether the deformations become plastic hinges on the scalar-valued yield function $\Phi: \Msym \times \Reel^m \rightarrow \Reel$, which is a continuous and convex function of the logarithmic stress tensor $\stressT$ and the thermodynamic forces $\IF$. Letting $\mathcal{A} := \left\lbrace (\stressT, \IF) \in \Msym \times \Reel^m \, \vert \, \Phi(\stressT, \IF) \leq 0 \right\rbrace$ be the convex set of admissible states, the elastic domain $\mathcal{A}^e$ is composed of all the pairs $ (\stressT, \IF)$ such that $\Phi(\stressT, \IF) < 0 $, and the yield surface $\partial \mathcal{A}$ of all the pairs $ (\stressT, \IF)$ such that $\Phi(\stressT, \IF) = 0 $.
\begin{hypothesis}[Yield function]\label{hyp_yield}
The yield function $\Phi: \Msym \times \Reel^m \rightarrow \Reel$ satisfies the following properties: \textup{(i)} $\Phi$ is a piecewise analytical function; \textup{(ii)} the point $(\matrice{0}, \vecteur{0})$  lies in the elastic domain, i.e., $\Phi(\matrice{0}, \vecteur{0}) < 0$; and \textup{(iii)} $\Phi$ is differentiable at all points on the yield surface $\partial\mathcal{A}$.
\end{hypothesis}
Finally, the incremental dissipation function $D :\IVGS \rightarrow \Reel$ is a convex function which is positively homogeneous of degree one and is defined as follows:
\begin{equation}\label{eq::dissipation}
D(\textrm{d} \IVG) = \sup_{(\stressT, \IF) \in \mathcal{A}} \left( \stressT : \textrm{d} \pElog - \IF \SCAL \textrm{d} \IV  \right),
\end{equation}
where $\textrm{d} \IVG, \textrm{d} \pElog$, and $\textrm{d} \IV $ are the finite increments of $ \IVG, \pElog$, and $ \IV $, respectively.
\subsection{Finite elastoplasticity model problem in incremental form}\label{ss:mech_model}
We are interested in finding the quasi-static evolution in the pseudo-time interval $I=[0,t_F]$, $t_F > 0$, of the elastoplastic material body $\mathcal{B}_0$. The pseudo-time interval $I$ is discretized into $N$ subintervals such that $t^0=0 < t^1 < \cdots < t^N = t_F$. The evolution occurs under the action of a body force $\loadext:\Omega_0 \times I \rightarrow \Rd $, a traction force $\Tn: \Bn \times I \rightarrow \Rd$ exerted on the Neumann boundary $\Bn$, and a displacement $\vu{}_{\textrm{D}}: \Bd \times I \rightarrow \Rd$ imposed on the Dirichlet boundary $\Bd$. We denote by $V_{\textrm{D}}^n$, resp. $V_0$, the set of all kinematically admissible displacements which satisfy the Dirichlet conditions, resp. homogeneous Dirichlet conditions on $\Bd$: 
\begin{equation}
 V_{\textrm{D}}^n = \left\lbrace \vv \in H^1(\Omega_0; \Rd) \: | \: \vv = \vu{}_{\textrm{D}}(t^n) \: \mbox{ on } \Bd \right\rbrace, \; V_0 = \left\lbrace \vv \in H^1(\Omega_0; \Rd) \: | \: \vv = \vecteur{0} \: \mbox{ on } \Bd \right\rbrace.
\end{equation}
Following \cite{Miehe2002}, we define for any pseudo-time step $1 \leq n \leq N$, the incremental pseudo-energy density $\Psi: \Rdd_{+} \times \IVGS \rightarrow \Reel$ acting on the pair $ (\Fdef, \IVG)$ such that 
\begin{equation}\label{eq::pseudo-energy}
\Psi(\Fdef, \IVG) = \left\lbrace \big( \psi^e(\eElog) + \psi^p(\IV) \big) - \big( \psi^e(\eElog(\vu^{n-1}))  + \psi^p(\IV^{n-1}) \big) \right\rbrace + D(\IVG -\IVG^{n-1}),
\end{equation}
where  $\vu^{n-1} \in V_{\textrm{D}}^{n-1}$ and $\IVG^{n-1} \in  L^2(\Omega_0; \IVGS)$ are given from the previous pseudo-time step or the initial condition. Note that the second term in \eqref{eq::pseudo-energy}, which is evaluated at $t^{n-1}$, is irrelevant for minimization purposes. It is added so that the pseudo-energy is in a time-incremental form. This allows us to define the energy functional $\mathcal{E}^n: V_{\textrm{D}}^n \times L^2(\Omega_0; \IVGS)  \rightarrow \Reel$ such that, for all kinematically admissible displacement fields  $\vv \in V_{\textrm{D}}^n$ and all generalized internal variables $\IVG \in L^2(\Omega_0; \IVGS)$.
\begin{equation}\label{eq:def_calE}
\mathcal{E}^n (\vv, \IVG) = \int_{\Omega_0} \Psi(\Fdef(\vv), \IVG) \dV - \int_{\Omega_0} \loadext^n\SCAL\vv \dV - \int_{\Bn} \Tn^n \SCAL\vv \dA.
\end{equation}
The quasi-static equilibrium of the elastoplastic body $\mathcal{B}_0$ is determined at each pseudo-time step $1 \leq n \leq N$ by finding a displacement field $\vu^n \in V_{\textrm{D}}^n$ and generalized internal variables $\IVG^n \in L^2(\Omega_0; \IVGS)$ which minimize the energy functional $\mathcal{E}^n$ in \eqref{eq:def_calE}, i.e.,
\begin{equation}\label{argmin}
(\vu^n, \IVG^n) \in \argmin_{\vv \in V_{\textrm{D}}^n, \, \IVG \in L^2(\Omega_0; \IVGS) }\mathcal{E}^n (\vv, \IVG).
\end{equation}
On the one hand, the first variation of $\mathcal{E}^n$ with respect to the displacement field leads to:
\begin{equation} \label{weak_form}
0 = D \mathcal{E}^n(\vu^n, \IVG^n)[\vv] =  \int_{\Omega_0} \PK^n : \gradX\vv \,d\Omega_0 - \int_{\Omega_0} \loadext^n \SCAL \vv \,d\Omega_0 - \int_{\Bn} \Tn^n \SCAL \vv \,d\Gamma, \quad  \textrm{for all } \vv\in V_0,
\end{equation}
where $\PK^n := \partial_{\Fdef} \Psi(\Fdef(\vu^n), \IVG^n)$ is the first Piola--Kirchhoff stress tensor. On the other hand, the first variation of $\mathcal{E}^n$ with respect to the generalized internal variables leads to the following incremental nonlinear equations (see \cite{Miehe2002, Simo1992b} for example):
\begin{equation}\label{normal_rules}
\pElogn - \pElogpn = \Lambda(\stressT^n, \IF^n) \, \partial_{\stressT} \Phi(\stressT^n, \IF^n), \quad \textrm{and} \quad  \IV^n - \IV^{n-1} = -  \Lambda(\stressT^n, \IF^n) \, \partial_{\IF} \Phi(\stressT^n, \IF^n),
\end{equation}
where the plastic multiplier $\Lambda(\stressT^n, \IF^n)$ verifies the Kuhn--Tucker conditions
\begin{equation}\label{consistency_cond}
\Lambda(\stressT^n, \IF^n) \geq 0, \quad \Phi(\stressT^n, \IF^n) \leq 0, \quad \textrm{and} \quad \Lambda(\stressT^n, \IF^n) \,\Phi(\stressT^n, \IF^n) = 0.
\end{equation}
We assume additionally that the incremental pseudo-energy density $\Psi$ is polyconvex so that local minimizers of the energy functional \eqref{eq:def_calE} exist (cf. e.g \cite{Miehe2002, Ball1976}).
Thus, the minimization problem \eqref{argmin}  can be reformulated, in a more classical way as follows: For all $1 \leq n \leq N$, given $\vu^{n-1} \in V_{\textrm{D}}^{n-1}$ and $\IVG^{n-1} \in  L^2(\Omega_0; \IVGS)$ from the previous pseudo-time step or the initial condition, find $\vu^n \in V_{\textrm{D}}^{n}$ and $\IVG^n \in  L^2(\Omega_0; \IVGS)$ such that
\begin{subequations}\label{weak_form_cont}
\begin{equation} \label{weak_equil}
  \int_{\Omega_0} \PK^n : \gradX\vv \,d\Omega_0 = \int_{\Omega_0} \loadext^n \SCAL \vv \,d\Omega_0 + \int_{\Bn} \Tn^n \SCAL \vv \,d\Gamma, \quad  \textrm{for all } \vv\in V_0,
\end{equation}
\textrm{and}
\begin{equation}\label{weak_plas}
( \IVG^n, \PK^n, \epmodulePKn) =  \textrm{FINITE\_PLASTICITY}(\IVG^{n-1}, \Fdef(\vu^{n-1}), \Fdef(\vu^n) ).
\end{equation}
\end{subequations}
The procedure $\textrm{FINITE\_PLASTICITY}$ allows one to compute the new values of the generalized internal variables $\IVG$, the first Piola--Kirchhoff stress tensor $\PK$ and the consistent nominal elastoplastic tangent modulus $\epmodulePK$ at each pseudo-time step. This procedure is detailed in Section~\ref{ss:flow_rules}.
\subsection{Algorithmic aspects}\label{ss:flow_rules}
The incremental elastoplasticity problem that has to be solved is to find the new value, after incrementation, of the generalized internal variables $\IVG^{\textrm{new}} = (\Elog^{p,\textrm{new}}, \IV^{\textrm{new}}) \in \IVGS$, the first Piola--Kirchhoff stress tensor $\PK^{\textrm{new}} \in \Msym$, and the consistent nominal elastoplastic tangent modulus $\epmodulePKnew$, given the generalized internal variables $\IVG \in \IVGS$, the deformation gradient $\Fdef \in \Rdd_{+}$, and the new value of the deformation gradient $\Fdef^{\textrm{new}} \in \Rdd_{+}$. Solving this problem is denoted as previously
\begin{equation}
(\IVG^{\textrm{new}}, \PK^{\textrm{new}}, \epmodulePKnew)= \textrm{FINITE\_PLASTICITY}(\IVG, \Fdef, \Fdef^{\textrm{new}}).
\end{equation}
The procedure to compute $(\IVG^{\textrm{new}}, \PK^{\textrm{new}}, \epmodulePKnew)$ is described in Algorithm~\ref{algo::plasticity} and is composed of three different steps. Firstly, a geometric pre-processing is applied in order to compute the logarithmic strain tensors $\Elog$ and $\Elog^{\textrm{new}}$. Secondly, the procedure $\textrm{SMALL\_PLASTICITY}$ is used to solve the nonlinear incremental problem \eqref{normal_rules}-\eqref{consistency_cond} so as to compute $(\IVG^{\textrm{new}}, \stressT^{\textrm{new}}, \epmoduleTnew)$. The resolution of \eqref{normal_rules}-\eqref{consistency_cond} inside the procedure $\textrm{SMALL\_PLASTICITY}$ requires to solve a constrained nonlinear problem which is the same as in the case of plasticity with small deformations and thus makes it possible to extend the procedures already developed for small deformations to finite deformations without modifications (further details about the procedure $\textrm{SMALL\_PLASTICITY}$ can be found in \cite[Sect.~2.3]{AbErPi2018a}). One significant example of such a procedure is the standard radial return mapping \cite{Simo1992b, Simo1998}. Finally, a geometric post-processing step is applied to compute the new values of the first Piola--Kirchhoff stress tensor $\PK^{\textrm{new}}$ and the consistent nominal elastoplastic tangent modulus $\epmodulePKnew$ from the logarithmic stress tensor $\stressT^{\textrm{new}}$ and the consistent logarithmic elastoplastic tangent modulus $\epmoduleTnew$. Detailed explanations to compute the pre- and post-processing steps are given in \cite[Box.~4]{Miehe2002}.
\begin{algorithm}[htpb]
\caption{Computation of $(\IVG^{\textrm{new}}, \PK^{\textrm{new}}, \epmodulePKnew)$}\label{algo::plasticity}
\begin{algorithmic}[1] 
        \Procedure{FINITE\_PLASTICITY}{$\IVG, \Fdef, \Fdef^{\textrm{new}}$}
            \State Set $\Elog = \frac{1}{2}\ln(\Fdef^T \Fdef)$, $\Elog^{\textrm{new}} = \frac{1}{2}\ln(\Fdef^{\textrm{new},T} \Fdef^{\textrm{new}})$ and $\mathrm{d}\Elog := \Elog^{\textrm{new}}- \Elog$
             \State  Compute  $(\IVG^{\textrm{new}}, \stressT^{\textrm{new}}, \epmoduleTnew)= \textrm{SMALL\_PLASTICITY}(\IVG, \Elog, \mathrm{d}{\Elog})$.
            \State Compute $\PK^{\textrm{new}} =  \stressT^{\textrm{new}} : (\partial_{\Fdef} \Elog)^{\textrm{new}}$ and $ \epmodulePKnew = (\partial_{\Fdef} \Elog)^{\textrm{new},T} : \epmoduleTnew : (\partial_{\Fdef} \Elog)^{\textrm{new}} + \stressT^{\textrm{new}} : (\partial_{\Fdef \Fdef} \Elog)^{\textrm{new}}$
            \State \Return $(\IVG^{\textrm{new}} , \PK^{\textrm{new}}, \epmodulePKnew)$
            \EndProcedure
    \end{algorithmic}
\end{algorithm}
Note that $\epmodulePKnew := \partial_{\Fdef \Fdef} \Psi(\Fdef^{\textrm{new}}, \IVG^{\textrm{new}})$ is the consistent elastoplastic tangent modulus and is a fourth-order tensor having only the major symmetries contrary to $\epmoduleTnew$ which has the major and minor symmetries. For a finite incremental strain, the consistent elastoplastic tangent modulus generally differs from the so-called continuous elastoplastic tangent modulus which is obtained by letting the incremental strain tend to zero \cite{Simo1985a}.
\subsection{Example: nonlinear isotropic hardening with a von Mises yield criterion}\label{ss:nl_hardening}
An illustration of the plasticity model defined above is the nonlinear isotropic hardening model with a von Mises criterion. The elastic part of the free energy is such that
\begin{equation}
\psi^e(\eElog) = \frac12 \eElog : \elasticmodule:\eElog,
\end{equation}
where the elastic modulus is $\elasticmodule = 2\mu \tenseur{4}{I}^{s} + \lambda \matrice{I} \otimes \matrice{I}$, with $\mu>0$, $3\lambda + 2 \mu > 0$, $(\tenseur{4}{I}^{s})_{ij,kl}= \frac{1}{2}(\delta_{ik}\delta_{jl} + \delta_{il}\delta_{jk})$, and $(\matrice{I} \otimes \matrice{I})_{ij,kl}=\delta_{ij}\delta_{kl}$. The internal variable is $\IV := p$, where $p \geq 0$ is the equivalent plastic strain. The plastic part of the free energy is such that
\begin{equation}
\psi^p(\IV) = \sigma_{y,0}p +   \frac{H}{2} p^2 + (\sigma_{y,\infty} -\sigma_{y,0})(p - \frac{1 - e^{-\delta p}}{\delta}), 
\end{equation}
where $H \geq 0$ is the isotropic hardening modulus, $ \sigma_{y,0} >0$, resp. $ \sigma_{y,\infty} \geq 0$, is the initial, resp. infinite, yield stress and $\delta \geq 0$ is the saturation parameter. The associated thermodynamic force $\IF = \sigma_{y,0} + Hp + (\sigma_{y,\infty} -\sigma_{y,0})(1 - e^{-\delta p})$ is called the internal stress. Concerning the yield function, we consider a $J_2$-plasticity model with a von Mises criterion:
\begin{equation}
\Phi(\stressT, \IF)  = \sqrt{\frac{3}{2}} \Vert \dev( \stressT) \Vert_{\matrice{\ell}^2} - \IF,
\end{equation}
where  $\dev(\matrice{\tau}) := \matrice{\tau} - \frac{1}{d} \trace(\matrice{\tau})\matrice{I}$ is the deviatoric operator, and the Frobenius norm is defined as $\Vert\matrice{\tau}\Vert_{\matrice{\ell}^2} = \sqrt{ \matrice{\tau}  : \matrice{\tau} }$, for all $\matrice{\tau} \in \Reel^{d \times d}$. Moreover, the above model describes with a reasonable accuracy the behaviour of metals \cite{Lemaitre1994}. This model is used for the numerical examples presented in Section~\ref{sec::sec_numexp}.
\section{The Hybrid High-Order method}
\label{sec:HHO} 
\subsection{Discrete setting}\label{ss::discrete_setting}
We consider a mesh sequence  $(\Th)_{h>0}$, where for each $h>0$, the mesh $\Th$ is composed of nonempty disjoint open polyhedra with planar faces such that $\overline{\Omega}_0 = \bigcup_{T\in\Th} \overline{T}$. The mesh-size is $h = \max_{T\in\Th} h_T$, where $h_T$ stands for the diameter of the cell $T$. A closed subset $F$ of $\overline{\Omega}_0$ is called a mesh face if it is a subset with nonempty relative interior of some affine hyperplane $H_F$ and \textup{(i)} if either there exist  two distinct mesh cells $T_-, \: T_+ \in \Th$ such that $F = \partial T_- \cap \partial T_+ \cap H_F$ (and $F$ is called an interface) or \textup{(ii)} there exists one mesh cell $T\in\Th$ such that $F = \partial T \cap \Gamma \cap H_F$ (and $F$ is called a boundary face). The mesh faces are collected in the set $\Fh$ which is further partitioned into the subset $\Fhi$ which is the collection of the interfaces and the subset $\Fhb$ which is the collection of the boundary faces. We assume that the mesh is compatible with the partition of the boundary $\Gamma$ into $\Bd$ and $\Bn$, so that we can further split the set $\Fhb$ into the disjoint subsets $\Fhbd$ and $\Fhbn$ with obvious notation. For all $T \in \Th$, $\FT$ is the collection of the mesh faces that are subsets of $\partial T$ and $\nT$ is the unit outward normal to $T$. We assume that the mesh sequence $(\Th)_{h>0}$ is shape-regular  in the sense specified in \cite{DiPEr:2015}, i.e., there is a matching simplicial submesh of $\Th$ that belongs to a shape-regular family of simplicial meshes in the usual sense of Ciarlet \cite{Ciarlet1978} and such that each mesh cell $T \in \Th$ (resp., mesh face $F \in \Fh$) can be decomposed in a finite number of sub-cells (resp., sub-faces) which belong to only one mesh cell (resp., to only one mesh face or to the interior of a mesh cell) with uniformly comparable diameter.

\begin{figure}
 \centering
        \subfloat[$(k,l)=(1,1)$]{
        \centering
       \includegraphics[scale=0.4]{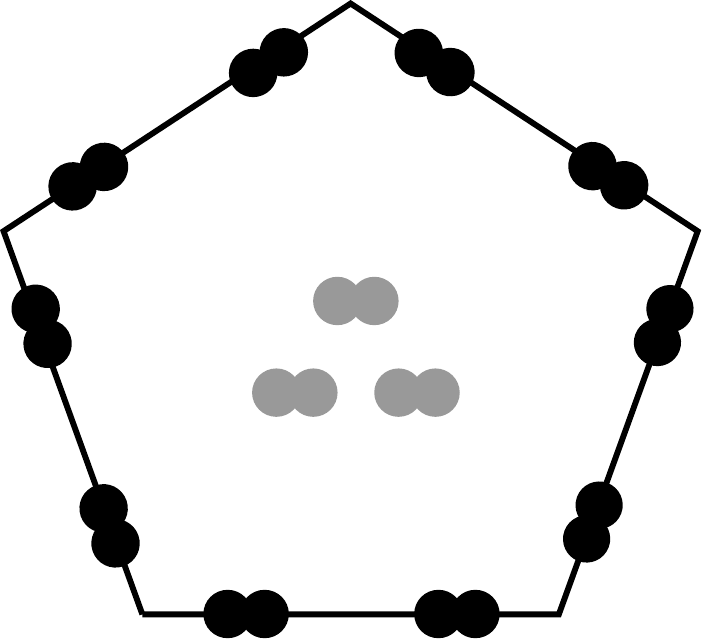} 
  }
    ~ 
        \subfloat[$(k,l)=(1,2)$]{
        \centering
       \includegraphics[scale=0.4]{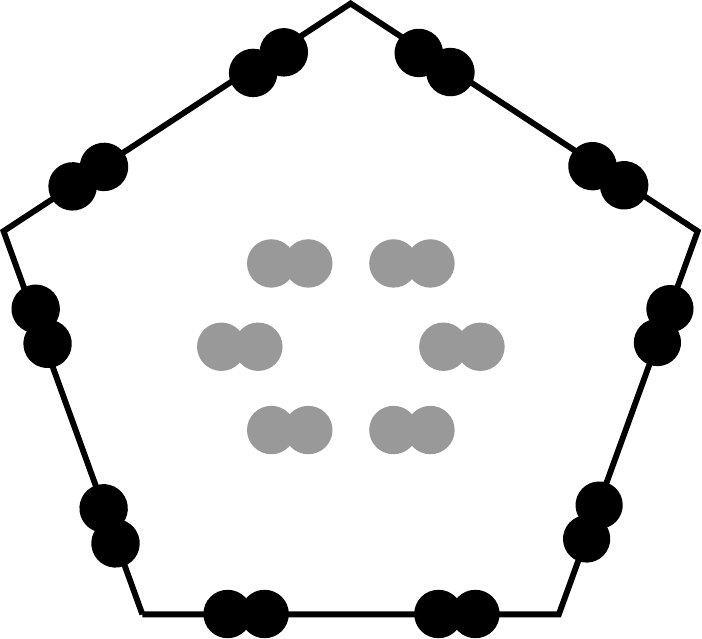} 
  }
    ~ 
        \subfloat[$(k,l)=(2,2)$]{
        \centering
       \includegraphics[scale=0.4]{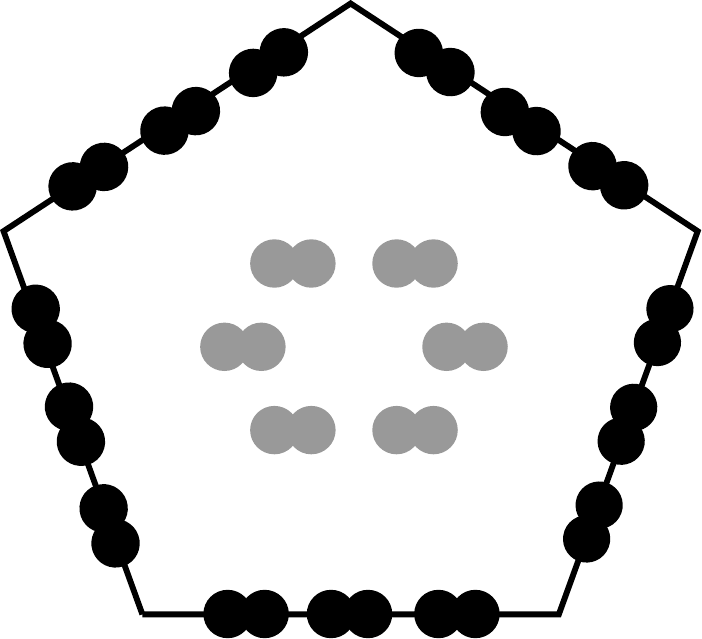} 
  }
    ~ 
    \subfloat[$(k,l)=(2,3)$]{
        \centering
        \includegraphics[scale=0.4]{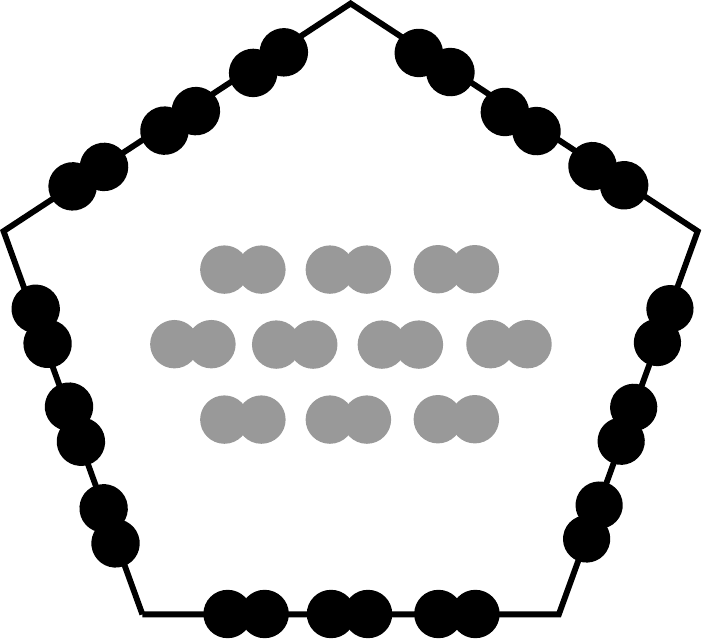} 
  } 
\caption{Face (black) and cell (gray) degrees of freedom in $\UklT$ for different values of the pair $(k,l)$ in the two-dimensional case (each dot represents a degree of freedom which is not necessarily a point evaluation).}
\label{fig_HHO_dofs}
\end{figure}

Let $k \geq 1$  be a fixed polynomial degree and $l \in \lbrace k, k+1 \rbrace$. In each mesh cell $T\in\Th$, the local HHO unknowns consist of a pair $(\vT,\vdT)$, where the cell unknown  $\vT\in \Pld(T; \Rd)$ is a vector-valued $d$-variate polynomial of degree at most $l$ in the mesh cell $T$, and $\vdT \in \PkF(\FT; \Rd) = \bigtimes_{F \in\FT} \PkF(F; \Rd)$ is a piecewise, vector-valued $(d-1)$-variate polynomial of degree at most $k$ on each face $F\in\FT$. We write more concisely that
\begin{equation}\label{local_dof}
(\vT,\vdT) \in \UklT := \Pld(T;\Rd) \times \PkF(\FT;\Rd).
\end{equation}
We write the superscript $k$ first since $k$ is the value that determines the convergence rates of the approximation. The degrees of freedom are illustrated in Fig.~\ref{fig_HHO_dofs}, where a dot indicates one degree of freedom (which is not necessarily computed as a point evaluation) and the geometric shape of the cell is only illustrative. We equip the space $\UklT$ with the following local discrete strain semi-norm:
\begin{equation} \label{eq:snorme}
\snorme[1,T]{( \vT, \vdT)}^2 := \normem[T]{\gradX \vT}^2 + \normev[\dT]{\gamma_{\dT}^{\frac12}(\vT-\vdT)}^2,
\end{equation}
with the piecewise constant function $\gamma_{\dT}$ such that $\gamma_{\dT|F}=h_F^{-1}$ for all $F\in\FT$, where $h_F$ is the diameter of $F$. We notice that  $\snorme[1,T]{( \vT, \vdT)}=0$ implies that $\vT$ is a constant and that $\vdT$ is the trace of $\vT$ on $\partial T$.
\subsection{Local gradient reconstruction and stabilization}
\label{sec:grad_rec}
The first key ingredient in the devising of the HHO method is a local gradient reconstruction in each mesh cell $T \in \Th$. This reconstruction is materialized by an operator $\GkT : \UklT \rightarrow \Pkd(T;\Rdd)$ mapping onto the space composed of $\Rdd$-valued polynomials in $T$. The main reason for reconstructing the gradient in a larger space than the space $\gradX \Pkpd(T; \Rd)$ (as for the linear elasticity problem \cite{DiPEr:2015} ) is that the reconstructed gradient of a test function acts against a discrete stress tensor which is not in gradient form, see \cite[Section~4]{DiPietro2018} for further insight. For all $(\vT,\vdT) \in \UklT$, the reconstructed gradient $\GkT(\vT,\vdT) \in \Pkd(T;\Rdd)$ is obtained by solving the following local problem: For all $\matrice{\tau} \in \Pkd(T;\Rdd)$,
\begin{equation}\label{eq_reconstruction_grad}
\psm[T]{\GkT( \vT,\vdT)}{\matrice{\tau}} = \psm[T]{\gradX{\vT}}{\matrice{\tau}} + \psv[\dT]{\vdT - \vT{}_{|\dT}}{\matrice{\tau} \SCAL \nT}.
\end{equation}
Solving this problem entails choosing a basis of the polynomial space $\Pkd(T;\Reel)$ and inverting the associated mass matrix for each component of the tensor $\GkT(\vT,\vdT)$. The second key ingredient in the HHO method is a local stabilization operator that enforces weakly the matching between the faces unknowns and the trace of the cell unknowns. Following \cite{DiPEL:2014,DiPEr:2015}, the stabilization operator $\SdTk : \PlF(\FT;\Rd) \rightarrow \PkF(\FT;\Rd)$ acts on the difference $\vecteur{\theta} = \vdT - \vT{}_{|\dT} \in \PlF(\FT;\Rd)$, and in the mixed-order case $l=k+1$ is such that, for all  $\vecteur{\theta}\in \PkpF(\FT;\Rd)$,
\begin{equation} \label{eq:stabHDG}
\SdTk(\vecteur{\theta}) = \PikdTv\big(\vecteur{\theta}\big),
\end{equation}
where $\PikdTv$ denotes the $L^2$-orthogonal projectors onto $\PkF(\FT;\Rd)$, and in the equal-order case $l=k$ is such that, for all  $\vecteur{\theta}\in \PkF(\FT;\Rd)$,
\begin{equation} \label{eq:stabHHO}
\SdTk(\vecteur{\theta}) = \PikdTv\big(\vecteur{\theta}-(\matrice{I}-\PikTv)\DTkp(\vecteur{0},\vecteur{\theta})_{|\dT}\big),
\end{equation}
where $\PikTv$ denotes the $L^2$-orthogonal projectors onto $\Pkd(T;\Rd)$. 
The local displacement reconstruction operator $\DTkp : \UklT \rightarrow \Poly_d^{k+1} (T; \Rd)$ is such that, for all $(\vT,\vdT) \in \UklT$, $\DTkp(\vT,\vdT) \in \Poly_d^{k+1}(T;\Rd)$ is obtained by solving the following local Neumann problem: For all $\vecteur{w} \in \Poly_d^{k+1} (T;\Rd)$,
\begin{equation}\label{eq_reconstruction_depl}
\psm[T]{\gradX{\DTkp(\vT,\vdT)}}{\gradX{\vecteur{w}}} = \psm[T]{\gradX{\vT}}{\gradX{\vecteur{w}}} + \psv[\dT]{\vdT - \vT{}_{|\dT}}{\gradX{\vecteur{w}} \SCAL \nT}.
\end{equation}
together with the mean-value conditions $\int_T \DTkp(\vT,\vdT) \:dT= \int_T \vT \:dT$. Comparing with~\eqref{eq_reconstruction_grad}, one readily sees that $\gradX{\DTkp(\vT,\vdT)}$ is the $L^2$-orthogonal projection of $\GkT(\vT,\vdT)$ onto the subspace $\gradX \Poly^{k+1}_d(T;\Rd)$. Note also that the right-hand side of~\eqref{eq:stabHHO} can be rewritten as $\PikdTv(\vdT - \vT{}_{|\dT}-(\matrice{I}-\PikTv)\DTkp(\vT,\vdT)_{|\dT})$.  Adapting \cite[Lemma 4]{DiPEr:2015}, it is straightforward to establish the following stability and boundedness properties (the proof is omitted for brevity).
\begin{lemma}[Boundedness and stability]\label{lemma_stability_stab}
Let the gradient reconstruction operator be defined by~\eqref{eq_reconstruction_grad} and the stabilization operator be defined by~\eqref{eq:stabHDG} or~\eqref{eq:stabHHO}. Let $\gamma_{\dT}$ be defined below~\eqref{eq:snorme}. Then, we have the following properties: \textit{(i)} Boundedness: there exists $\alpha_\sharp< +\infty$, uniform w.r.t.~$h$, such that, for all $T \in \Th$ and for  $(\vT,\vdT) \in \UklT$,
\begin{equation}
\bigg(\normem[T]{\GkT (\vT, \vdT)}^2 + \normev[\dT]{\gamma_{\dT}^{\frac12}\SdTk(\vdT - \vT{}_{|\dT})}^2\bigg)^{\frac12} \leq \alpha_\sharp \snorme[{1,T}]{(\vT, \vdT)}.
\end{equation}
\textit{(ii)} Stability: there exists $\alpha_\flat > 0$, uniform w.r.t.~$h$, such that, for all $T \in \Th$ and all $(\vT,\vdT) \in \UklT$,
\begin{equation}
\alpha_\flat \snorme[{1,T}]{(\vT, \vdT)} \leq \bigg(\normem[T]{\GkT (\vT, \vdT)}^2 + \normev[\dT]{\gamma_{\dT}^{\frac12}\SdTk(\vdT - \vT{}_{|\dT})}^2\bigg)^{\frac12}.
\end{equation}
\end{lemma}
As shown in \cite{DiPEr:2015}, the following important commuting property holds true:
\begin{equation}\label{eq:commuting}
\GkT(\vecteur{I}_{T,\dT}(\vecteur{v})) = \matrice{\Pi}^k_T(\gradX\vecteur{v}),\qquad \forall \vecteur{v}\in H^1(T;\Rd),
\end{equation}
with the reduction operator $\vecteur{I}_{T,\dT} : H^1(T;\Rd)\rightarrow \UklT$ defined as $\vecteur{I}_{T,\dT}(\vecteur{v})=(\PilTv(\vecteur{v}),\PikdTv(\vecteur{v}{}_{|\dT}))$. Taking the trace in~\eqref{eq:commuting}, we infer that 
\begin{equation} \label{eq:trace_stab}
\trace\big(\GkT(\vecteur{I}_{T,\dT}(\vecteur{v}))\big) = \Pi_T^k(\divergenceX{\vecteur{v}}), \qquad \forall \vecteur{v}\in H^1(T;\Rd),
\end{equation}
which is the key commuting property to prove robustness for quasi-incompressible linear elasticity, see \cite{DiPEr:2015}. This absence of volumetric locking is confirmed in the numerical experiments performed in Section~\ref{sec::sec_numexp} in the nonlinear setting of finite elastoplasticity. Finally, proceeding as in \cite[Thm.~8]{DiPEr:2015}, one can show that for the linear elasticity problem and smooth solutions, the energy error converges as $h^{k+1}|\vecteur{u}|_{\vecteur{H}^{k+2}(\Omega_0)}$. 
\begin{remark}[HDG-type stabilization]
The stabilization operator \eqref{eq:stabHHO} is essential to prove the above mentioned convergence rates in the equal-order case for linear problems and smooth solutions on general meshes. In general, HDG methods use the stabilization operator $\SdTk(\vecteur{\theta}) = \vecteur{\theta}$ in the equal-order case which differs from the stabilization operator \eqref{eq:stabHHO} and allows one to show only that the energy error converges as $h^{k}|\vecteur{u}|_{\vecteur{H}^{k+1}(\Omega_0)}$ for linear problems and smooth solutions on general meshes. In the mixed-order case, the stabilization operator \eqref{eq:stabHDG} has been initially introduced in \cite{Lehrenfeld:10} and the same convergence rates as for the HHO method are obtained.
\end{remark}
\subsection{Global discrete problem}
Let us now devise the global discrete problem. We set $\Pld(\Th;\Rd) :=\bigtimes_{T \in\Th} \Pld(T; \Rd)$ and $\PkF(\Fh;\Rd) := \bigtimes_{F \in\Fh} \PkF(F; \Rd)$. The global space of discrete HHO unknowns is defined as
\begin{equation} \label{eq:def_Ukh}
\Uklh := \Pld(\Th;\Rd) \times \PkF(\Fh;\Rd).
\end{equation}
For an element $\vh \in \Uklh$, we use the generic notation $\vh = (\vTh,\vFh)$. For any mesh cell $T \in \Th$, we denote by $(\vT,\vdT)\in \UklT$ the local components of $\vh$ attached to the mesh cell $T$ and to the faces composing its boundary $\dT$, and for any mesh face $F\in\Fh$, we denote by $\vF$ the component of $\vh$ attached to the face $F$. The Dirichlet boundary condition on the displacement field can be enforced explicitly on the discrete unknowns attached to the boundary faces in $\Fhbd$. Letting $\PikFv$ denote the $L^2$-orthogonal projector onto $\PkF(F;\Rd)$, we set
\begin{subequations}
\begin{align}
\Uklnhd &:= \left\lbrace (\vTh, \vFh ) \in \Uklh \: \vert \: \vF = \PikFv(\vu_{\textrm{D}}(t^n)), \; \forall F \in \Fhbd   \right \rbrace, \\
\Uklhz &:= \left\lbrace (\vTh, \vFh ) \in \Uklh \: \vert \: \vF = \vecteur{0}, \; \forall F \in \Fhbd   \right \rbrace.
\end{align}
\end{subequations}
Note that the map $\vh\mapsto (\sum_{T\in\Th} \snorme[1,T]{( \vT, \vdT)}^2)^{\frac12}$ defines a norm on $\Uklhz$.

A key feature of the present HHO method is that the discrete generalized internal variables are computed only at the quadrature points in each mesh cell. We introduce for all $T \in \Th$, the quadrature points $\Qp_T= (\Qp_{T,j})_{1 \leq j \leq m_Q}$, with $\Qp_{T,j} \in T$ for all $1 \leq j \leq m_Q$, and the quadrature weights $\Wp_T= (\Wp_{T,j})_{1 \leq j \leq m_Q}$, with $\Wp_{T,j} \in \Reel$ for all $1 \leq j \leq m_Q$. We denote by $k_Q$ the order of the quadrature. Then, the discrete internal variables are sought in the space
\begin{equation}
\IVGTh := \bigtimes_{T \in\Th}  \IVGS^{m_Q},
\end{equation} 
that is, for all $T\in\Th$, the internal variables attached to $T$ form a vector $\IVGT=(\IVGT(\Qp_{T,j}))_{1\le j\le m_Q}$ with $\IVGT(\Qp_{T,j})\in \IVGS$ for all $1\le j\le m_Q$.

We can now formulate the global discrete problem. We use the following notation for two tensor-valued functions $(\matrice{p}, \matrice{q})$ defined on $T$:
\begin{equation}\label{eq_prods}
\psmd[T]{\matrice{p}}{\matrice{q}} := \sum_{j=1}^{m_Q} \Wp_{T,j} \, \matrice{p}(\Qp_{T,j}) : \matrice{q}(\Qp_{T,j}).
\end{equation}
We also need to consider the case where we know the tensor $\matrice{\tilde p}$ only at the quadrature nodes (we use a tilde to indicate this situation), i.e., we have $\matrice{\tilde p} = (\matrice{\tilde p}(\Qp_{T,j}))_{1\le j\le m_Q} \in (\Rdd)^{m_Q}$. In this case, we slightly abuse the notation by denoting again by $\psmd[T]{\matrice{\tilde p}}{\matrice{q}}$ the quantity equal to the right-hand side of~\eqref{eq_prods}.
The discrete energy functional $\mathcal{E}_h^n: \Uklnhd \times \IVGTh \rightarrow \Reel$ is defined for any pseudo-time step $ 1 \leq n \leq N$ by
\begin{align}\label{eq::Eh}
\mathcal{E}_h^n\big((\vTh, \vFh), \IVGh\big) =& \sum_{T\in\Th}  \left\lbrace \psd[T]{\dQP{ \stressPotW}(\FkT(\vT, \vdT), \IVGT)}{1} - \psv[T]{\loadext^n}{\vT} \right\rbrace - \sum_{F\in\Fhbn} \psv[F]{ \Tn^n}{\vF} \nonumber \\
&+  \sum_{T\in\Th} \frac{\beta}{2} \normev[\dT]{\gamma_{\dT}^{\frac12}\SdTk(\vdT- \vT{}_{|\dT})}^2,
\end{align}
for all $(\vTh, \vFh) \in \Uklnhd$ and $\IVGh \in \IVGTh$, with the local deformation gradient operator $\FkT:\UklT \rightarrow \Pkd(T;\Rdd)$ such that $\FkT(\vT, \vdT) := \matrice{I} + \GkT(\vT, \vdT)$. Moreover, in the second line of~\eqref{eq::Eh}, the stabilization employs a weight of the form $\beta=2\mu \beta_0$ with $\beta_0>0$. In the original HHO method for linear elasticity \cite{DiPEr:2015}, the choice $\beta_0=1$ is considered. In the present setting, the choice for $\beta_0$ is further discussed in Section~\ref{sec:newton} and in Section~\ref{ss:stab}.  The global discrete problem consists in seeking for any pseudo-time step $ 1 \leq n \leq N$, a stationary point of the discrete energy functional: Find the pair of discrete displacements $(\uThn, \uFhn) \in \Uklnhd$ and the discrete internal variables $\IVGh^n \in \IVGTh$ such that, for all $(\dvTh, \dvFh) \in \Uklhz$,
\begin{align}\label{discrete_problem_ptv}
&\sum_{T\in\Th} \psmd[T]{\tilde{\PK}^{n}}{\GkT(\dvT,\dvdT)}
+  \sum_{T\in\Th} \beta \psv[\dT]{\gamma_{\dT}\SdTk(\udTn- \uTn{}_{|\dT})}{\SdTk(\dvdT - \dvT{}_{|\dT})} \nonumber \\
&= \sum_{T\in\Th}  \psv[T]{ \loadext^n}{\dvT} + \sum_{F\in\Fhbn} \psv[F]{\Tn^n}{\dvF},
\end{align}
where for all $T \in \Th$ and all $1 \leq j \leq m_Q$,
\ifCM
\begin{equation}\label{discrete_problem_plast}
( \IVGT^n(\Qp_{T,j}), \tilde{\PK}^n(\Qp_{T,j}), \depmodulePKn(\Qp_{T,j}))  = 
\textrm{FINITE\_PLASTICITY}( \IVGT^{n-1}(\Qp_{T,j}), \FkT(\uT^{n-1},\udT^{n-1})(\Qp_{T,j}),\FkT(\uT^{n},\udT^{n})(\Qp_{T,j})),
\end{equation}%
\else
\begin{multline}\label{discrete_problem_plast}
( \IVGT^n(\Qp_{T,j}), \tilde{\PK}^n(\Qp_{T,j}), \depmodulePKn(\Qp_{T,j}))  = \\
\textrm{FINITE\_PLASTICITY}( \IVGT^{n-1}(\Qp_{T,j}), \FkT(\uT^{n-1},\udT^{n-1})(\Qp_{T,j}),\FkT(\uT^{n},\udT^{n})(\Qp_{T,j})),
\end{multline}%
\fi
with $(\uTh^{n-1}, \uFh^{n-1}) \in \Uklpnhd$ and $\IVGh^{n-1} \in \IVGTh$ given
either from the previous pseudo-time step or the initial condition.
\subsection{Discrete principle of virtual work}
The discrete problem~\eqref{discrete_problem_ptv} expresses the principle of virtual work at the global level, and adapting the ideas introduced in \cite{CoDPE:2016} (see also \cite{AbErPi2018a, BoDPS:2017}), it is possible to infer a local principle of virtual work in terms of face-based discrete tractions that comply with the law of action and reaction. 

Let $\SdTks : \PkF(\FT;\Rd) \rightarrow \PkF(\FT;\Rd)$ be the adjoint operator of $\SdTk$ with respect to the $L^2(\dT;\Rd)$-inner product so that we have $\psv[\dT]{\gamma_{\dT}\SdTk(\vecteur{\theta})}{\SdTk(\vecteur{\zeta})} = \psv[\dT]{\SdTks(\gamma_{\dT}\SdTk(\vecteur{\theta}))}{\vecteur{\zeta}}$ (recall that the weight $\gamma_{\dT}$ is piecewise constant on $\dT$). Let $\PikTmd: (\Rdd)^{m_Q} \rightarrow \Pkd(T;\Rdd )$ denote the $L^2_Q$-orthogonal projector such that for all $\tilde{\matrice{p}} \in (\Rdd)^{m_Q}$, $\psm[T]{\PikTmd(\tilde{\matrice{p}})}{\matrice{q}} = \psmd[T]{\tilde{\matrice{p}}}{\matrice{q}}$ for all $\matrice{q} \in \Pkd(T; \Rdd)$. Finally, for any pseudo-time step $1\le n\le N$ and all $T\in\Th$, let us define the discrete traction:
\begin{equation}\label{eq:traction_stab}
\vecteur{T}_{T}^{n} := \PikTmd(\dPK_T^{n})\SCAL\nT
+ \beta \SdTks(\gamma_{\dT}\SdTk(\udTn- \uTn{}_{|\dT}))\in \PkF(\FT; \Rd),
\end{equation} 
where $\dPK_T^n =  (\dPK_{T}^n(\Qp_{T,j}))_{1 \leq j \leq m_Q} \in( \Rdd)^{m_Q}$ and $(\uTn, \udTn) \in \UklT$.
\begin{lemma}[Equilibrated tractions] \label{lem:equil_sHHO}
Assume that $k_Q\ge 2k$. Then, for any pseudo-time step $1 \leq n \leq N$, the following local principle of virtual work holds true for all $T\in\Th$:
\begin{equation} \label{eq:pwk}
\psmd[T]{\dPK^n_T}{\gradX{\dvT}}-
\psv[\dT]{\vecteur{T}_{T}^n}{\dvT{}_{|\dT}} = \psv[T]{ \loadext^n}{\dvT},
\qquad \forall \dvT \in \Pld(T;\Rd),
\end{equation}
where the discrete tractions $\vecteur{T}_{T}^n$ defined by~\eqref{eq:traction_stab} satisfy the following law of action and reaction for all $F\in\Fhi\cup\Fhbn$:
\begin{subequations}\label{eq:balance}
\begin{alignat}{2}
&\vecteur{T}_{T_-|F}^n + \vecteur{T}_{T_+|F}^n = \vecteur{0},
&\quad&\text{if $F\in\Fhi$ with $F=\partial T_- \cap \partial T_+ \cap H_F$},\\
&\vecteur{T}_{T|F}^n  = \PikFv(\Tn^n),&\quad&\text{if $F\in\Fhbn$ with $F=\partial T\cap \Bn\cap H_F$}.
\end{alignat}
\end{subequations}  
\end{lemma}
\subsection{Nonlinear solver and implementation}
\label{sec:newton}
The nonlinear problem \eqref{discrete_problem_ptv}-\eqref{discrete_problem_plast} arising at any pseudo-time step $1 \leq n \leq N$ is solved using a Newton's method. Given  $(\uTh^{n-1}, \uFh^{n-1}) \in \Uklpnhd$ and $\IVGh^{n-1} \in \IVGTh$ from the previous pseudo-time step or the initial condition, the Newton's method is initialized by setting $(\uTh^{n,0}, \uFh^{n,0})=(\uTh^{n-1}, \uFh^{n-1})$, up to the update of the Dirichlet condition, and $\IVGh^{n,0}= \IVGh^{n-1}$. Then, at each Newton's step $i\ge0$, one computes the incremental displacement $(\duTh^{n,i}, \duFh^{n,i}) \in \Uklhz$ and updates the discrete displacement as $(\uTh^{n,i+1} \uFh^{n,i+1}) = (\uTh^{n,i}, \uFh^{n,i})+(\duTh^{n,i}, \duFh^{n,i})$. The linear system of equations to be solved is
\begin{align}\label{eq_stiffness_matrix}
& \hphantom{+} \sum_{T\in\Th} \psmd[T]{\depmodulePKni: \GkT(\duT^{n,i},\dudT^{n,i})}{\GkT(\dvT,\dvdT)} \nonumber \\ 
&+  \sum_{T\in\Th} \beta \psv[\dT]{\gamma_{\dT}\SdTk(\dudT^{n,i} - 
\duT^{n,i}{}_{|\dT})}{\SdTk(\dvdT - \dvT{}_{|\dT})} \nonumber \\ 
&= -R_h^{n,i}(\dvTh,\dvFh),
\end{align}
for all $(\dvT,\dvdT)\in\Uklhz$, where for all $T \in \Th$ and all $1 \leq j \leq m_Q$,
\begin{equation}
 (\IVGT^{n,i}(\Qp_{T,j}), \dPK^{n,i}(\Qp_{T,j}), \depmodulePKni(\Qp_{T,j})) = \textrm{FINITE\_PLASTICITY}(\IVG_{T,j}^{n-1},\Fdef_{T,j}^{n-1},\Fdef_{T,j}^{n,i}),
\end{equation}
with $\IVG_{T,j}^{n-1} := \IVGT^{n-1}(\Qp_{T,j})$, $\Fdef_{T,j}^{n,i} := \FkT(\uT^{n,i},\udT^{n,i})(\Qp_{T,j})$, 
$\Fdef_{T,j}^{n-1} := \FkT(\uT^{n-1},\udT^{n-1})(\Qp_{T,j})$, and the residual term
\begin{align}
R_h^{n,i}(\dvTh, \dvFh) 
={}& \sum_{T\in\Th} \psmd[T]{\dPK^{n,i}}{\GkT(\dvT,\dvdT)} - \sum_{T\in\Th}  \psv[T]{ \loadext^n}{\dvT} - \sum_{F\in\Fhbn} \psv[F]{\Tn^n}{\dvF}  \nonumber \\
&+  \sum_{T\in\Th} \beta \psv[\dT]{\gamma_{\dT}\SdTk(\udT^{n,i} - \uT^{n,i}{}_{|\dT})}{\SdTk(\dvdT - \dvT{}_{|\dT})}.
\end{align}
The assembling of the stiffness matrix resulting from the left-hand side of~\eqref{eq_stiffness_matrix} is local (and thus fully parallelizable). The discrete internal variables $\IVGh^{n} \in \IVGTh$ are updated at the end of each pseudo-time step. Moreover, since the consistent elastoplastic tangent modulus $\epmodulePK$ has major symmetries, its eigenvalues are real. Let us set $\theta_{\Th, Q} := \min_{(T,j) \in \Th\times \{1,\ldots,m_Q\}} \theta^{\min}(\depmodulePK(\Qp_{T,j}))$, where $\theta^{\min}(\tenseur{4}{{\mathbb{M}}})$ denotes the smallest eigenvalue of a symmetric fourth-order tensor $\tenseur{4}{{\mathbb{M}}}$. The following result shows that the linear system~\eqref{eq_stiffness_matrix} arising at each Newton's step is coercive under the simple choice $\beta_0>0$ on the stabilization parameter if $\depmodulePK$ is positive-definite. Notice that strain-hardening plasticity is not a sufficient condition for positive-definiteness of $\depmodulePK$ (only for $\depmoduleT$) since in finite elastoplasticity, geometrical nonlinearities also exist.
\begin{theorem}[Coercivity]\label{th::coer_newton}
Assume that $k_Q\ge 2k$ and that all the quadrature weights are positive. Moreover, assume that  the consistent elastoplastic tangent modulus $\epmodulePK$ is positive-definite, i.e., $\theta_{\Th, Q} >0$. Then, the linear system~\eqref{eq_stiffness_matrix} in each Newton's step is coercive for all $\beta_0 > 0$, i.e., there exists $C_{\rm ell}>0$, independent of $h$, such that for all $(\vTh, \vFh) \in \Uklhz$,
\begin{multline}\label{eq_stable}
\sum_{T\in\Th} \psmd[T]{\depmodulePK: \GkT(\vT,\vdT)}{\GkT(\vT,\vdT)}+  \sum_{T\in\Th} \beta \normev[\dT]{\gamma_{\dT}^{\frac12}\SdTk(\vdT- \vT{}_{|\dT})}^2 \\
\geq C_{\rm ell} \min\left(\beta_0, \frac{\theta_{\Th, Q}}{2\mu} \right) 2\mu \sum_{T\in\Th} \snorme[{1,T}]{(\vT, \vdT)}^2.
\end{multline}
\end{theorem}
The proof follows directly from \cite[Theorem 6]{AbErPi2018a}. Note that  Theorem~\ref{th::coer_newton} states that one iteration of the Newton's process is stable under a positive-definiteness assumption but does not state that a solution to the whole Newton's process exists. The existence of such a solution has been showed in the context of nonlinear elliptic equations for dG methods \cite{Yadav2013}. Moreover, if the consistent elastoplastic tangent modulus $\epmodulePK$ is no longer positive-definite for at least one quadrature point which is a possibility in finite elastoplasticity since geometrical nonlinearities also exist, then Theorem~\ref{th::coer_newton} is no longer valid (even if $\depmoduleT$ remains positive-definite). Moreover, a reasonable choice of the stabilization parameter appears to be $\beta_0 \geq \max(1,\frac{\theta_{\Th, Q}}{2\mu})$ because $\beta_0 = 1$ is a natural choice for linear elasticity \cite{DiPEr:2015} and the choice $\beta_0 \geq  \frac{\theta_{\Th, Q}}{2\mu}$ allows one to adjust the stabilization parameter if the evolution is plastic. We investigate numerically the choice of $\beta_0$ in
Section~\ref{ss:stab}.

From a numerical  point of view, as is classical with HHO methods \cite{DiPEr:2015, DiPEL:2014}, and more generally with hybrid approximation methods, the cell unknowns $\duT^{n,i}$ in~\eqref{eq_stiffness_matrix} can be eliminated locally by using a static condensation (or Schur complement) technique. This allows one to reduce~\eqref{eq_stiffness_matrix} to a linear system in terms of the face unknowns only. The reduced system is of size $N_{\Fh} \times  d{k+d-1\choose d-1}$, where $N_{\Fh}$ denotes the number of mesh faces (unknowns attached to Dirichlet boundary faces can be eliminated by enforcing the boundary condition explicitly). The implementation of HHO methods is realized using the open-source library \texttt{DiSk++} which provides generic programming tools for the implementation of HHO methods and is available online\footnote{ \texttt{https://github.com/wareHHOuse/diskpp}}.  We refer the reader to  \cite{CicDE:2018} and  \cite[Section 3.6]{AbErPi2018a} for further aspects about the implementation.
\section{Numerical examples}\label{sec::sec_numexp}
The goal of this section is to evaluate the proposed HHO method on two- and three-dimensional benchmarks from the literature: (\textit{i}) a necking of a 2D rectangular bar subjected to uniaxial extension, (\textit{ii}) a Cook's membrane subjected to bending, (\textit{iii}) a torsion of a square-section bar, and (\textit{iv}) a quasi-incompressible sphere under internal pressure. We compare our results to the analytical solution whenever available or to numerical results obtained using the industrial open-source FEM software \CA \cite{CodeAster}. In this case, we consider a linear, resp. quadratic, cG formulation, referred to as Q1, resp. T2 or Q2 when full integration is used, or,  Q2\_RI when reduced integration is used, depending on the mesh, and a three-field mixed formulation in which the unknowns are the displacement, the pressure and the volumetric strain fields referred to as UPG \cite{AlAkhrass2014}; in the UPG method, the displacement field is quadratic, whereas both the pressure and the volumetric strain fields are linear. The conforming Q1, T2 and Q2 methods with full integration, contrary to  Q2\_RI method with reduced integration in most of the situations, are known to present volumetric locking due to plastic incompressibility, whereas  the UPG method is known to be robust but costly. Numerical results obtained using the UPG method  are used as a reference solution whenever an analytical solution is not available. 

The nonlinear isotropic  plasticity model with a von Mises yield criterion described in Section~\ref{ss:nl_hardening} is used for the test cases. 
For the first three test cases, strain-hardening plasticity is considered with the following material parameters: Young modulus $E =206.9~\GPa$, Poisson ratio $\nu=0.29$, hardening parameter $H = 129.2~\MPa$, initial yield stress $\sigma_{y,0} = 450~\MPa$, infinite yield stress $\sigma_{y,\infty} = 715~\MPa$, and saturation parameter $\delta = 16.93$. For the fourth case, perfect plasticity is considered with the following material parameters: Young modulus $E =28.85~\MPa$, Poisson ratio $\nu=0.499$, hardening parameter $H =0~\MPa$, initial and infinite yield stresses $\sigma_{y,0} = \sigma_{y,\infty} = 6~\MPa$, and saturation parameter $\delta = 0$.  Moreover, for the two-dimensional test cases (\textit{i}) and (\textit{ii}), we assume additionally a plane strain condition. In the numerical experiments reported in this section, the stabilization parameter is taken to be $\beta = 2 \mu$ ($\beta_0=1$), and all the quadratures use positive weights. In particular, for the HHO method, we employ a quadrature of order $k_Q=2k$ for the behavior cell integration. We employ the notation HHO($k$, $l$) when using face polynomials of order $k$ and cell polynomials of order $l$.  

In Section~\ref{ss::num_inv}, we perform further numerical investigations to test other aspects of HHO methods such as the support of general meshes with possibly non-conforming interfaces, the possibility of considering the lowest-order case $k=0$, and the dependence on the stabilization parameter $\beta$. 
\subsection{Necking of a 2D rectangular bar}\label{ss::bar}
\begin{figure}
        \centering
         \subfloat[]{
        \centering
        \includegraphics[scale=0.47]{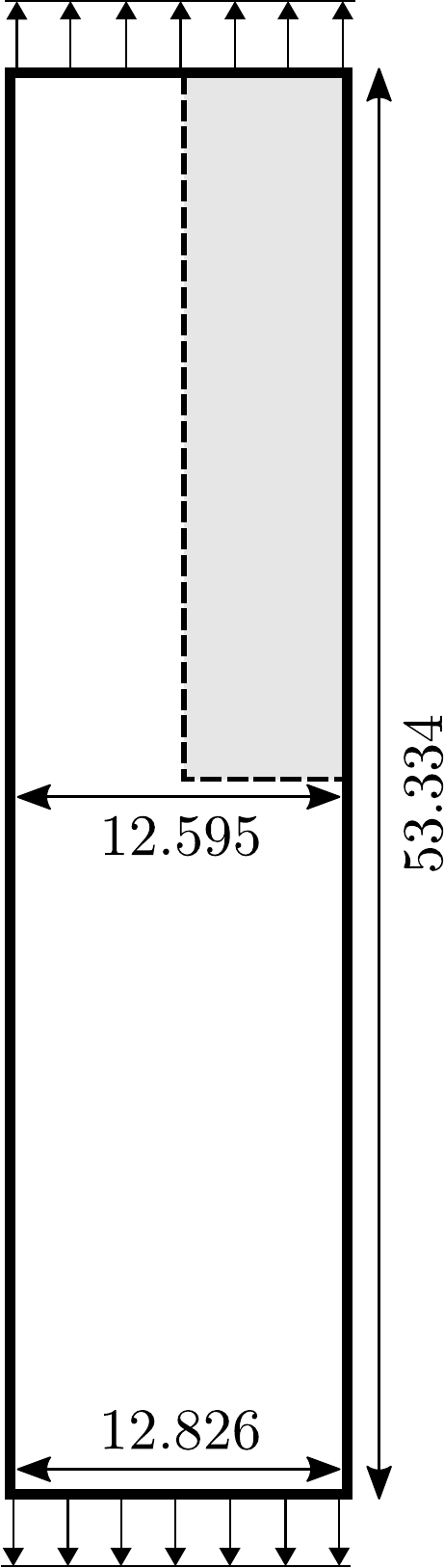}
         \label{fig::necking_geom}
  }
    ~ 
         \subfloat[]{
        \centering
       \includegraphics[scale=0.38, trim= 530 0 670 0, clip=true]{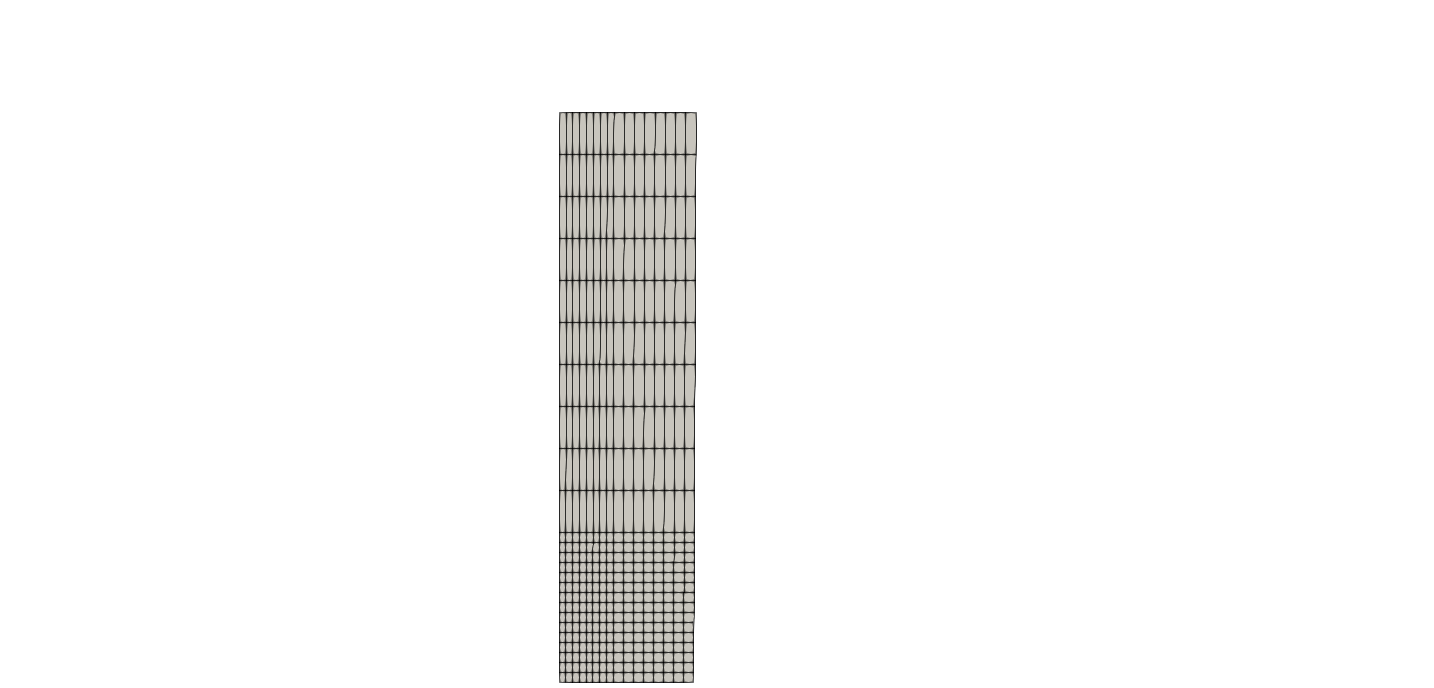}
         \label{fig::necking_mesh}
  }
    ~ 
    \subfloat[]{
        \centering
	    \includegraphics[scale=1.0]{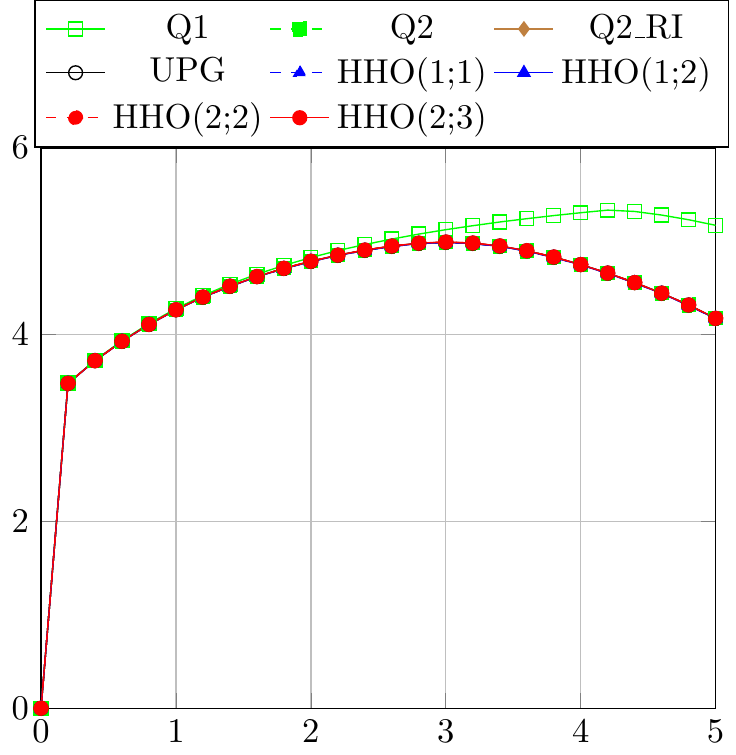}
	     \label{fig::necking_load}
    }
    \caption{Necking of a 2D rectangular bar: (a) Geometry and boundary conditions (dimensions in $\mm$). For symmetric reasons only the upper right-quarter of the bar is considered (b) Mesh composed of 400 quadrangles used for the computations. (c) Vertical reaction versus imposed displacement for the different methods (all the curves overlap except that for Q1) .}
\end{figure}

\begin{figure}
    \centering
    \subfloat[Q1]{
        \centering
        \includegraphics[scale=0.33, trim= 620 0 550 0, clip=true]{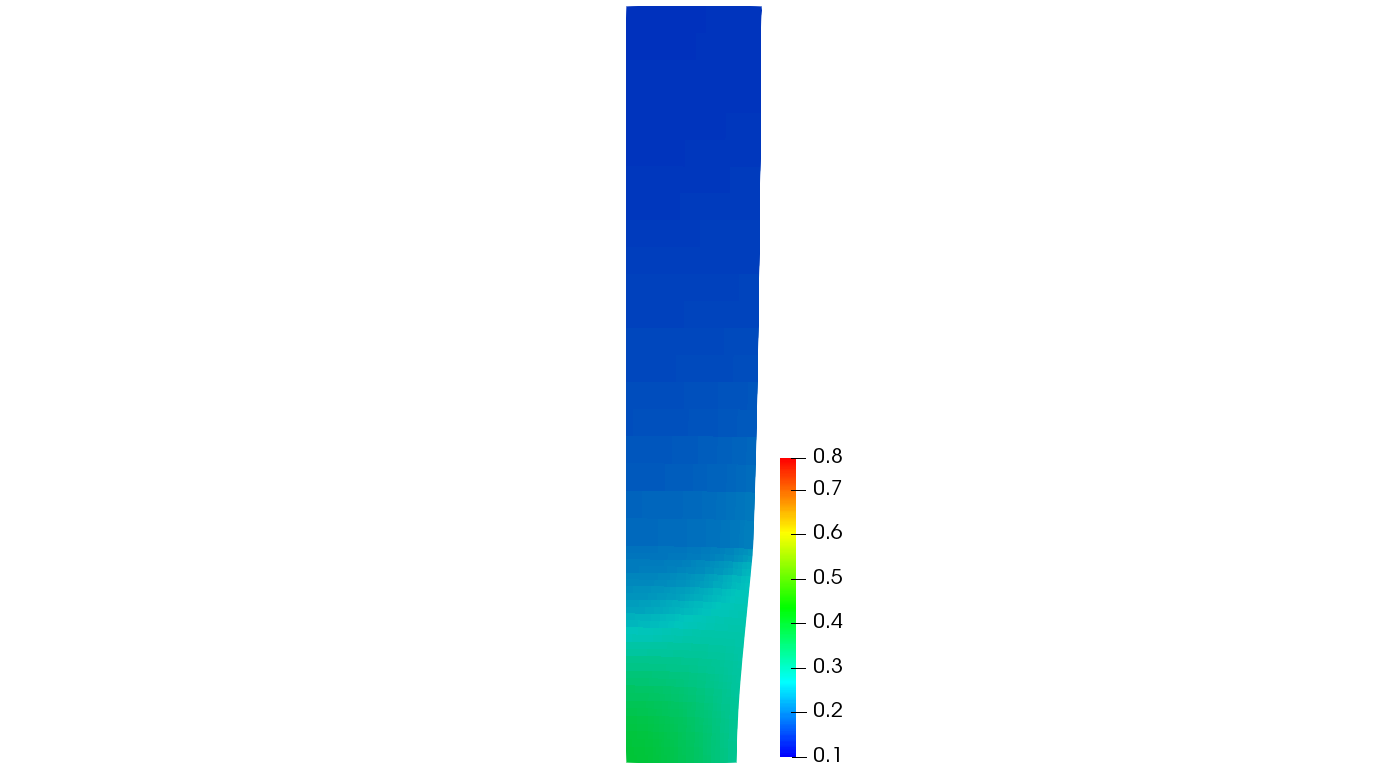}
  }
    ~ 
    \subfloat[Q2]{
        \centering
        \includegraphics[scale=0.33, trim= 620 0 550 0, clip=true]{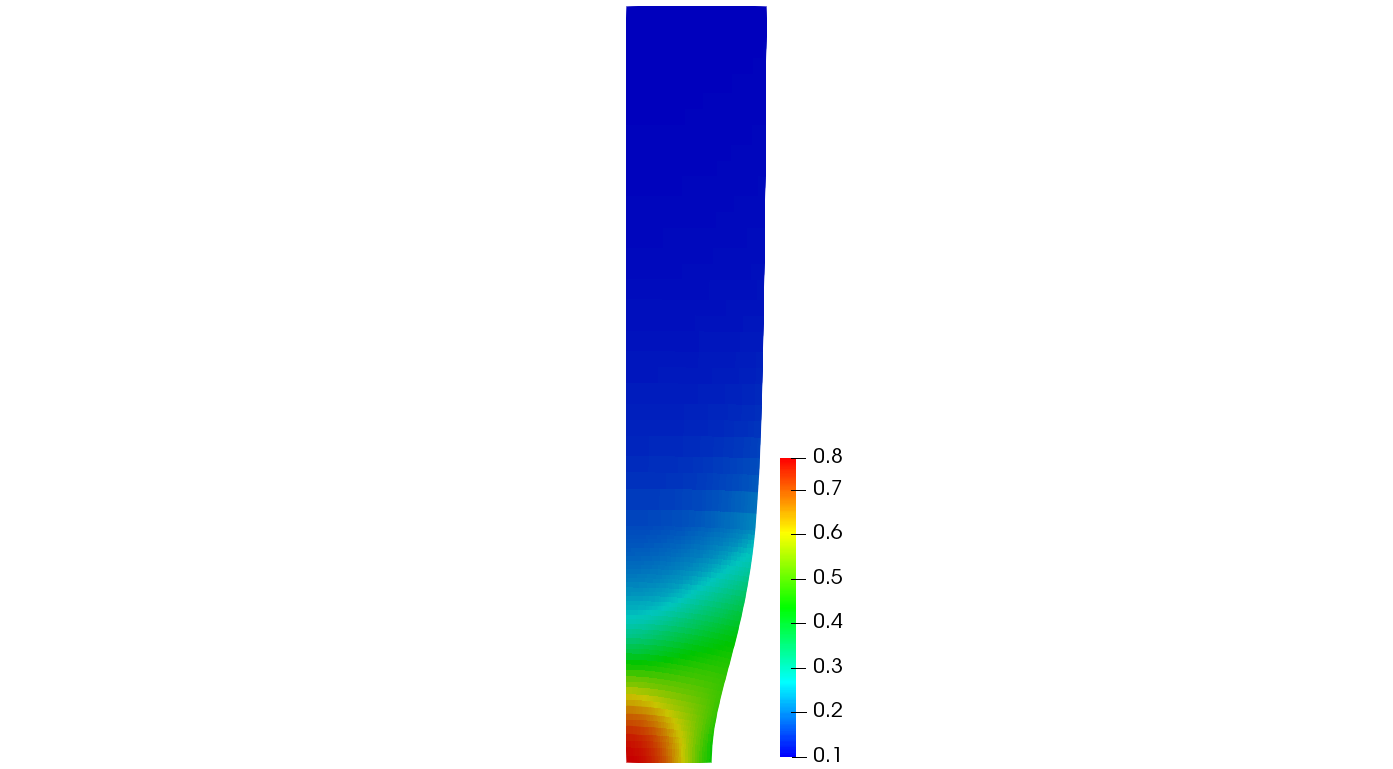}
  }
      ~ 
    \subfloat[Q2\_RI]{
        \centering
        \includegraphics[scale=0.33, trim= 620 0 550 0 clip=true]{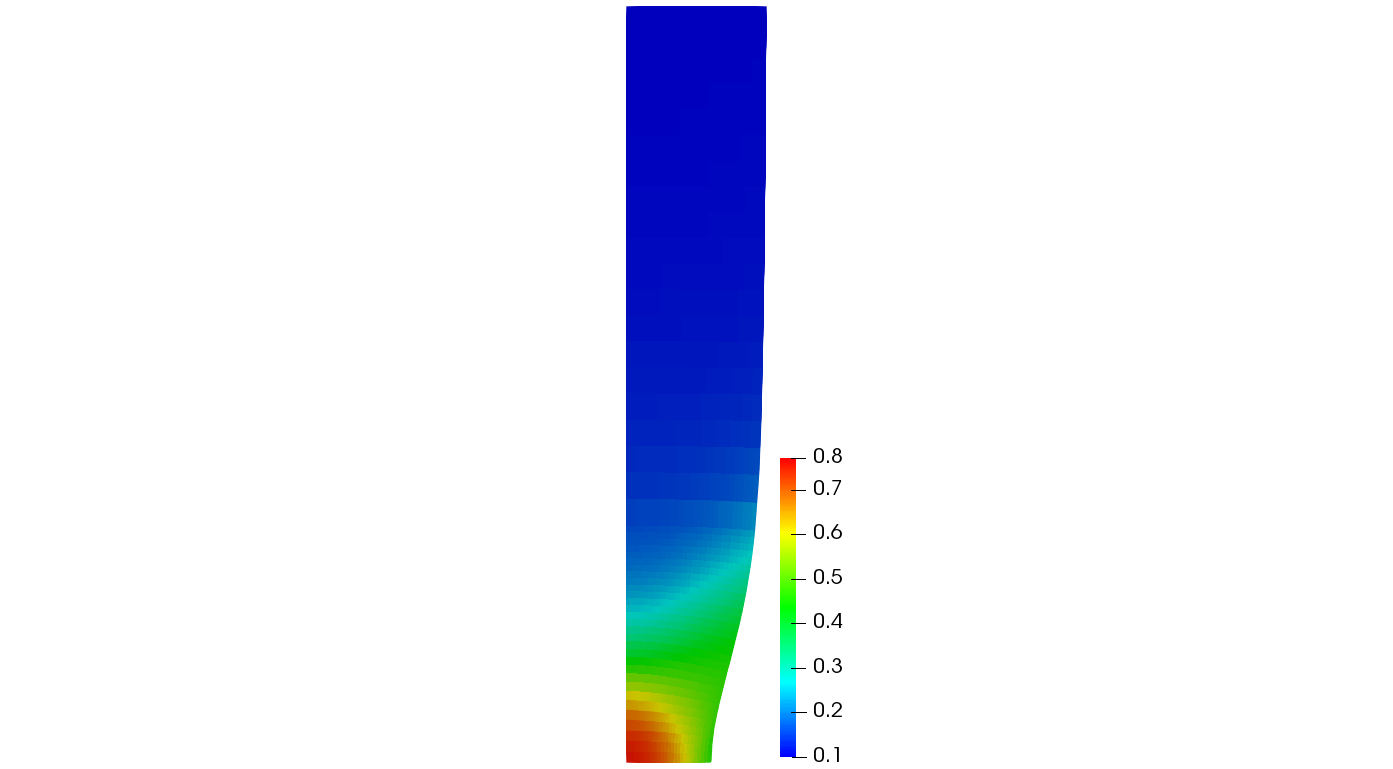}
  }
  ~ 
    \subfloat[UPG]{
        \centering
        \includegraphics[scale=0.33, trim= 620 0 550 0, clip=true]{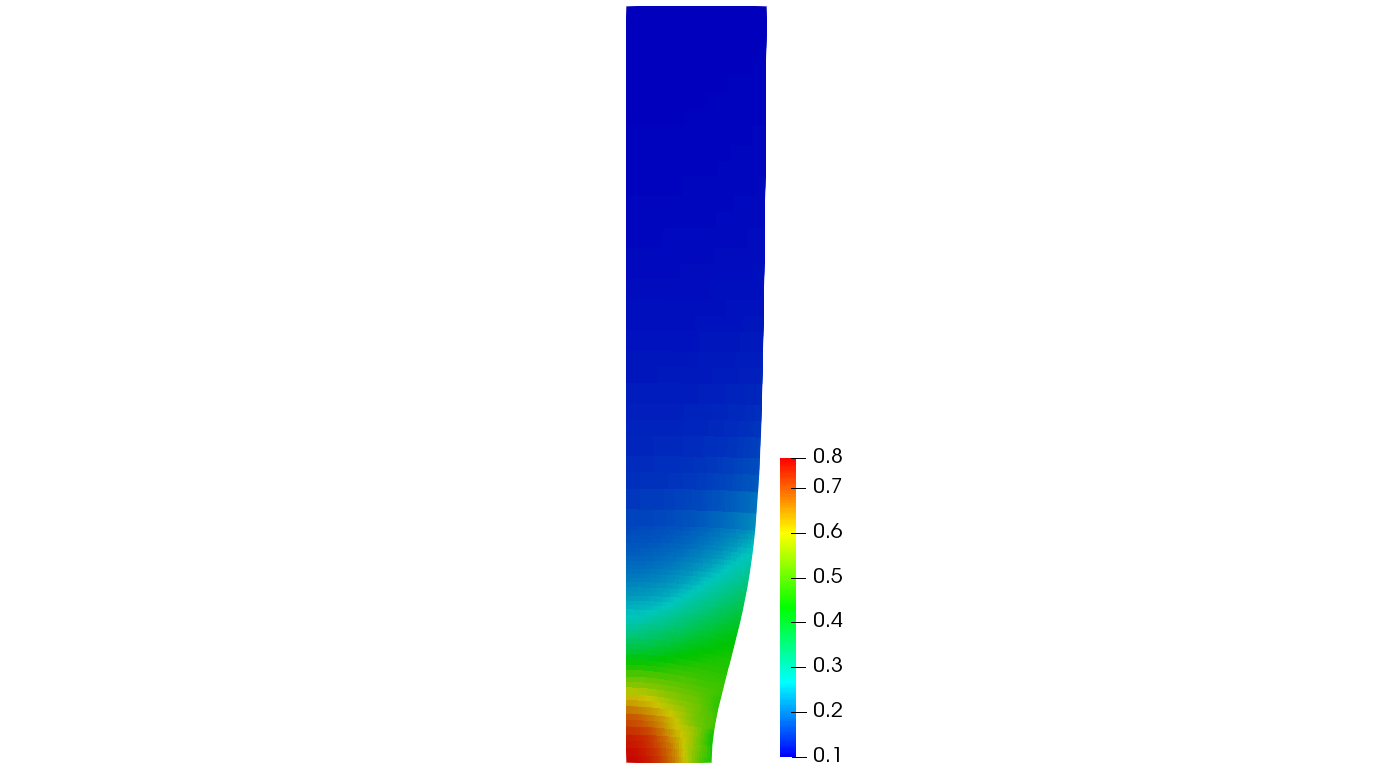}
  }
  
      \subfloat[HHO(1;1)]{
        \centering
        \includegraphics[scale=0.33, trim= 620 0 550 0, clip=true]{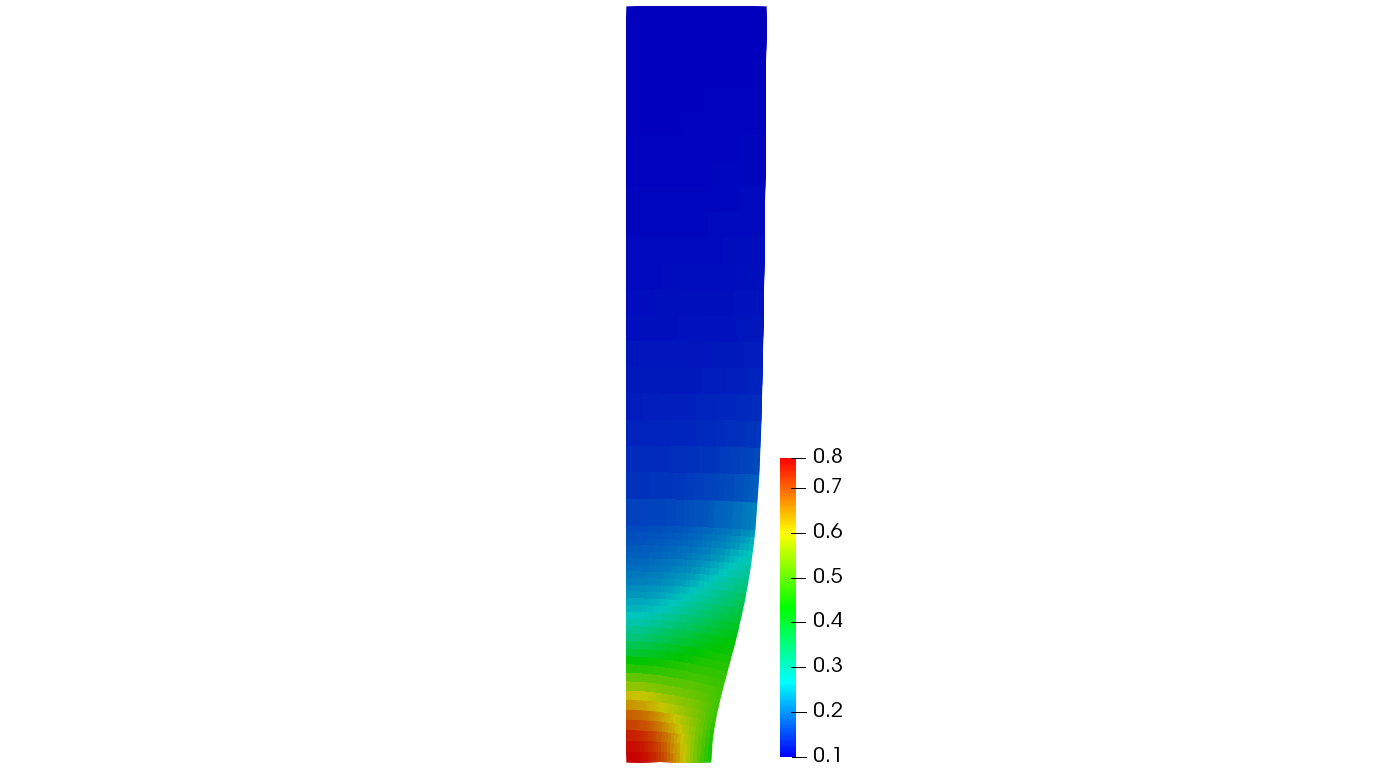}
  } 
 ~ 
    \subfloat[HHO(1;2)]{
        \centering
        \includegraphics[scale=0.33, trim= 620 0 550 0, clip=true]{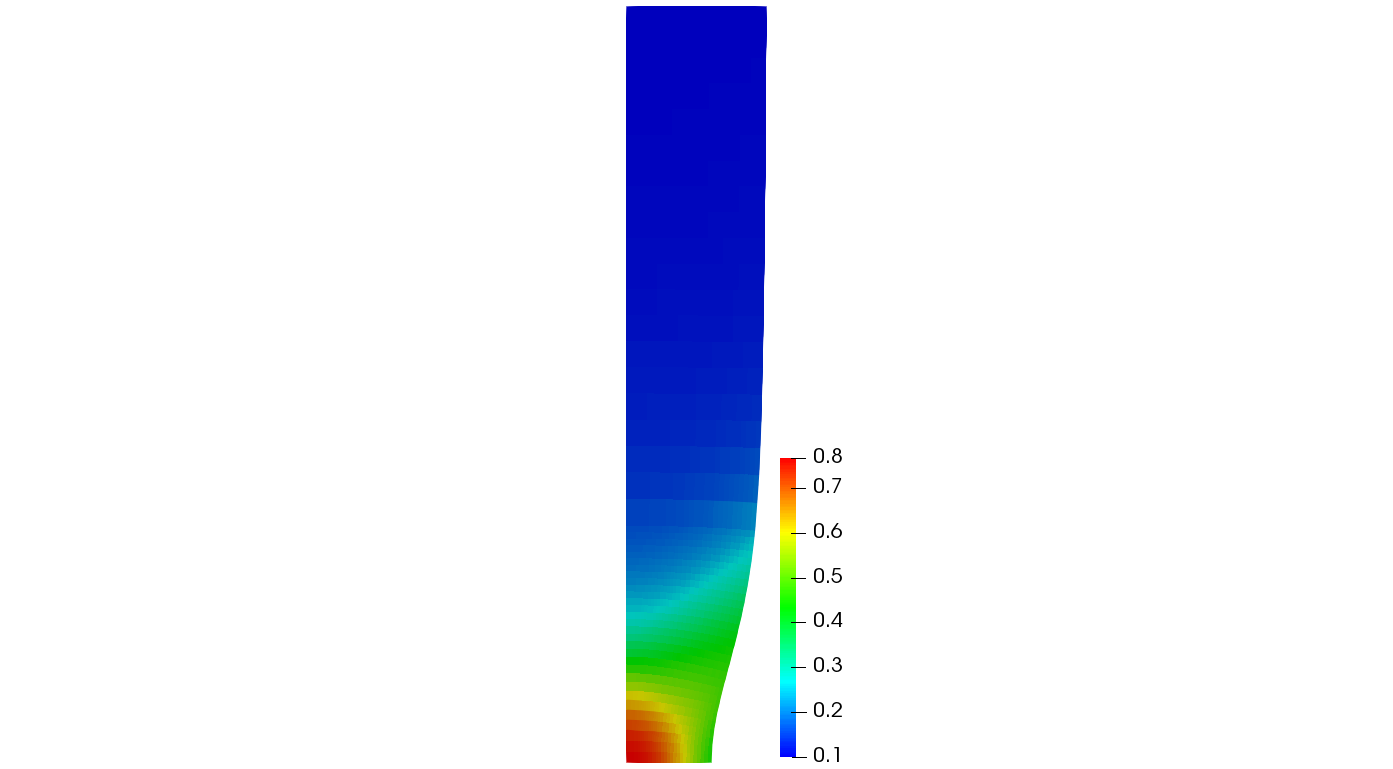}
  }
    ~ 
    \subfloat[HHO(2;2)]{
        \centering
        \includegraphics[scale=0.33, trim= 620 0 550 0, clip=true]{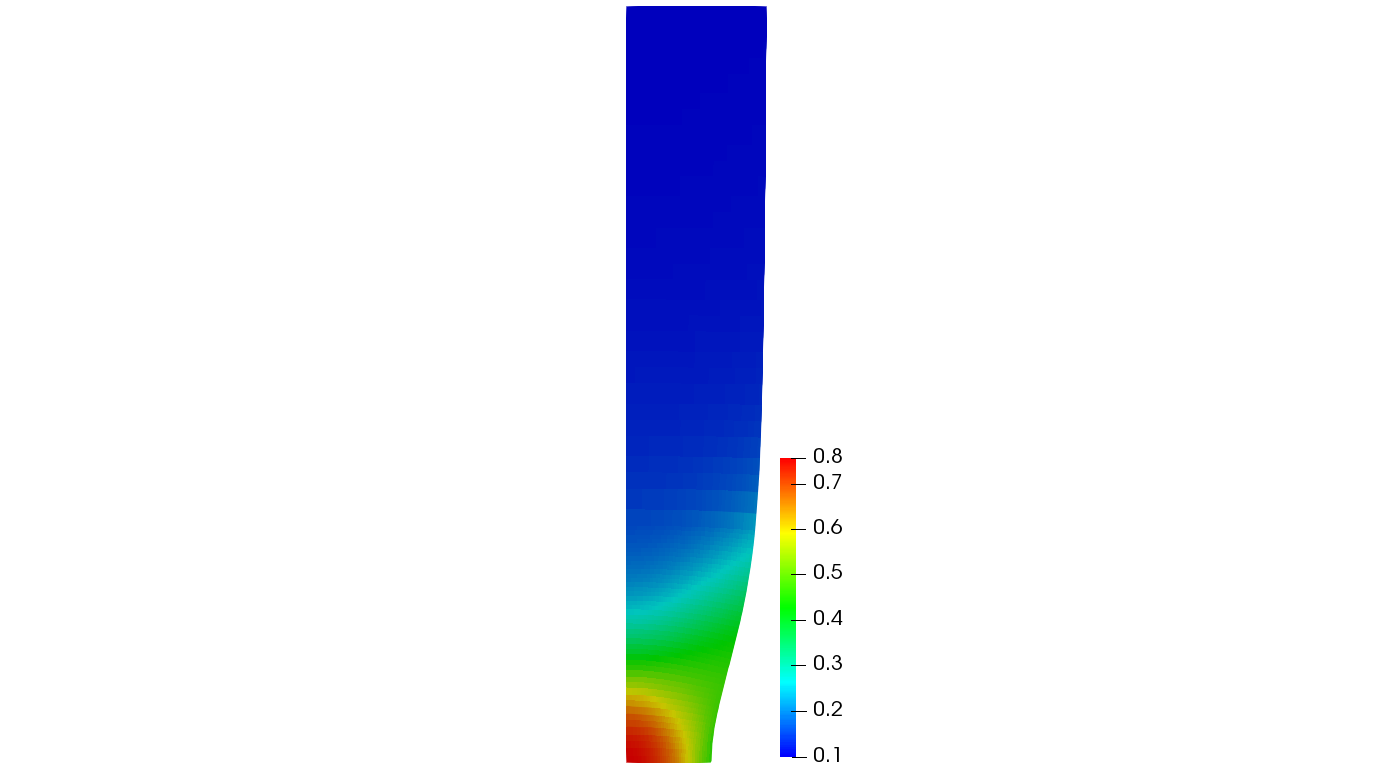}
  }
  ~ 
    \subfloat[HHO(2;3)]{
        \centering
        \includegraphics[scale=0.33, trim= 620 0 550 0, clip=true]{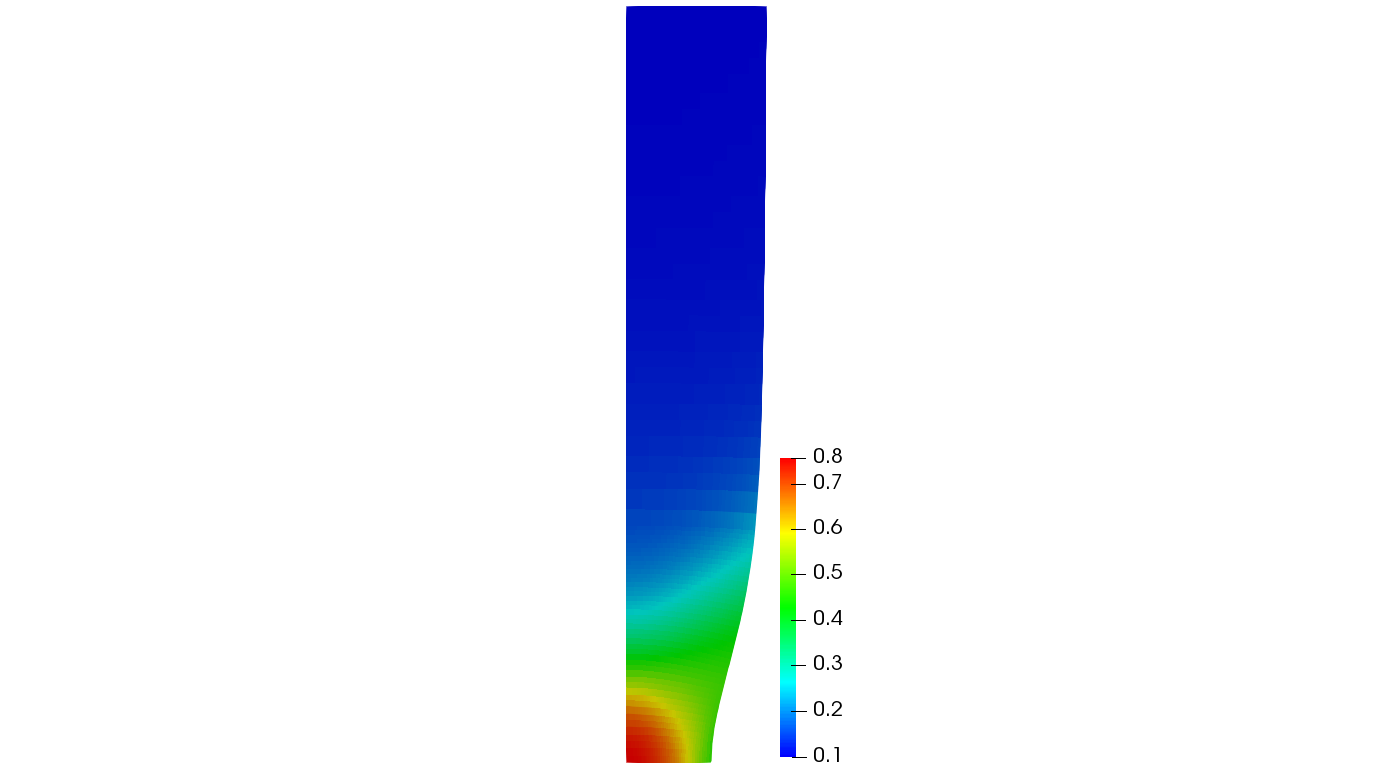}
  }
    \caption{Necking of a 2D rectangular bar: Equivalent plastic strain $p$ at the quadrature points on the final configuration for the different methods.}
        \label{fig::necking_p}
\end{figure}

\begin{figure}
    \centering
     \subfloat[Q1]{
        \centering
        \includegraphics[scale=0.33, trim= 620 0 525 0, clip=true]{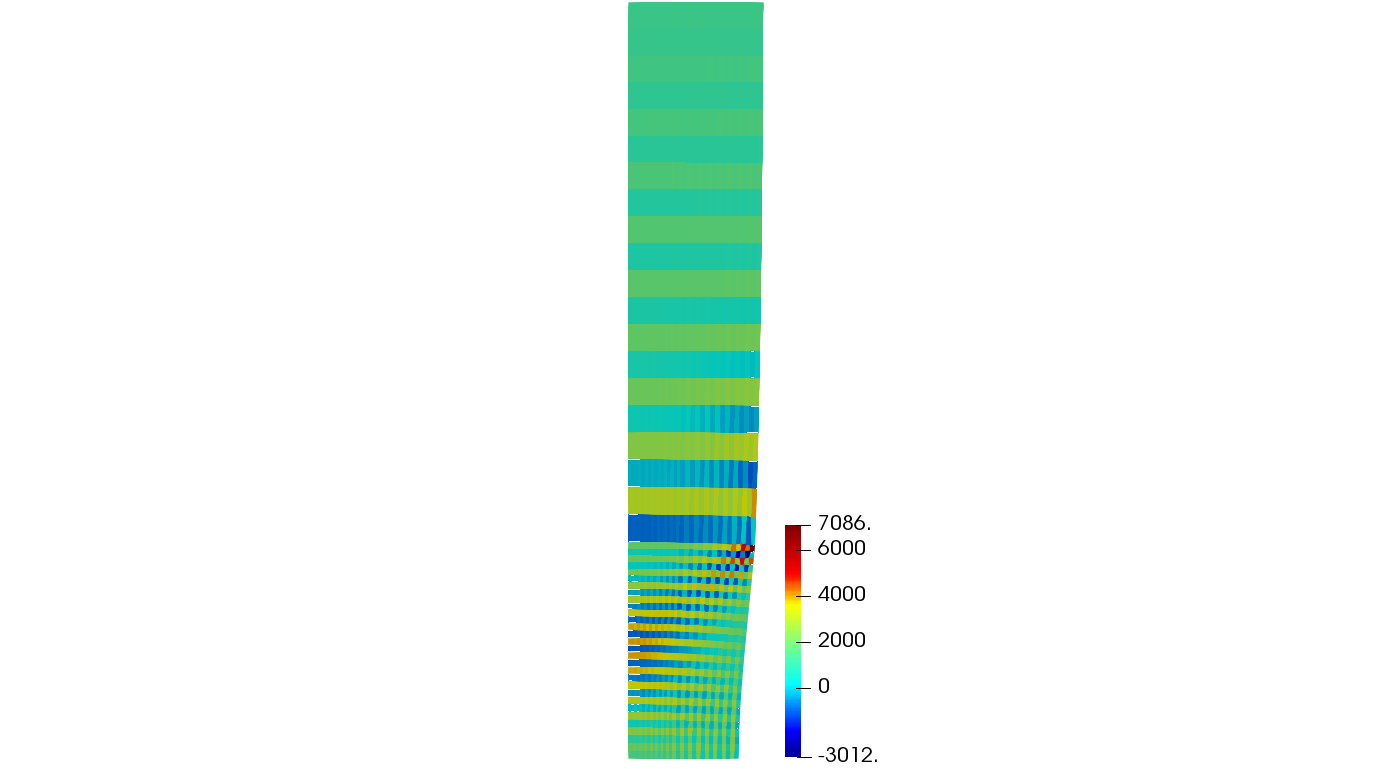}
  }
    ~ 
    \subfloat[Q2]{
        \centering
        \includegraphics[scale=0.33, trim= 620 0 525 0, clip=true]{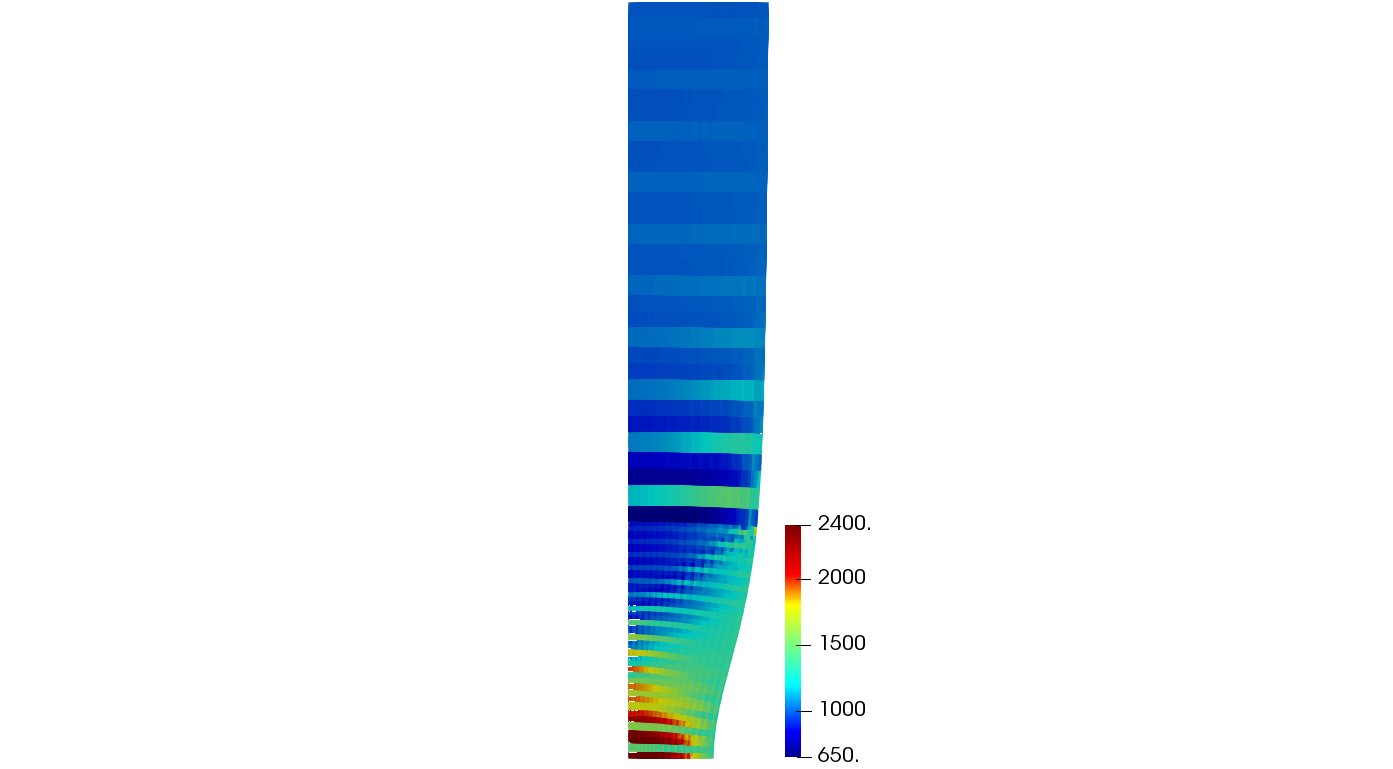}
  }
        ~ 
    \subfloat[Q2\_RI]{
        \centering
        \includegraphics[scale=0.33, trim= 620 0 525 0, clip=true]{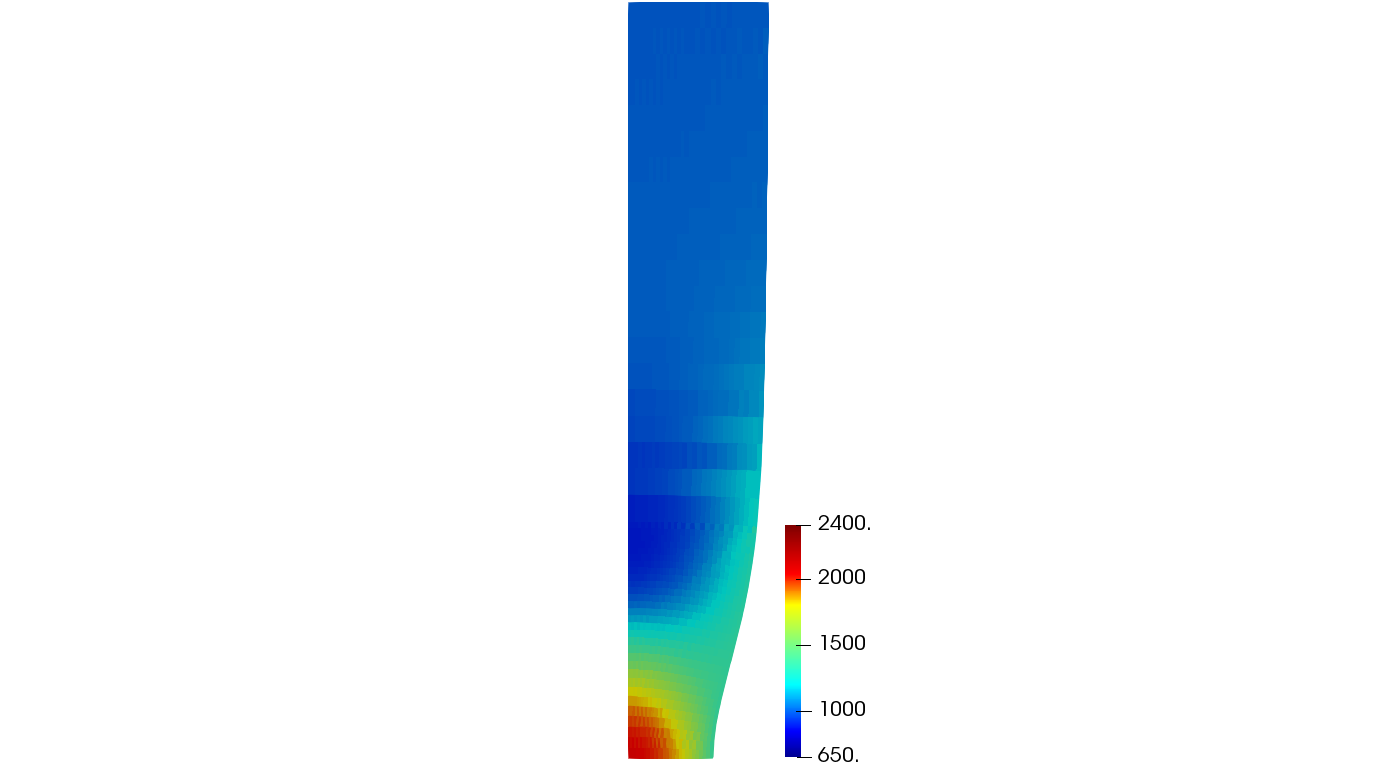}
  }
  ~ 
    \subfloat[UPG]{
        \centering
        \includegraphics[scale=0.33, trim= 620 0 525 0, clip=true]{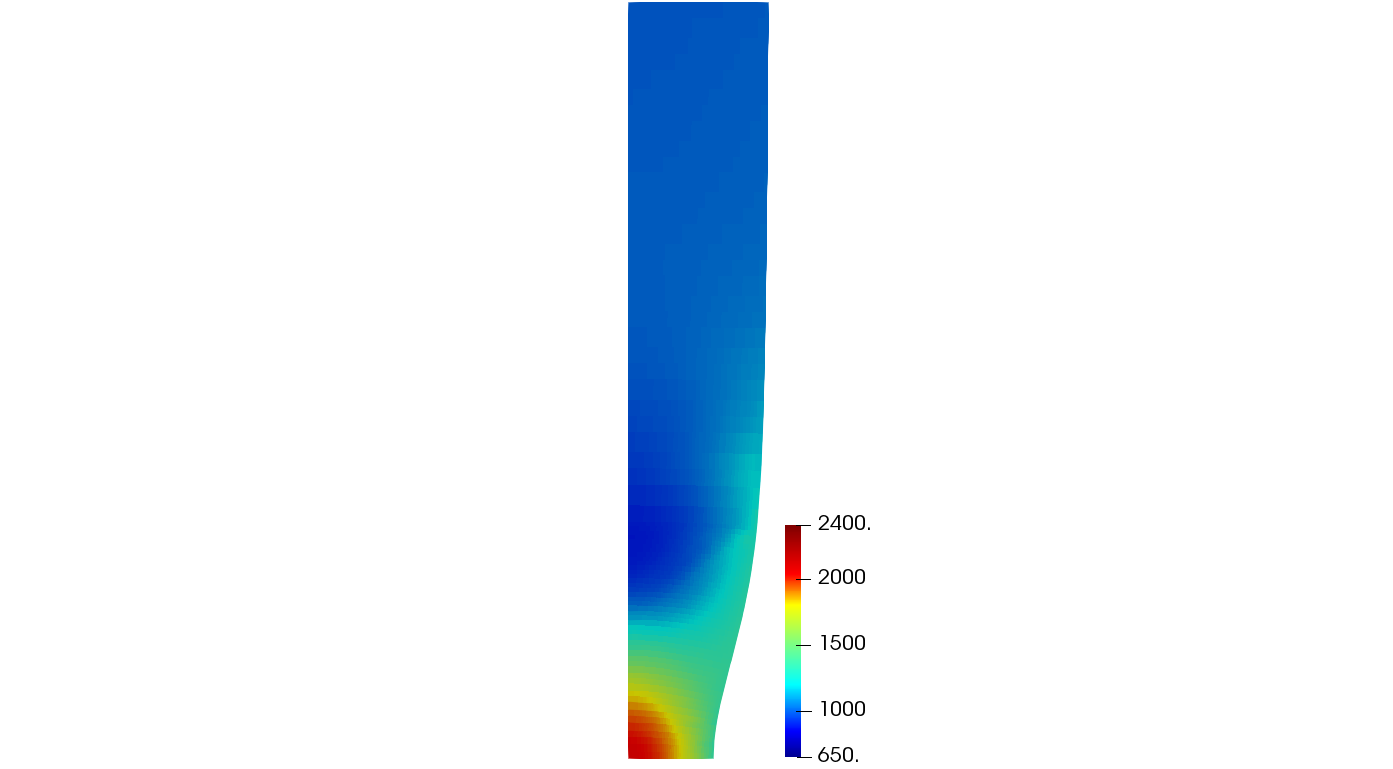}
  }
   
     \subfloat[HHO(1;1)]{
        \centering
        \includegraphics[scale=0.33, trim= 620 0 525 0, clip=true]{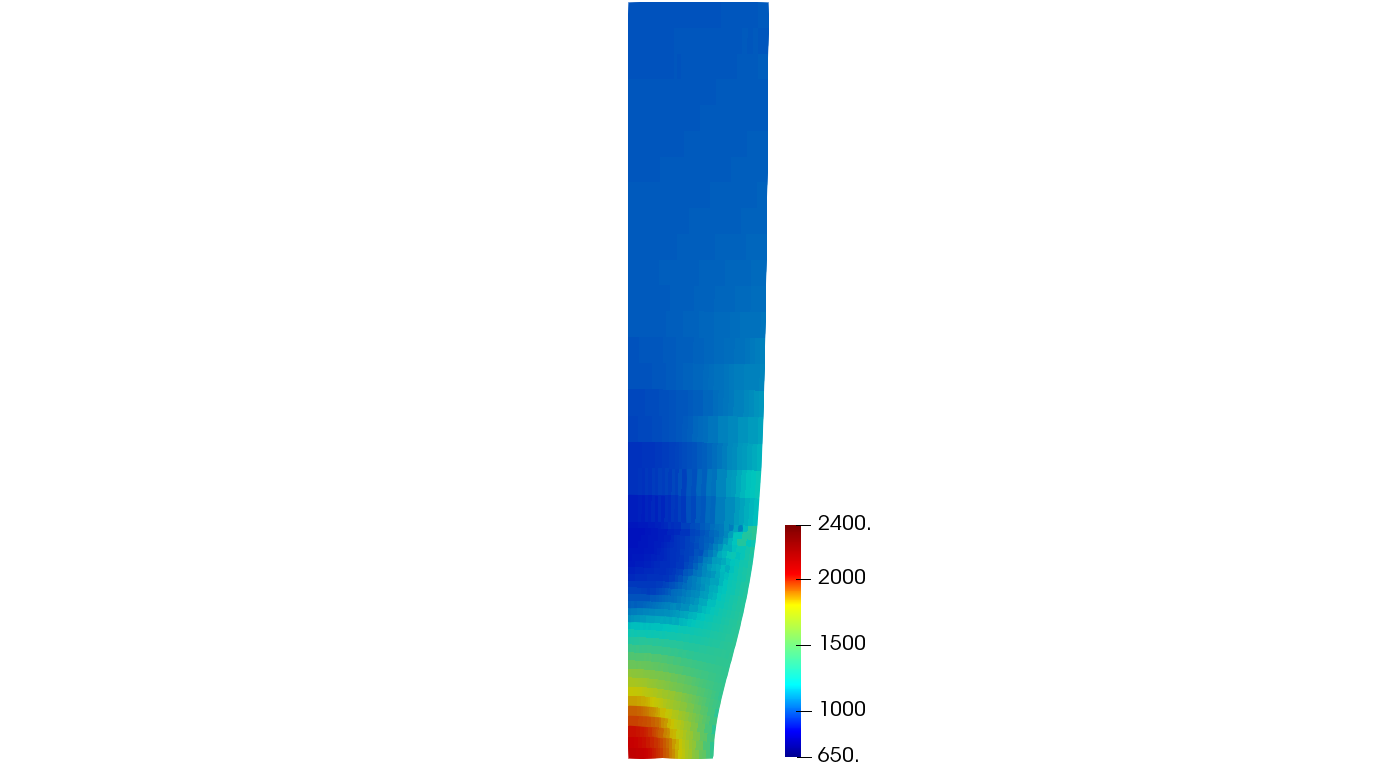}
  }
    ~ 
    \subfloat[HHO(1;2)]{
        \centering
        \includegraphics[scale=0.33, trim= 620 0 525 0, clip=true]{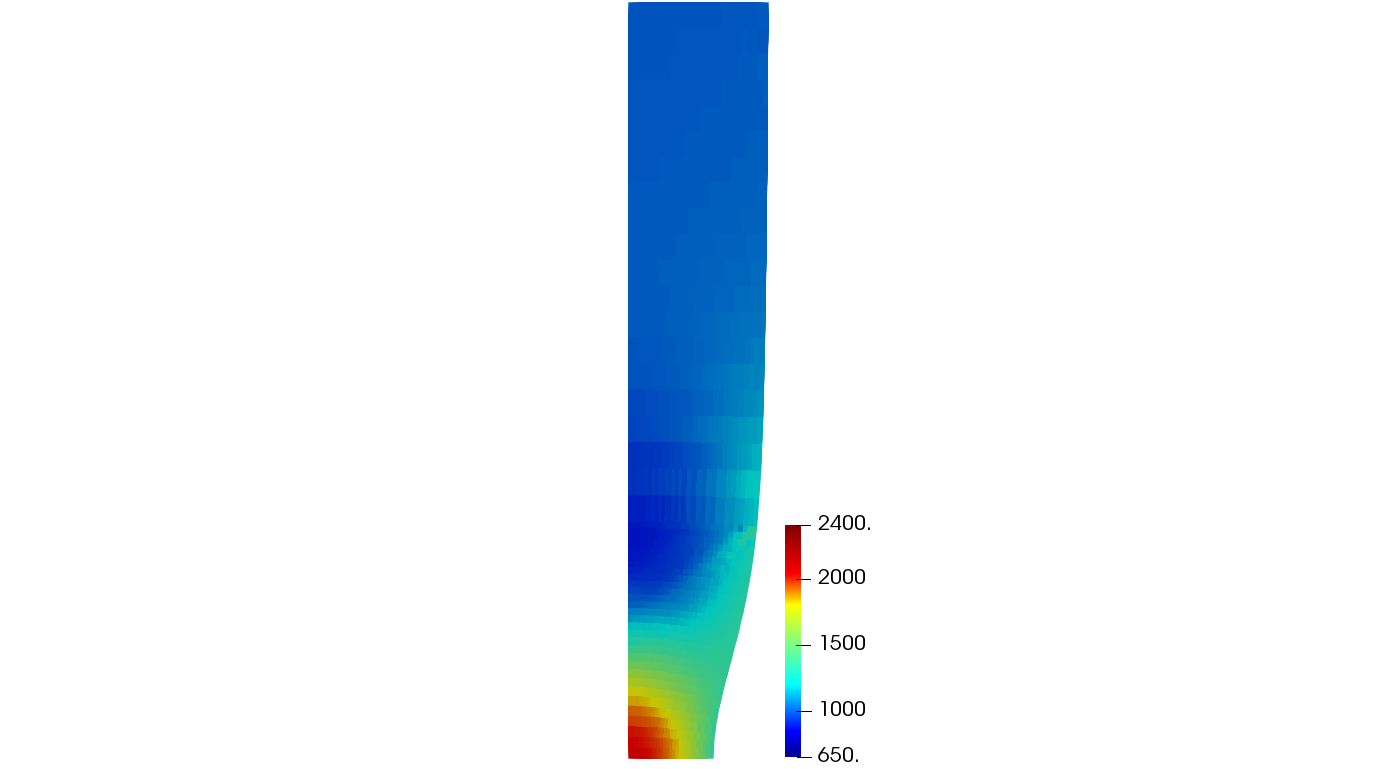}
  }
   ~ 
      \subfloat[HHO(2;2)]{
        \centering
        \includegraphics[scale=0.33, trim= 620 0 525 0, clip=true]{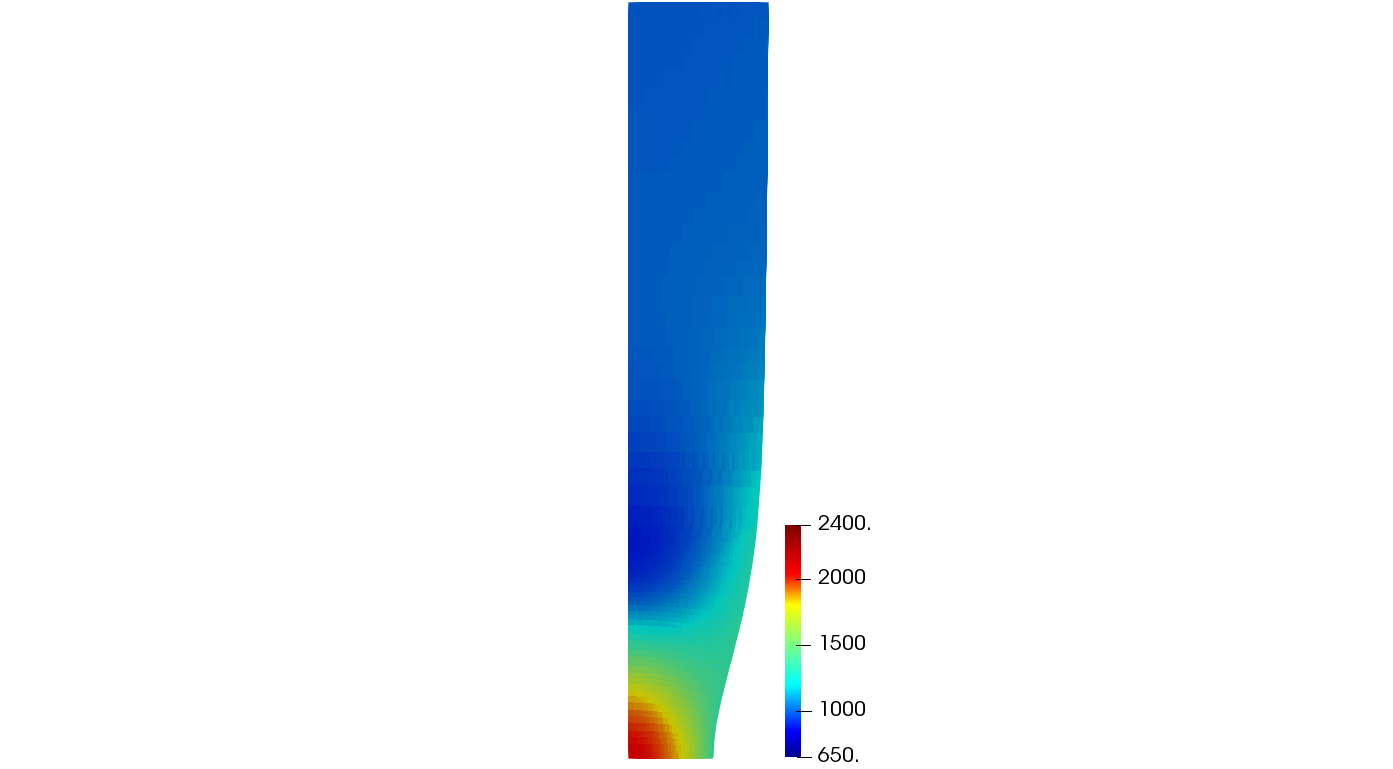}
  }
  ~ 
    \subfloat[HHO(2;3)]{
        \centering
        \includegraphics[scale=0.33, trim= 620 0 525 0, clip=true]{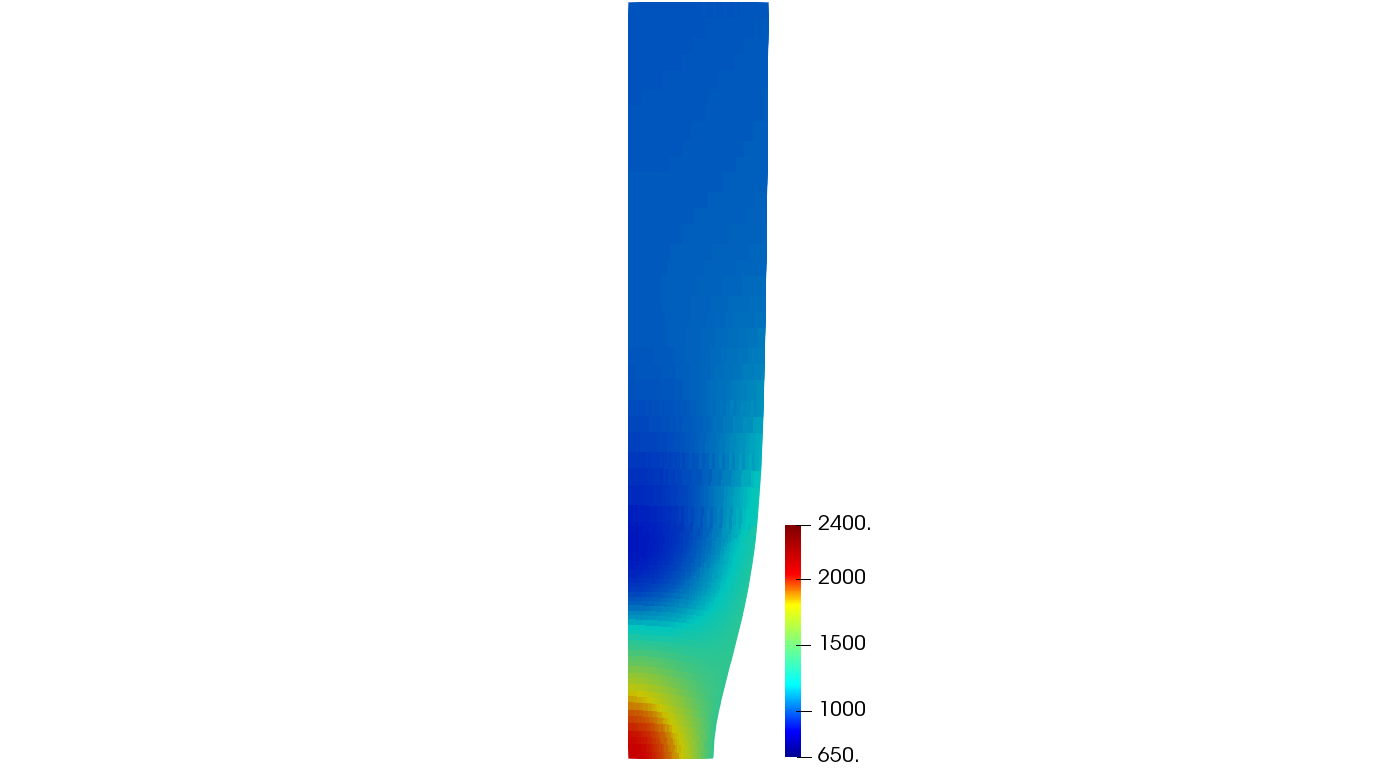}
  }
    \caption{Necking of a 2D rectangular bar: trace of the Cauchy stress tensor $\stress$ (in $\MPa$) at the quadrature points on the final configuration for the different methods.}
    \label{fig::necking_trace}
\end{figure}

In this first benchmark, we consider a 2D rectangular bar with an initial imperfection. The bar is subjected to uniaxial extension. This example has been studied previously by many authors as a necking problem \cite{Elguedj2014, Simo1992a, SouzaNeto2005, Saracibar2006, Wriggers2017a} and can be used to test the robustness of the different methods. The bar has a length of $53.334~\mm$ and a variable width from an initial width value of $12.826~\mm$ at the top to a width of $12.595~\mm$ at the center of the bar to create a geometric imperfection. A vertical displacement $u_y = 5~\mm$ is imposed at both ends, as shown in Fig.~\ref{fig::necking_geom}. For symmetry reasons, only one-quarter of the bar is discretized, and the mesh is composed of 400 quadrangles, see Fig.~\ref{fig::necking_mesh}. The load-displacement curve is plotted in Fig.~\ref{fig::necking_load}. We observe that except for Q1, all the other methods give very similar results. Moreover, the equivalent plastic strain $p$, respectively the trace of the Cauchy stress tensor $\stress$,  are shown in Fig.~\ref{fig::necking_p}, resp. in Fig.~\ref{fig::necking_trace}, at the quadrature points on the final configuration. A sign of locking is the presence of strong oscillations in the trace of the Cauchy stress tensor $\stress$. We notice that the cG formulations Q1 and Q2 lock, contrary to the HHO,  Q2\_RI, and UPG methods which deliver similar results. We remark however that the results for HHO(1;1), HHO(1;2), and Q2\_RI are slightly less smooth than for HHO(2;2), HHO(2;3), and UPG. The reason is that on a fixed mesh, the three former methods have less quadrature points than the three latter ones, see Table~\ref{tab::necking} (HHO(2;2), HHO(2;3), and UPG have the same number of quadrature points). Therefore, the stress is evaluated using less points in HHO(1;1), HHO(1;2), and Q2\_RI. It is sufficient to refine the mesh or to increase the order of the quadrature by two in HHO(1;1) and HHO(1;2) to retrieve similar results to those for the three other methods (not shown for brevity).
\begin{table}
\centering
\begin{tabular}{|c|c|c|c|c|c|c|c|c|}
\hline 
Method & Q1 & Q2 & \red{Q2\_RI} & UPG & HHO(1;1) & HHO(1;2) & HHO(2;2) & HHO(2;3) \\ 
\hline 
Number of dofs & 884 & 2566 & \red{2566} & 3450 & 3364 & 3364 & 5046 & 5046 \\ 
\hline 
Number of QPs & 1600 & 3600 & \red{1600} & 3600 & 1600 & 1600 & 3600 & 3600 \\ 
\hline 
\end{tabular}
\caption{Necking of a 2D rectangular bar: number of globally coupled degrees of freedom (dofs) and quadrature points (QPs) for the different methods.}\label{tab::necking}
\end{table}
\subsection{Cook's membrane} \label{ss::cook}
\begin{figure}
    \centering
    \subfloat[]{
    \label{fig::cook_geom}
        \centering
        \includegraphics[scale=0.37]{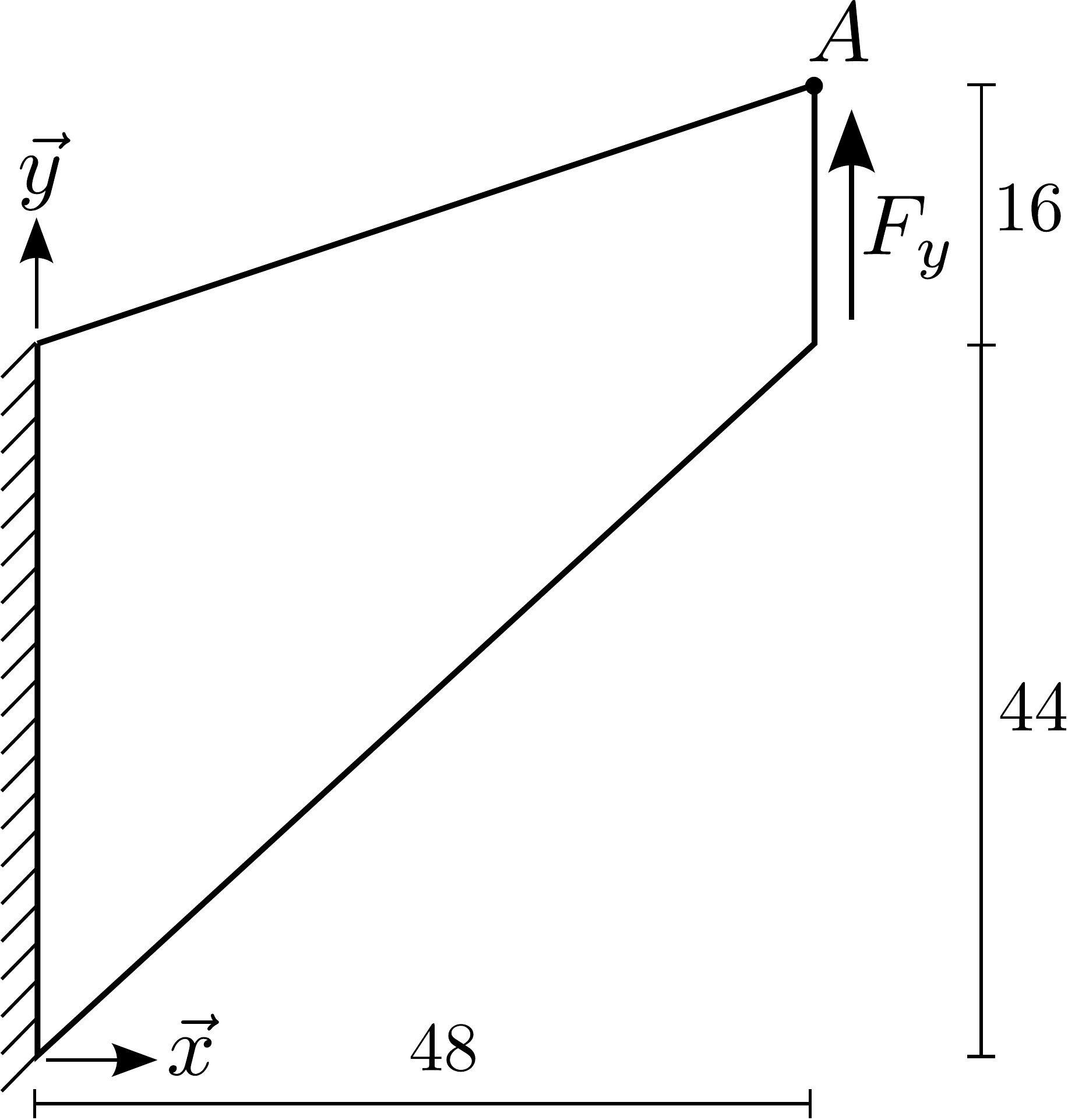}
  }
    ~ 
    \subfloat[]{
        \centering
	    \includegraphics[scale=1.0]{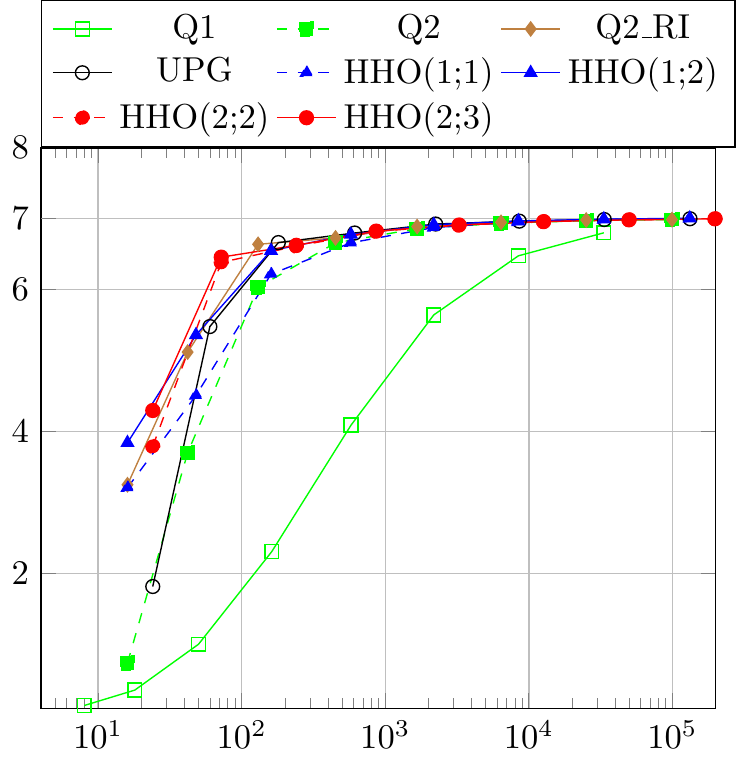}
	    \label{fig::cook_depl}
    }
    \caption{Cook's membrane: (a) Geometry and boundary conditions (dimensions in $\mm$). (b) Convergence of the vertical displacement of the point $A$ (in $\mm$) vs. the number of degrees of freedom for Q1, Q2, Q2\_RI, UPG, and HHO methods.}
\end{figure}

\begin{figure}
\centering
\subfloat[Q1]{
        \centering 
        \includegraphics[scale=0.31, trim= 43 42 27 62, clip=true]{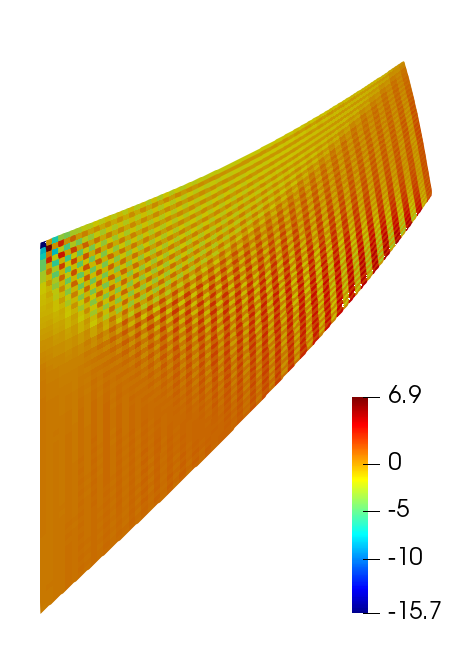} 
  }
    ~ 
    \subfloat[Q2]{
        \centering
	    \includegraphics[scale=0.30, trim= 41 45 36 51, clip=true]{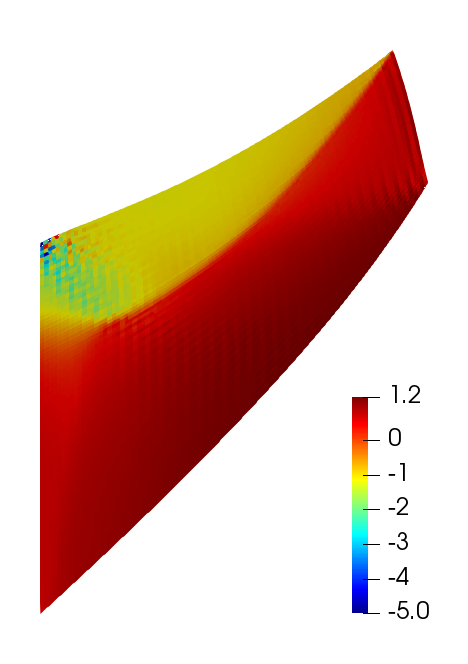}
    }
 ~ 
      \subfloat[Q2\_RI]{
        \centering 
        \includegraphics[scale=0.30, trim= 41 45 36 51, clip=true]{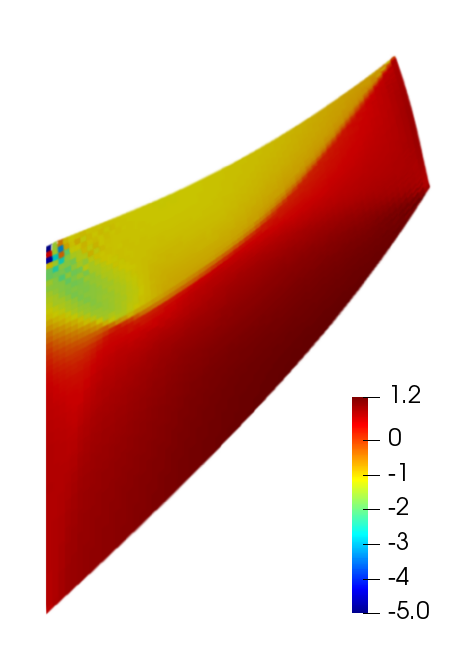} 
  }
  
       \subfloat[UPG]{
        \centering 
        \includegraphics[scale=0.30, trim= 41 45 27 51, clip=true]{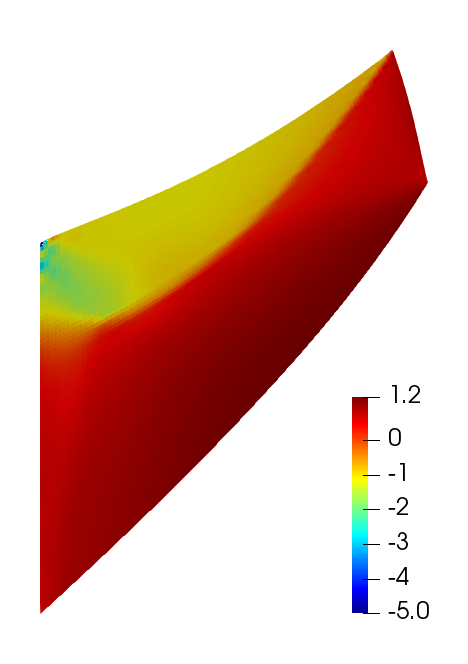} 
  }
   ~ 
  \subfloat[HHO(1;1)]{
        \centering
	    \includegraphics[scale=0.30, trim= 41 45 36 51, clip=true]{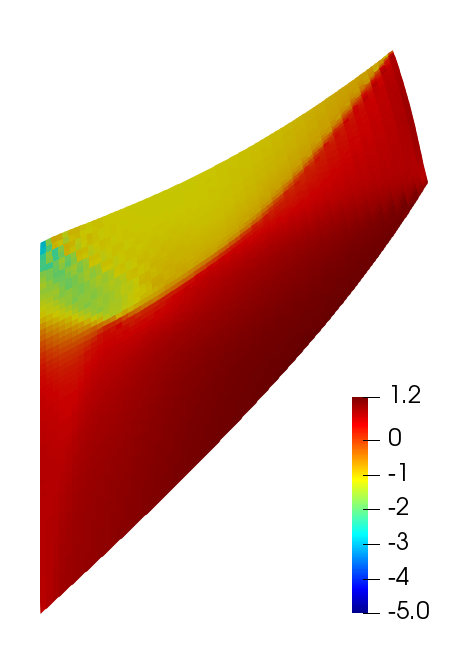}
    }
    ~ 
    \subfloat[HHO(1;2)]{
        \centering
	    \includegraphics[scale=0.30, trim= 41 45 36 51, clip=true]{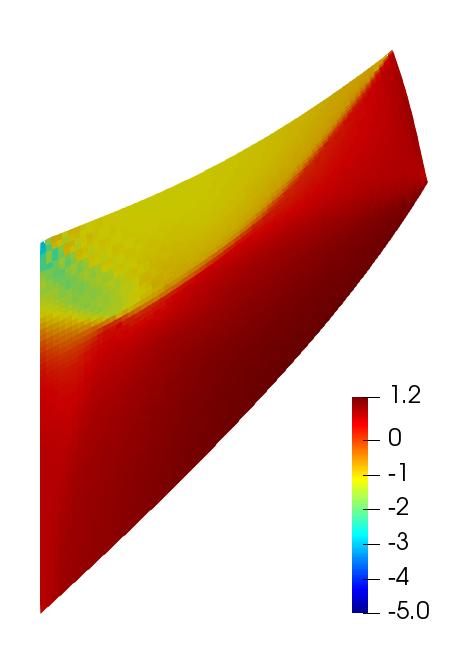}
    }
    
    \subfloat[HHO(2;2)]{
        \centering
	    \includegraphics[scale=0.30, trim= 41 45 36 51, clip=true]{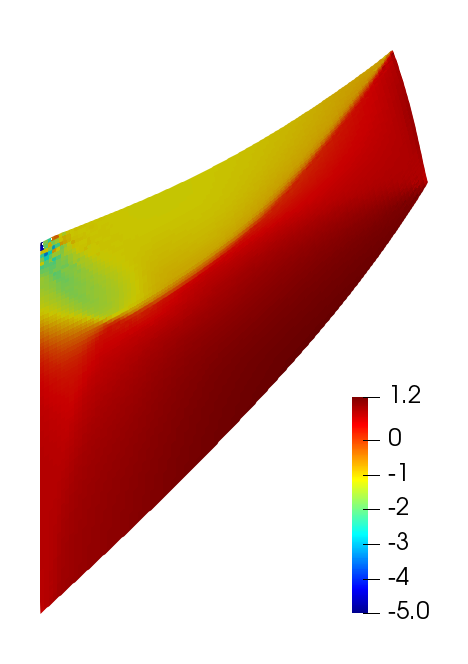}
    }
   ~ 
    \subfloat[HHO(2;3)]{
        \centering
	    \includegraphics[scale=0.30, trim= 41 45 36 51, clip=true]{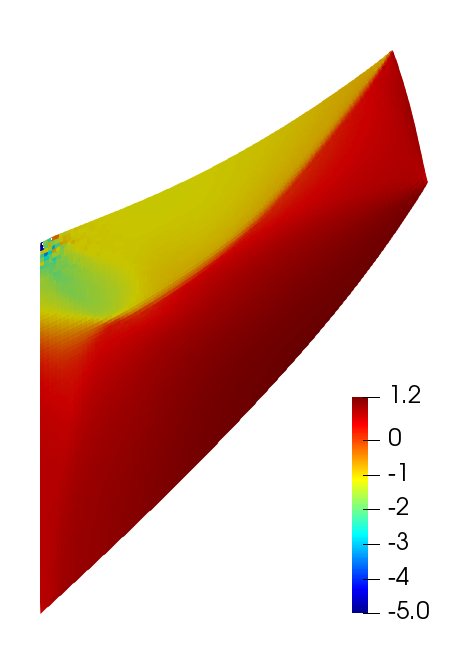}
    }
\caption{Cook's membrane: trace of the Cauchy stress tensor $\stress$ (in $\GPa$) at the quadrature points on the final configuration for a $32 \times 32$ quadrangular mesh and for the different methods.}
\label{fig::cook_trace}
\end{figure}

We consider the Cook's membrane problem which is a well known bending-dominated test case \cite{Simo1992a, AlAkhrass2014, Elguedj2008}. It consists of a tapered panel, clamped on one side, and subjected to a total vertical load $F_y = 5~\kN$ applied uniformly along all the opposite side, as shown in Fig.~\ref{fig::cook_geom}. The simulation is performed on a sequence of refined quadrangular meshes such that each side contains $2^N$ edges with $0\leq N \leq 6$. The vertical displacement of the point $A$ versus the number of degrees of freedom is plotted in Fig.~\ref{fig::cook_depl} for the different methods. As expected when comparing the number of degrees of freedom, the linear cG formulation Q1 has the slower convergence, HHO(1;2), Q2\_RI, and UPG converge slightly faster than HHO(1;1), Q2, whereas HHO(2;2) and HHO(2;3) outperform all the other methods and give almost the same results. Moreover, we show in Fig.~\ref{fig::cook_trace} the trace of the Cauchy stress tensor $\stress$ at the quadrature points on the final configuration. The cG formulations Q1 and Q2  present oscillations  that confirm the presence of volumetric locking, contrary to the HHO, Q2\_RI, and UPG methods which deliver similar and smooth results (even if the cG formulations Q1 and Q2 present volumetric locking in terms of stress, they converge in terms of displacement with mesh refinement). However, if we compare the trace of the Cauchy stress tensor $\stress$ for HHO(1;1) and HHO(1;2), we remark that the trace is slightly smoother near the upper-right corner for HHO(1;2) than for HHO(1;1). This can be explained by the presence of non-physical vertical localization bands of plastic deformations for HHO(1;1) and not for HHO(1;2). Localization bands constitute a well-known problem when the plasticity model is local. Computational practice with cG approximations indicates that increasing the order of the finite elements mitigates this issue. The same effect is observed here by increasing the degree of the cell unknowns (further numerical investigations are performed in Sect.~\ref{ss::poly_mesh} for HHO methods). An alternative is to use a non-local plasticity model \cite{McBride2009}.
\subsection{Torsion of a square-section bar}\label{ss:torsion}
\begin{figure}[htbp]
    \centering
    \subfloat[]{
    \label{fig::bar_geom}
        \centering 
        \includegraphics[scale=0.32]{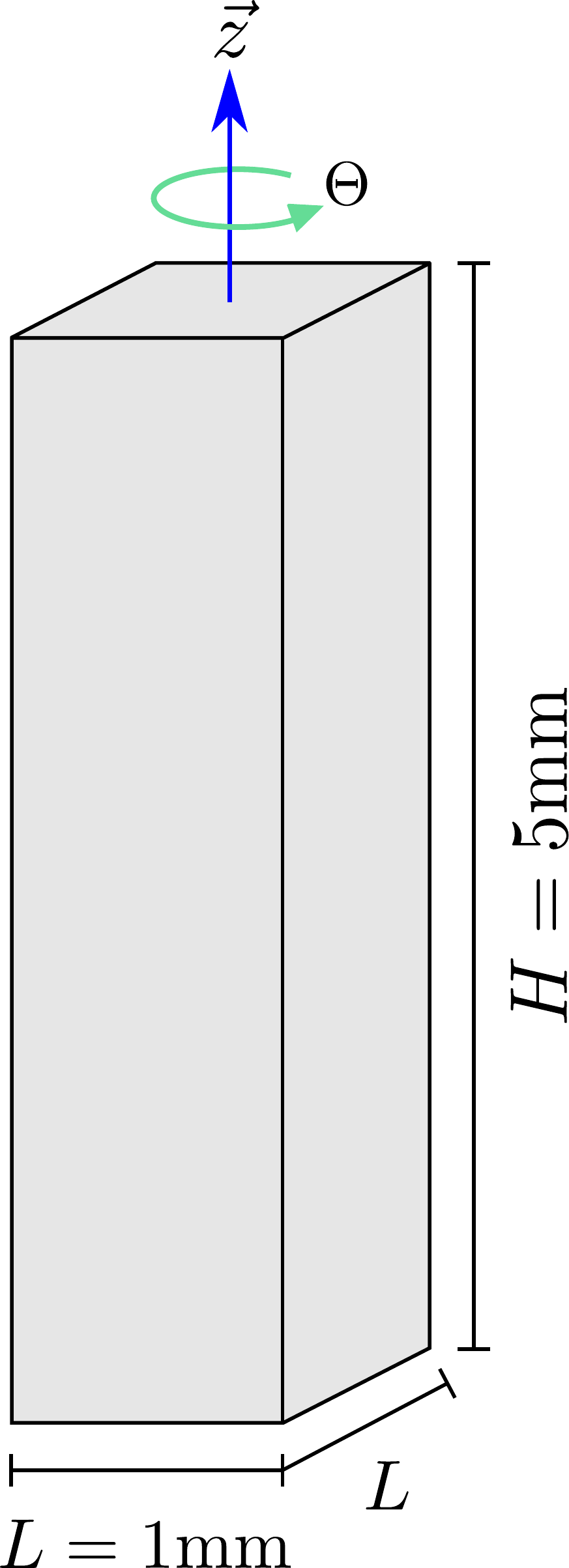} 
  }
    ~ 
    \subfloat[]{
        \centering
	    \includegraphics[scale=0.31]{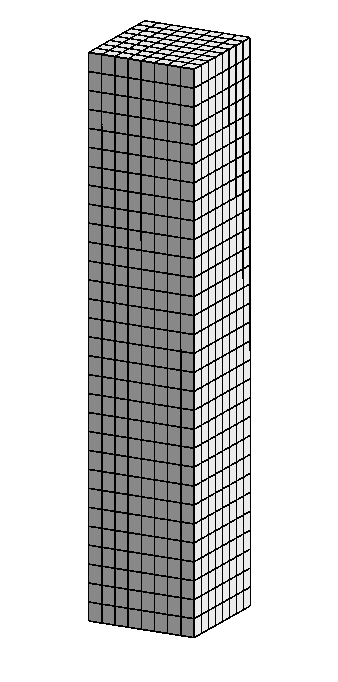}
	    \label{fig::bar_mesh}
    }
    \caption{Torsion test of a square-section bar: (a) Geometry and boundary conditions (dimensions in $\mm$) (b) Mesh in the reference configuration composed of 1920 hexahedra.}
\end{figure}

\begin{figure}[htbp]
    \centering
         \subfloat[$\Theta=0^{\circ}$]{
        \centering
        \includegraphics[scale=0.38, trim= 50 10 30 20, clip=true]{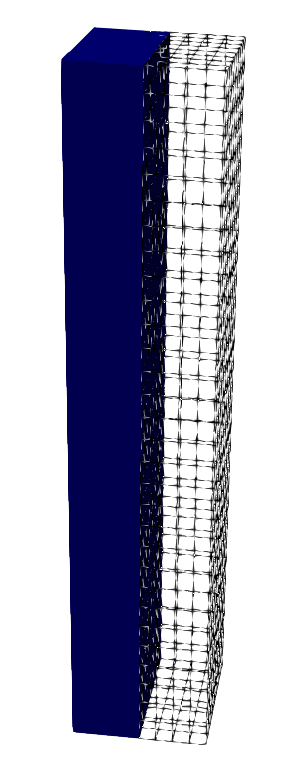}
  }
    ~ 
    \subfloat[$\Theta=90^{\circ}$]{
        \centering
        \includegraphics[scale=0.38, trim= 30 10 30 20, clip=true]{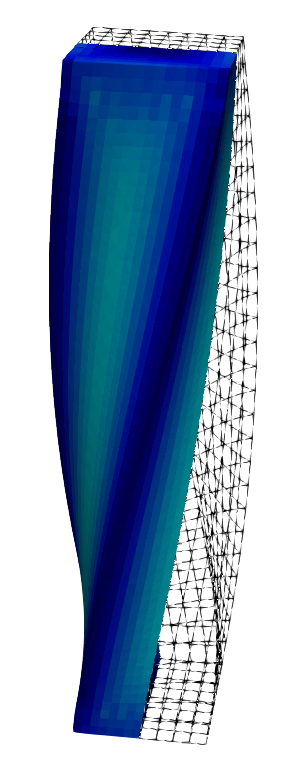}
  }
  ~ 
    \subfloat[$\Theta=180^{\circ}$]{
        \centering
        \includegraphics[scale=0.38, trim= 30 10 30 20, clip=true]{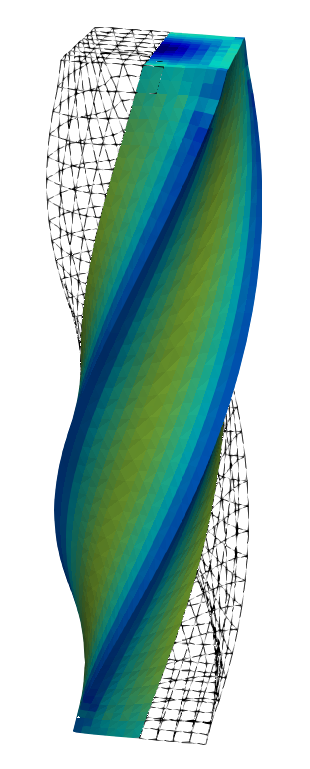}
  }
   ~ 
    \subfloat[$\Theta=270^{\circ}$]{
        \centering
        \includegraphics[scale=0.38, trim= 30 10 30 20, clip=true]{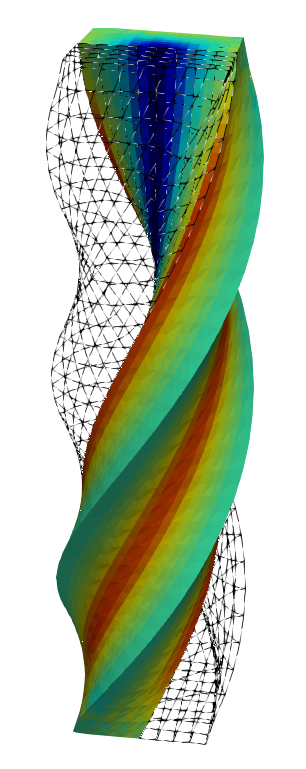}
  }
    ~ 
    \subfloat[$\Theta=360^{\circ}$]{
        \centering
        \includegraphics[scale=0.38, trim= 30 10 30 20, clip=true]{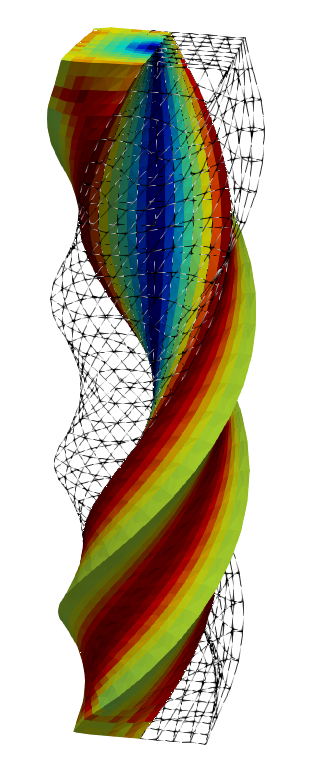}
  }
  ~ 
    \subfloat{
        \centering
        \includegraphics[scale=0.75, trim= 0 0 40 0, clip=true]{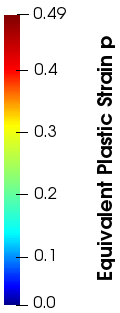}
  }
    \caption{Torsion of a square-section bar: Equivalent plastic strain $p$ for HHO(1;1) at the quadrature points for different rotation angles $\Theta$.}
        \label{fig::bar_p}
\end{figure}

\begin{figure}[htbp]
     \centering
    \subfloat[Q2]{
        \centering 
        \includegraphics[scale=0.33, trim= 10 0 141 150, clip=true]{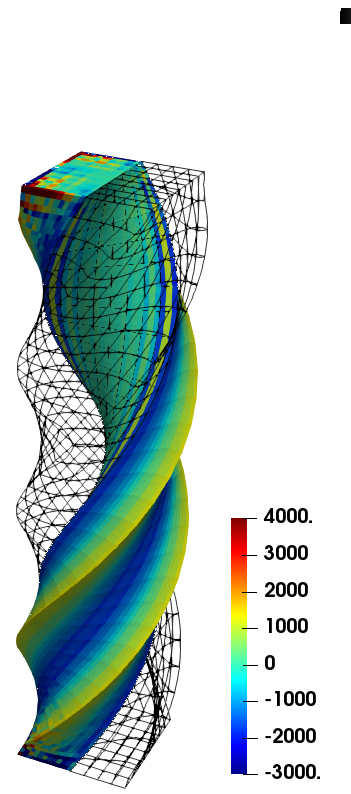} 
  }
    ~ 
      \centering
    \subfloat[UPG]{
        \centering 
        \includegraphics[scale=0.33, trim= 10 0 141 150, clip=true]{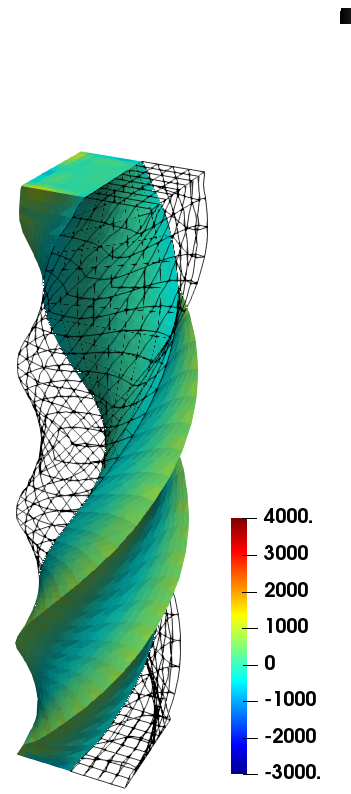} 
  }
    ~ 
      \centering
    \subfloat[HHO(1;1)]{
        \centering 
        \includegraphics[scale=0.33, trim= 10 0 141 150, clip=true]{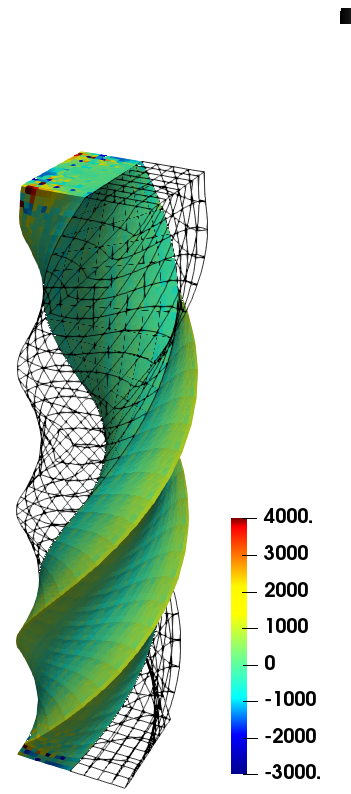} 
  }
      ~ 
      \centering
    \subfloat[HHO(1;2)]{
        \centering 
        \includegraphics[scale=0.33, trim= 10 0 141 150, clip=true]{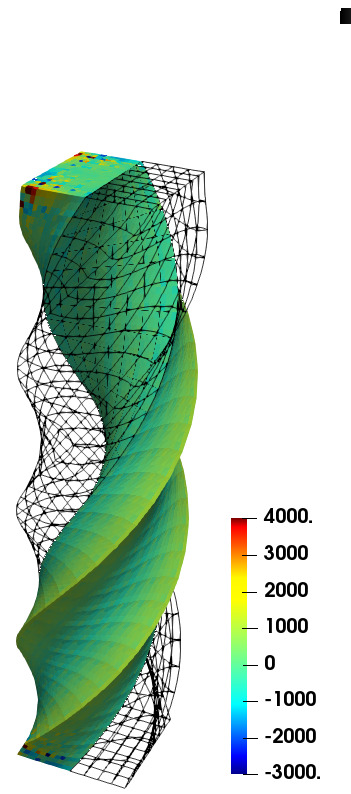} 
  }
      ~ 
      \centering
    \subfloat[HHO(2;2)]{
        \centering 
        \includegraphics[scale=0.33, trim= 10 0 141 150, clip=true]{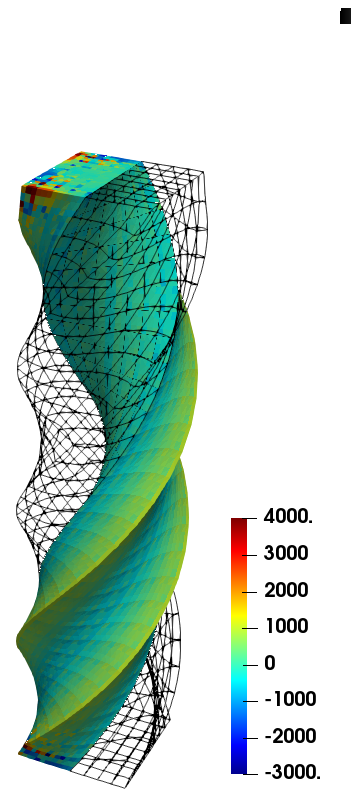} 
  }
    ~ 
      \centering
    \subfloat[HHO(2;3)]{
        \centering 
        \includegraphics[scale=0.33, trim= 10 0 141 150, clip=true]{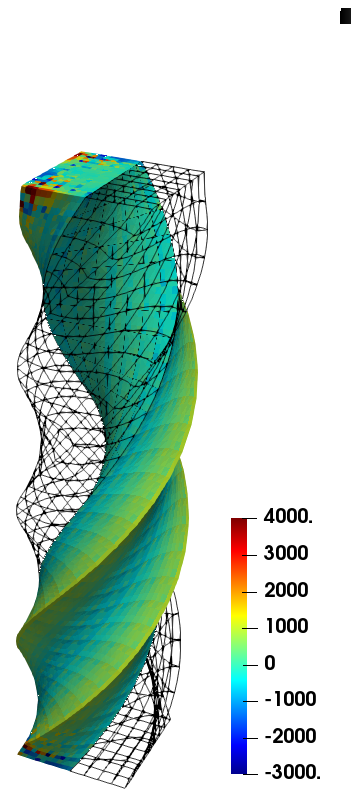} 
  }
      \centering
    \subfloat{
        \centering 
        \includegraphics[scale=0.4, trim= 233 0 33 500, clip=true]{Fig_torsion_hho232_trace.png}
  }
    \caption{Torsion of a square-section bar: trace of the Cauchy stress tensor $\stress$ (in $\MPa$) at the quadrature points for $\Theta=360^{\circ}$ and for the HHO, UPG, and Q2 methods .}
        \label{fig::bar_trace}
\end{figure}

This third benchmark \cite{Hudobivnik2018} allows one to test the robustness of HHO methods under large torsion. The bar has a square-section of length $L=1~\mm$ and a height of $H=5~\mm$ along the $z$-direction. The bottom end is clamped and the top end is subjected to a planar rotation of an angle $\Theta $ around its center along the $z$-direction and remains plane (the displacement of the top end along the $z$-direction is blocked),  see Fig.~\ref{fig::bar_geom}. The mesh is composed of 1900 hexahedra, see Fig.~\ref{fig::bar_mesh}. The equivalent plastic strain $p$ is plotted in Fig.~\ref{fig::bar_p} for HHO(1;1) and for different rotation angles $\Theta$. There is no sign of localization of the plastic deformations even for large rotations whatever the HHO variant is used. Moreover, the trace of the Cauchy stress tensor $\stress$ is plotted on the final configuration for $\Theta=360^{\circ}$ and for the Q2, UPG, and HHO methods in Fig.~\ref{fig::bar_trace}. As expected, there is no sign of volumetric locking for the HHO and UPG methods which give similar results contrary to Q2. The small oscillations observed at both ends are due to the imposed conditions on the displacement.
\subsection{Quasi-incompressible sphere under internal pressure}\label{ss::sphere}
\begin{figure}
    \centering
    \subfloat[]{
    \label{fig::sphere_mesh}
        \centering 
        \includegraphics[scale=0.45]{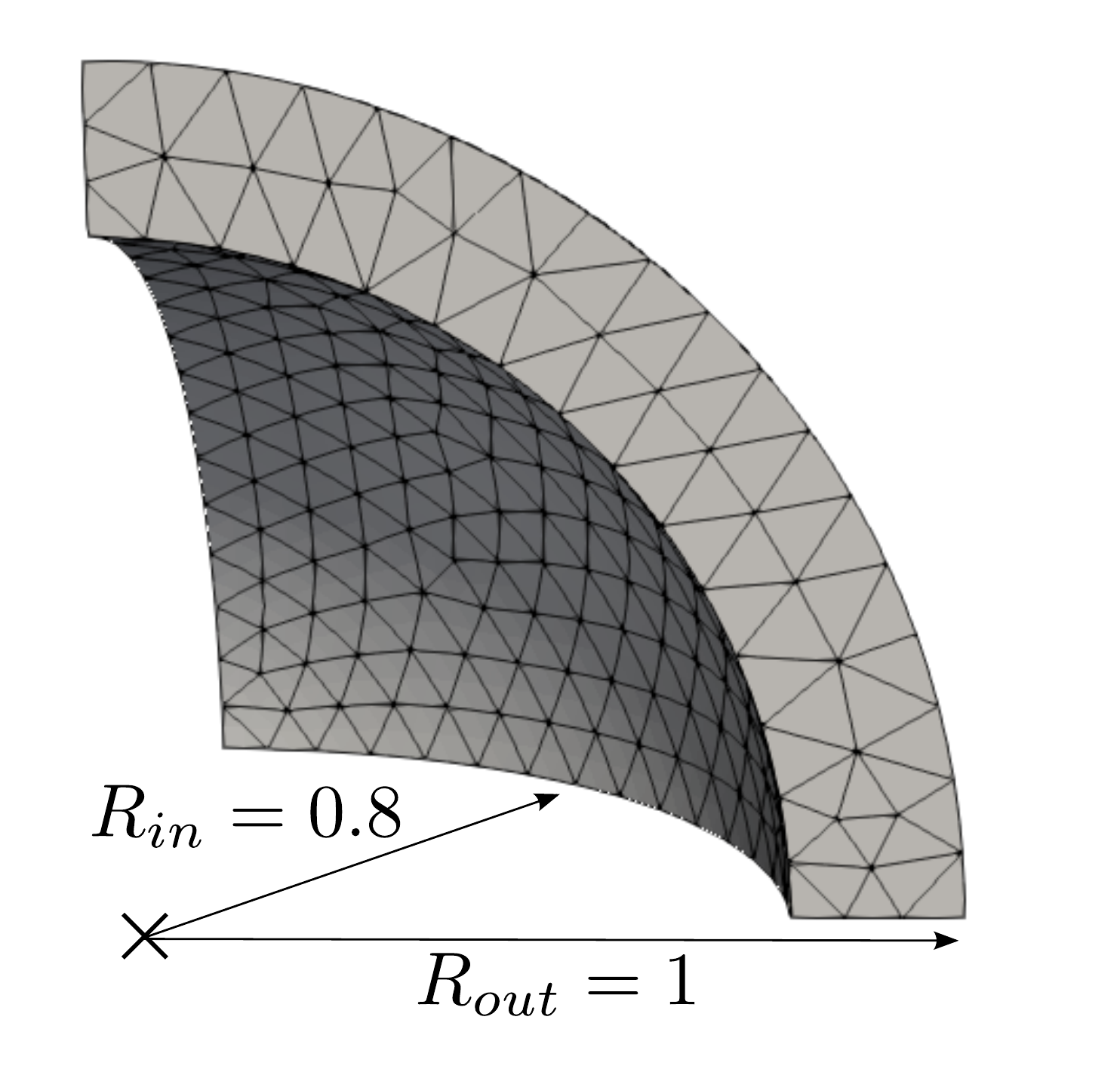} 
  }
    ~ 
    \subfloat[]{
        \centering
	     \raisebox{3.5mm}{\includegraphics[scale=0.36, trim= 350 0 150 40, clip=true]{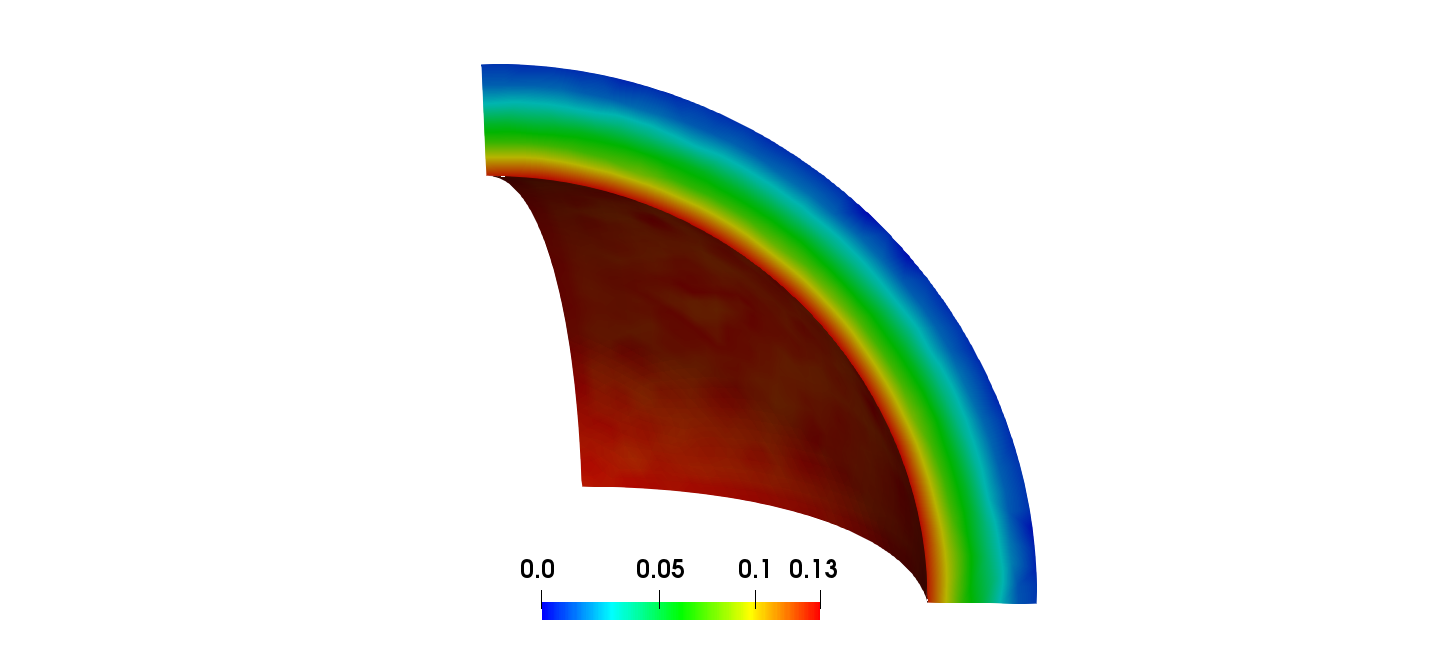}}
	    \label{fig::sphere_p}
    }
    \caption{Quasi-incompressible sphere under internal pressure: (a) Mesh in the reference configuration composed of 1580 tetrahedra (dimensions in $\mm$). (b) Equivalent plastic strain $p$ for HHO(1;2) on the final configuration.}
\end{figure}

\begin{figure}[htbp]
    \centering
    \subfloat[HHO(1;1)]{
        \centering 
        \includegraphics[scale=0.9]{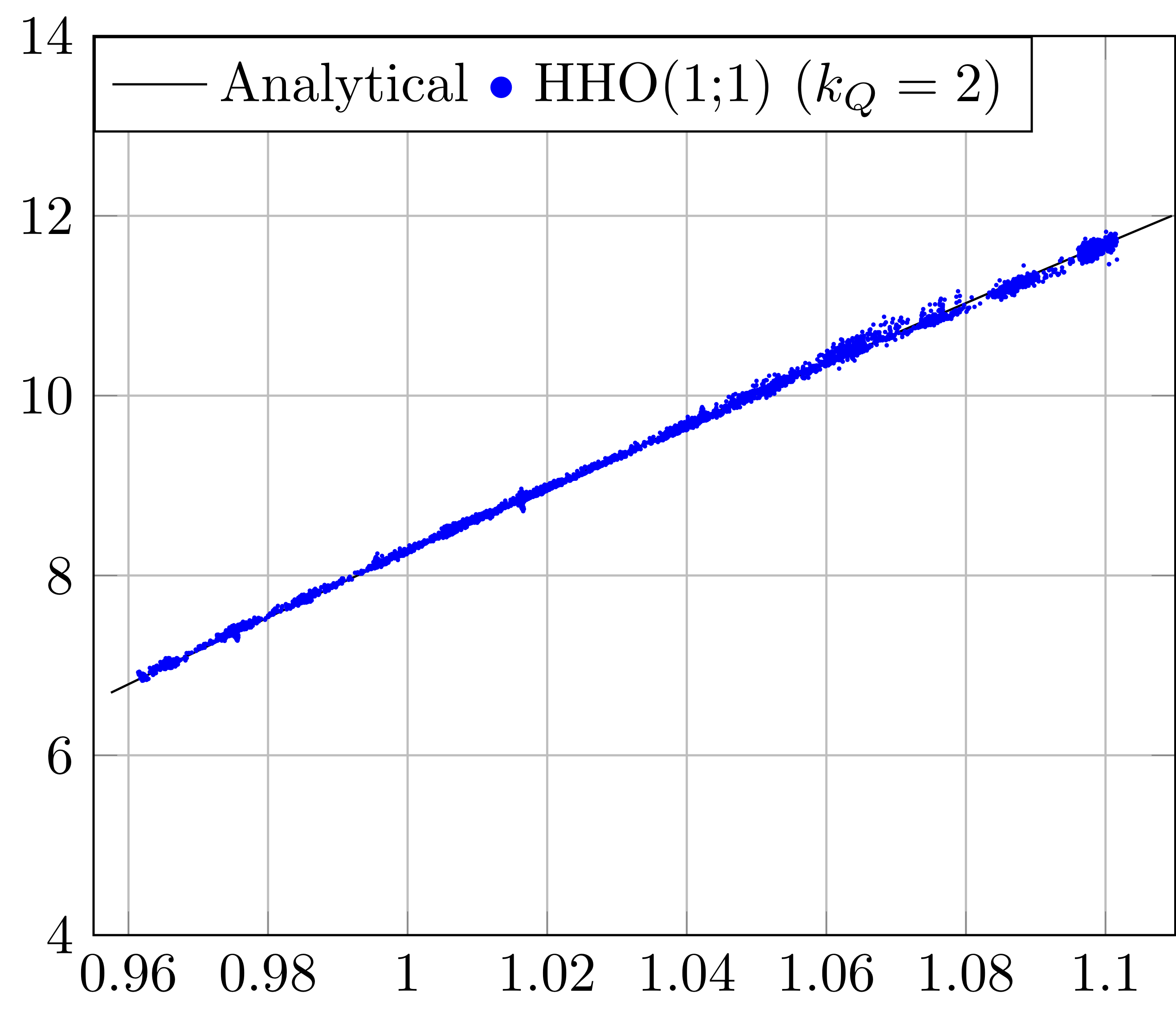} 
  }
    ~ 
    \subfloat[HHO(1;2;)]{
        \centering
	    \includegraphics[scale=0.9]{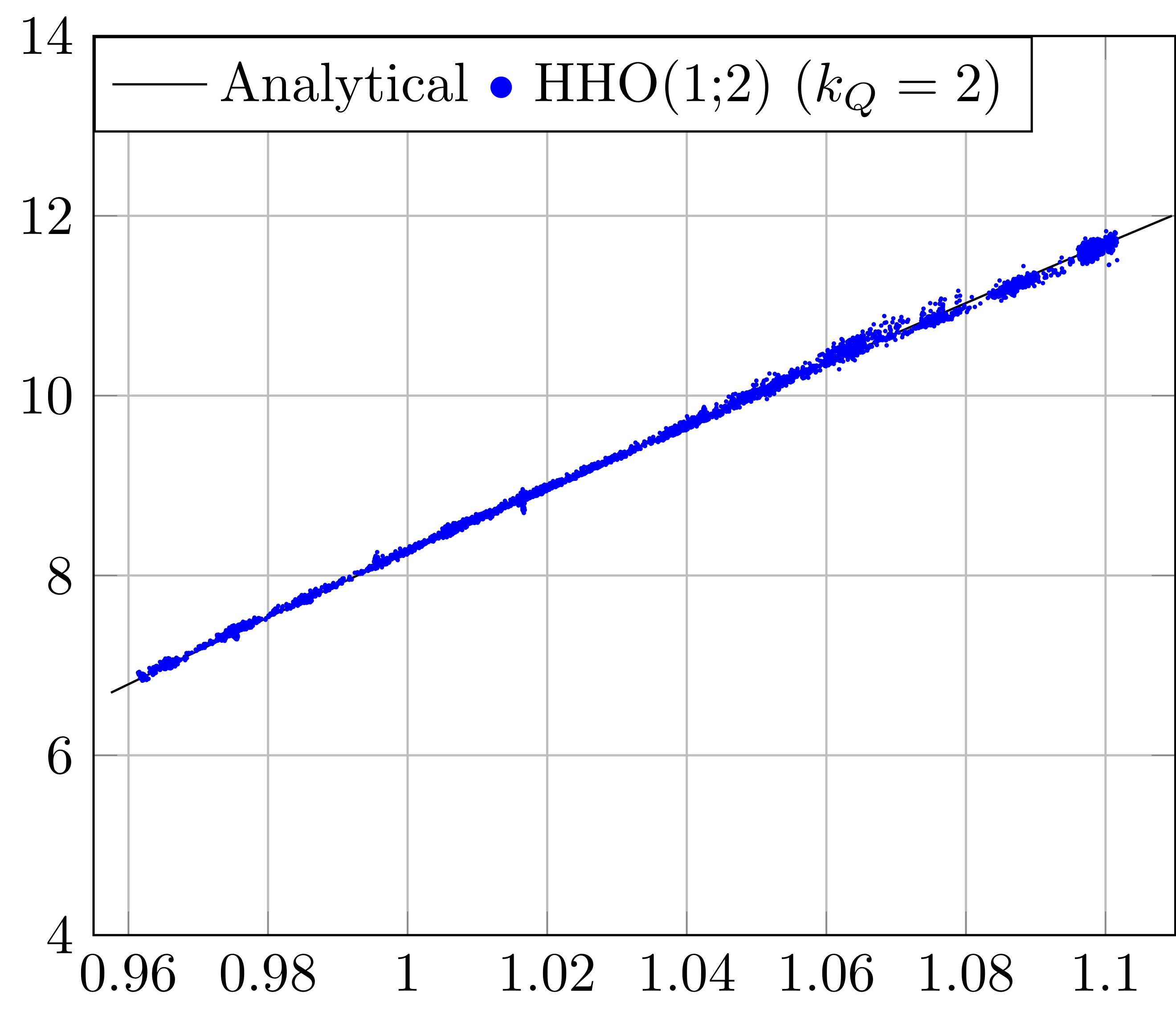}
    }
    
   \subfloat[HHO(2;2)]{
        \centering 
        \includegraphics[scale=0.9]{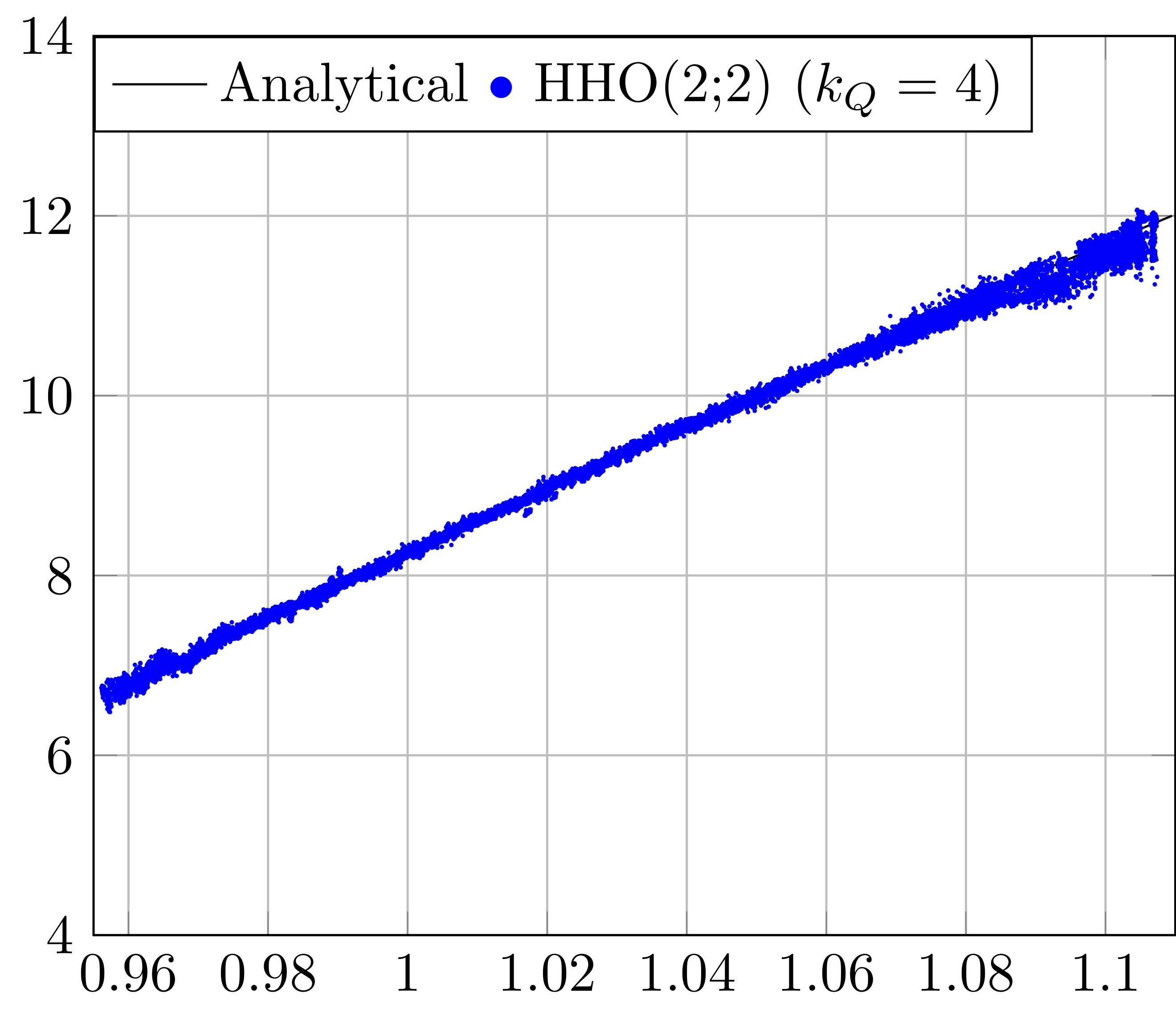} 
  }
    ~ 
    \subfloat[HHO(2;3)]{
        \centering
	    \includegraphics[scale=0.9]{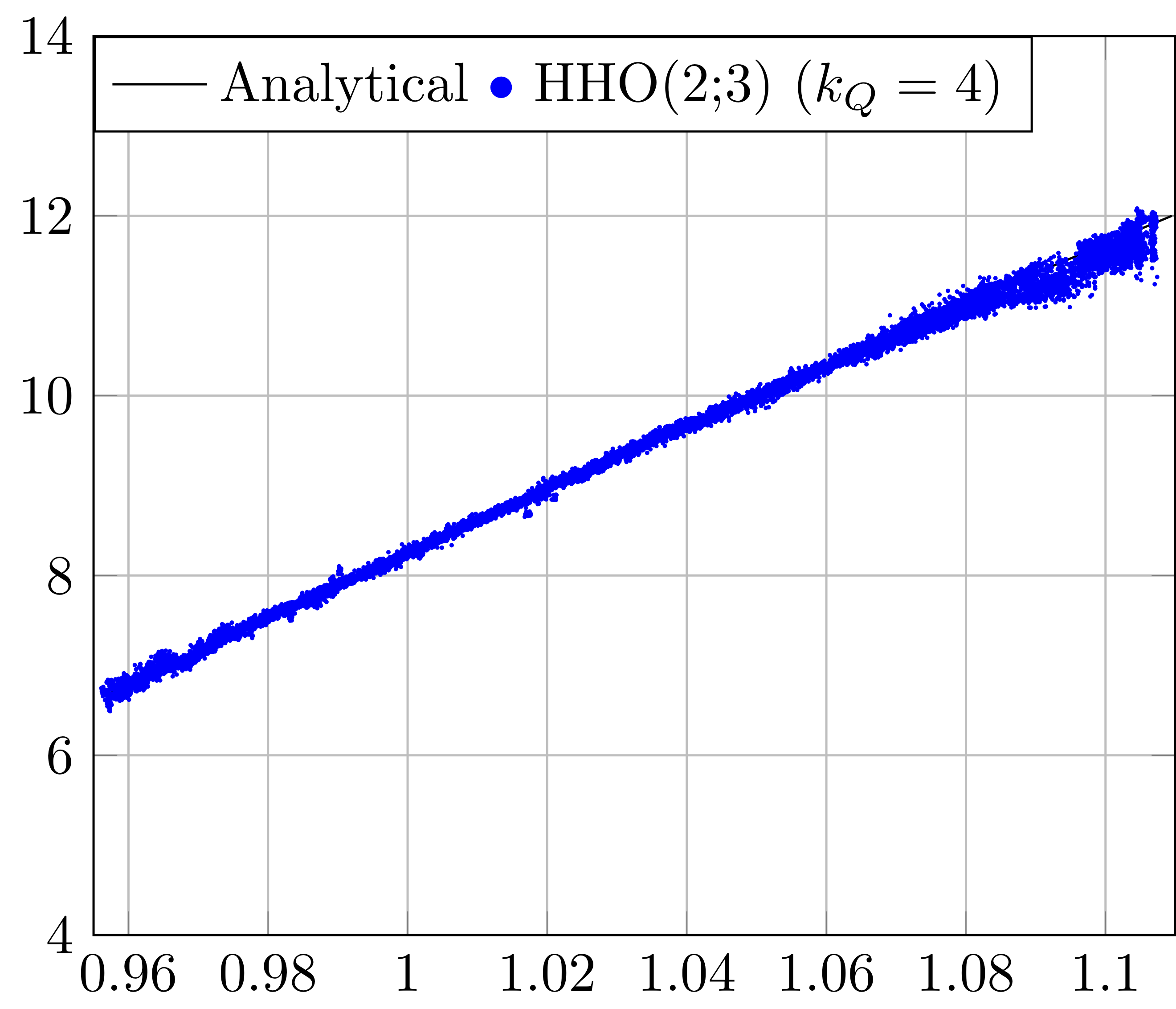}
    }  
    
      \subfloat[T2]{
        \centering 
        \includegraphics[scale=0.9]{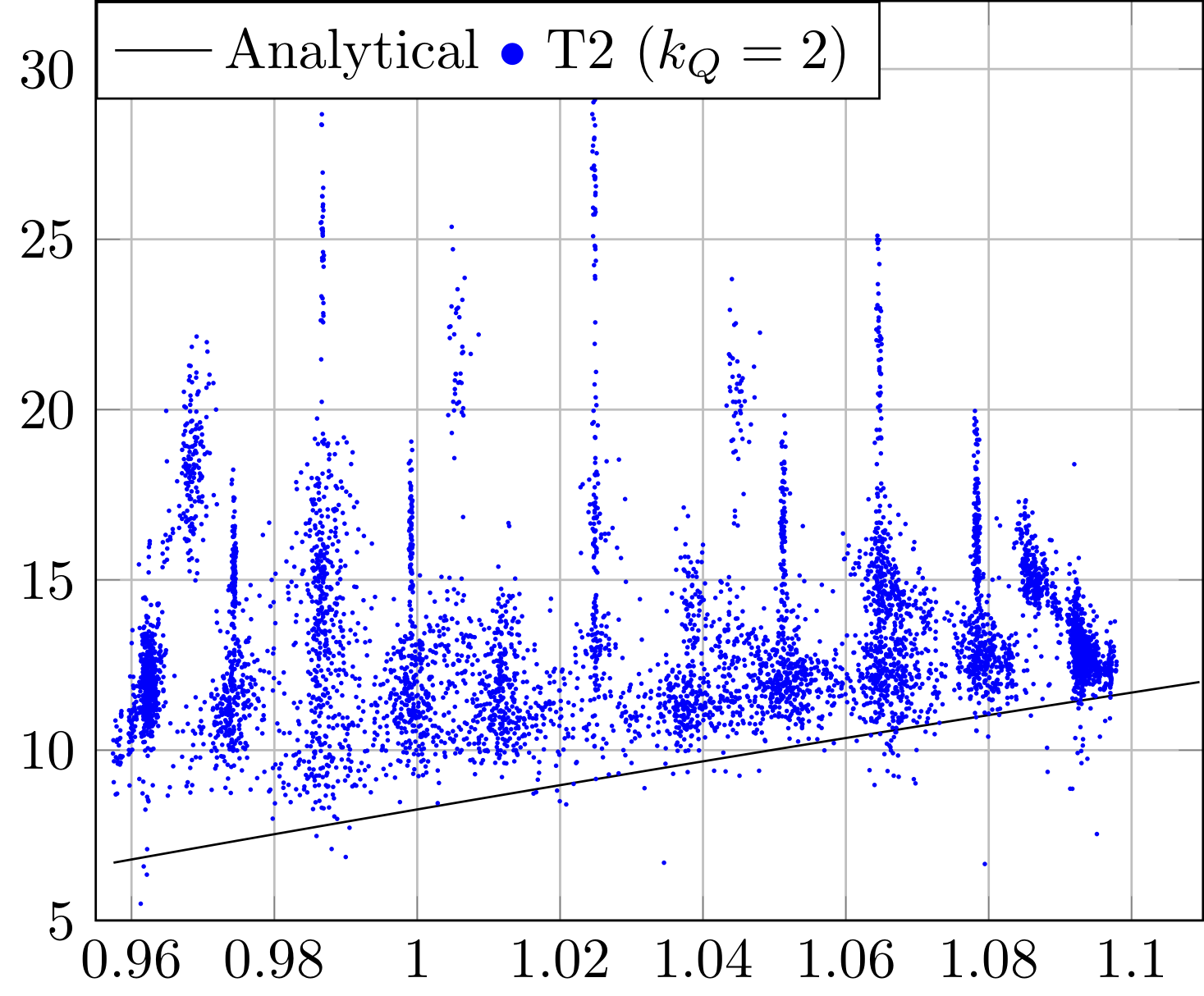} 
  }
    ~ 
    \subfloat[UPG]{
        \centering
	    \includegraphics[scale=0.9]{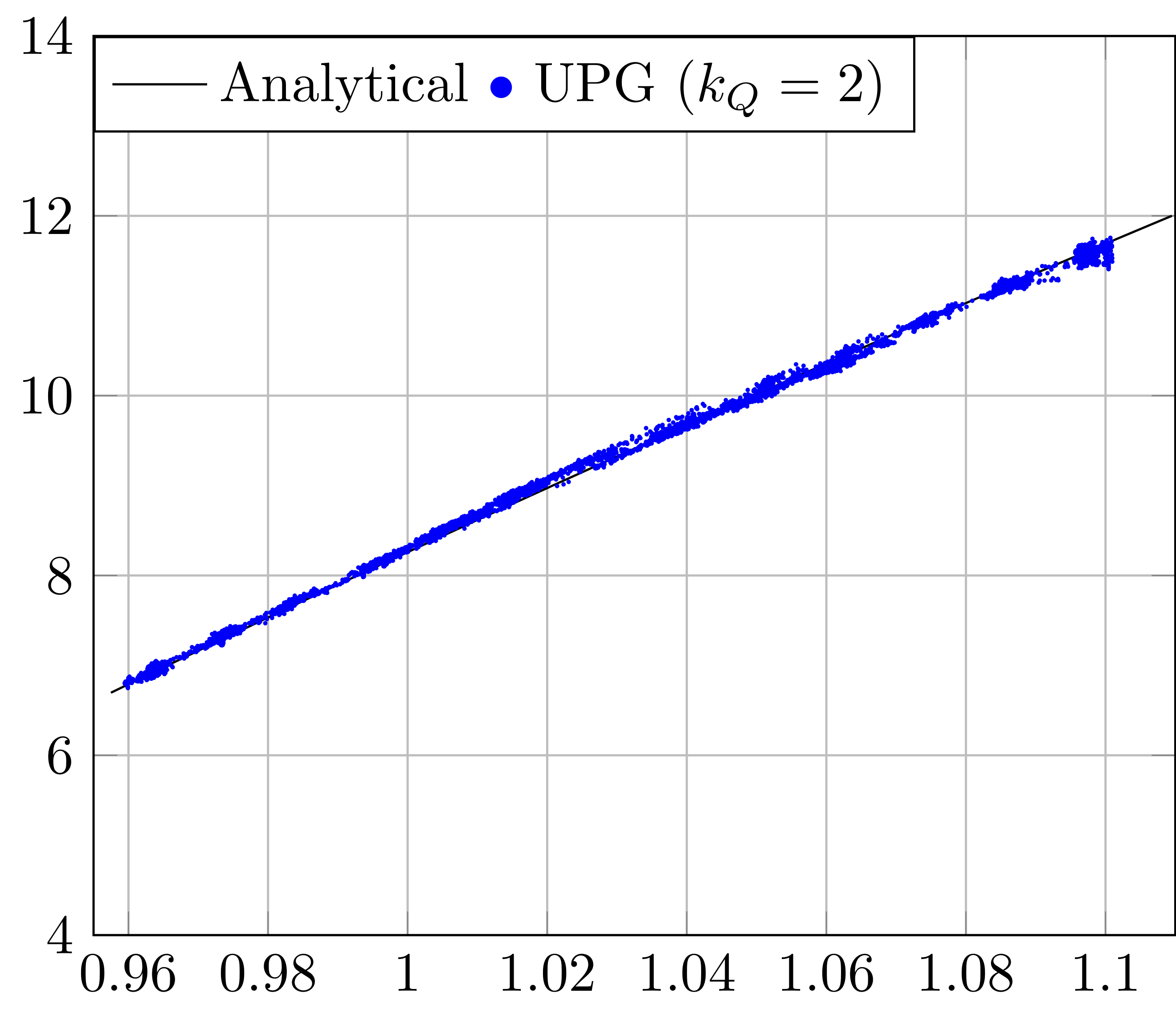}
    }  
    \caption{Quasi-incompressible sphere under internal pressure: trace of the Cauchy stress tensor $\stress$ (in $\MPa$)  vs. deformed radius $r$ (in $\mm$) for the different methods at all the quadrature points and for the limit load. }
    \label{fig::sphere_trace}
\end{figure}

This last benchmark \cite{AlAkhrass2014}  consists of a quasi-incompressible sphere under internal pressure for which an analytical solution is known when the entire sphere has reached a plastic state. This benchmark is particularly challenging compared to the previous ones since we consider here perfect plasticity. The sphere has an inner radius $R_{in}=0.8~\mm$ and an outer radius $R_{out}=1~\mm$. An internal radial pressure $P$ is imposed. For symmetry reasons, only one-eighth of the sphere is discretized, and the mesh is composed of 1580 tetrahedra, see Fig.~\ref{fig::sphere_mesh}. The simulation is performed until the limit load corresponding to an internal pressure $P_{lim} \simeq 2.54~\MPa$ is reached.
The equivalent plastic strain $p$ is plotted for HHO(1;2) in Fig.~\ref{fig::sphere_p}, and the trace of the Cauchy stress tensor $\stress$ is compared for HHO, UPG and T2 methods in Fig.~\ref{fig::sphere_trace} at all the quadrature points on the final configuration for the limit load. We notice that the quadratic element T2 locks, whereas HHO and UPG do not present any sign of locking and produce results that are very close to the analytical solution. However, the trace of the Cauchy stress tensor $\stress$ is slightly more dispersed around the analytical solution for HHO(2;2) and HHO(2;3) than for HHO(1;1) and HHO(1;2) near the outer boundary. For this test case, we do not expect that HHO(2;2) and HHO(2;3) will deliver more accurate solutions than HHO(1;1) and HHO(1;2) since the geometry is discretized using tetrahedra with planar faces.

\begin{figure}[htbp]
    \centering
    \subfloat[HHO]{
        \centering 
        \includegraphics[scale=0.9]{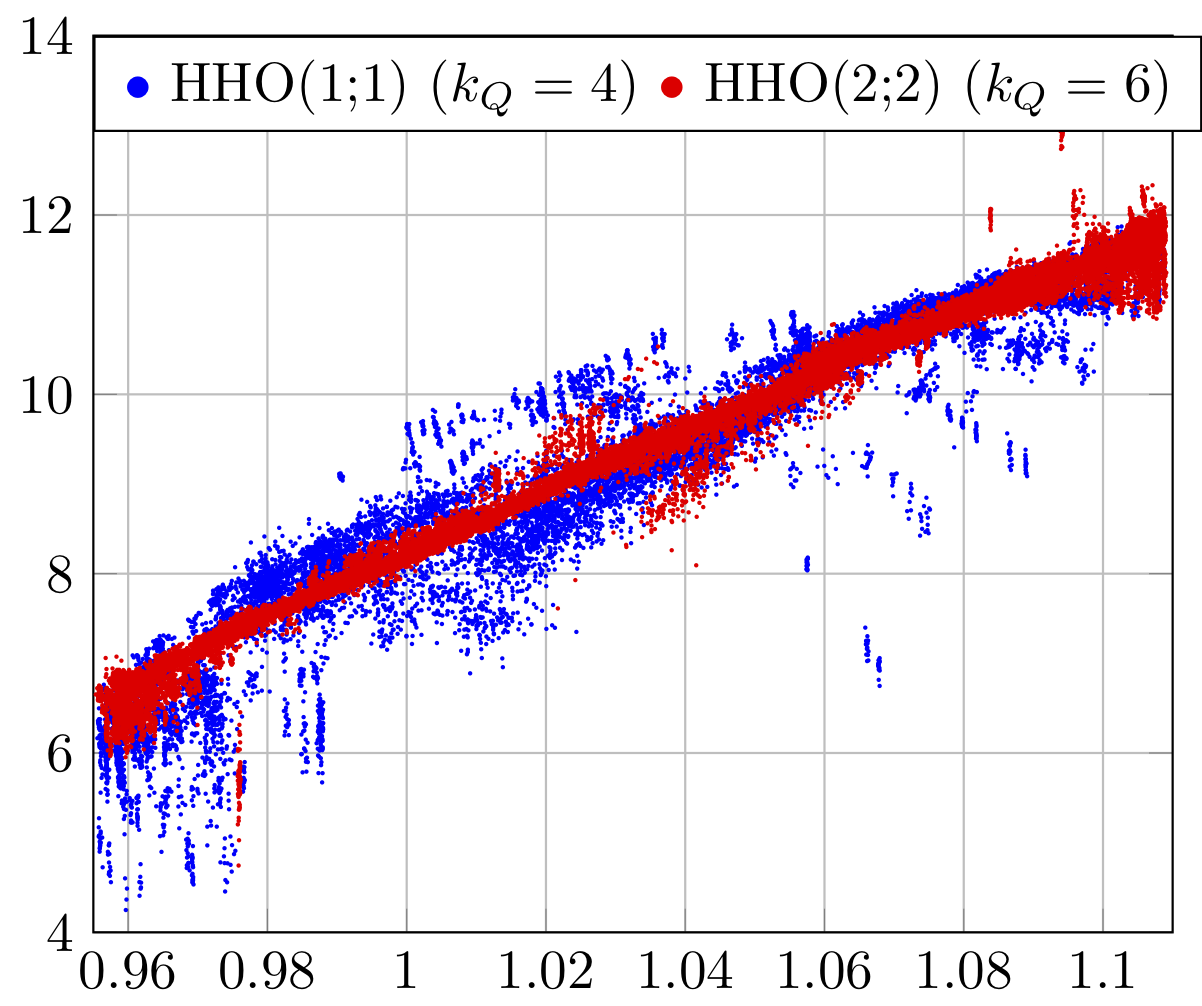} 
  }
    ~ 
    \subfloat[UPG]{
        \centering
	    \includegraphics[scale=0.9]{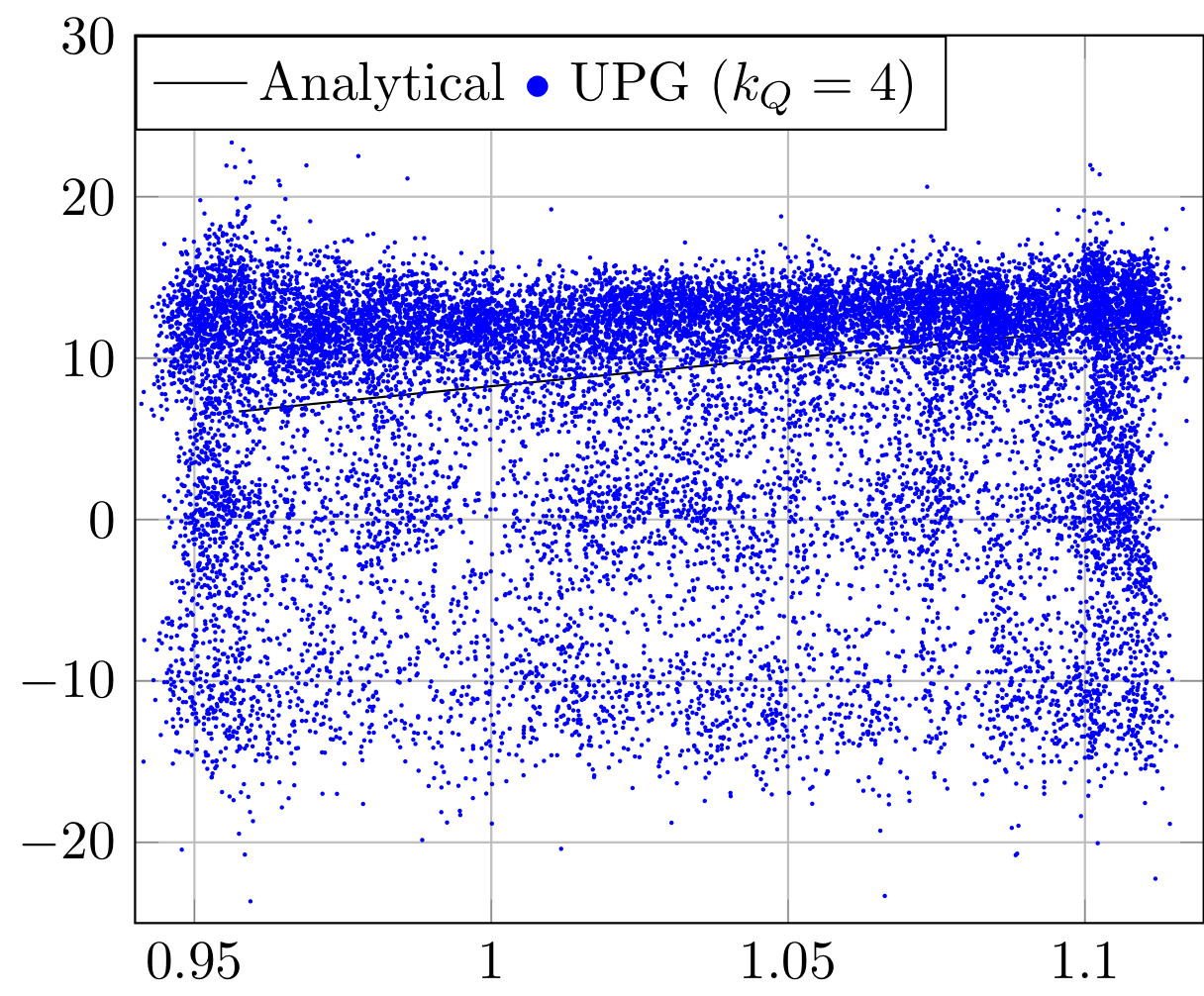}
    }  
    \caption{Quasi-incompressible sphere under internal pressure: trace of the Cauchy stress tensor $\stress$ (in $\MPa$)  vs. deformed radius $r$ (in $\mm$) for HHO(1;1), HHO(2;2) and UPG at all the quadrature points for the limit load and for a higher quadrature order $k_Q$.}
    \label{fig::sphere_trace_kq}
\end{figure}

\begin{figure}[htbp]
    \centering
    \subfloat[HHO with $k_Q = 2k$]{
        \centering 
        \includegraphics[scale=0.9]{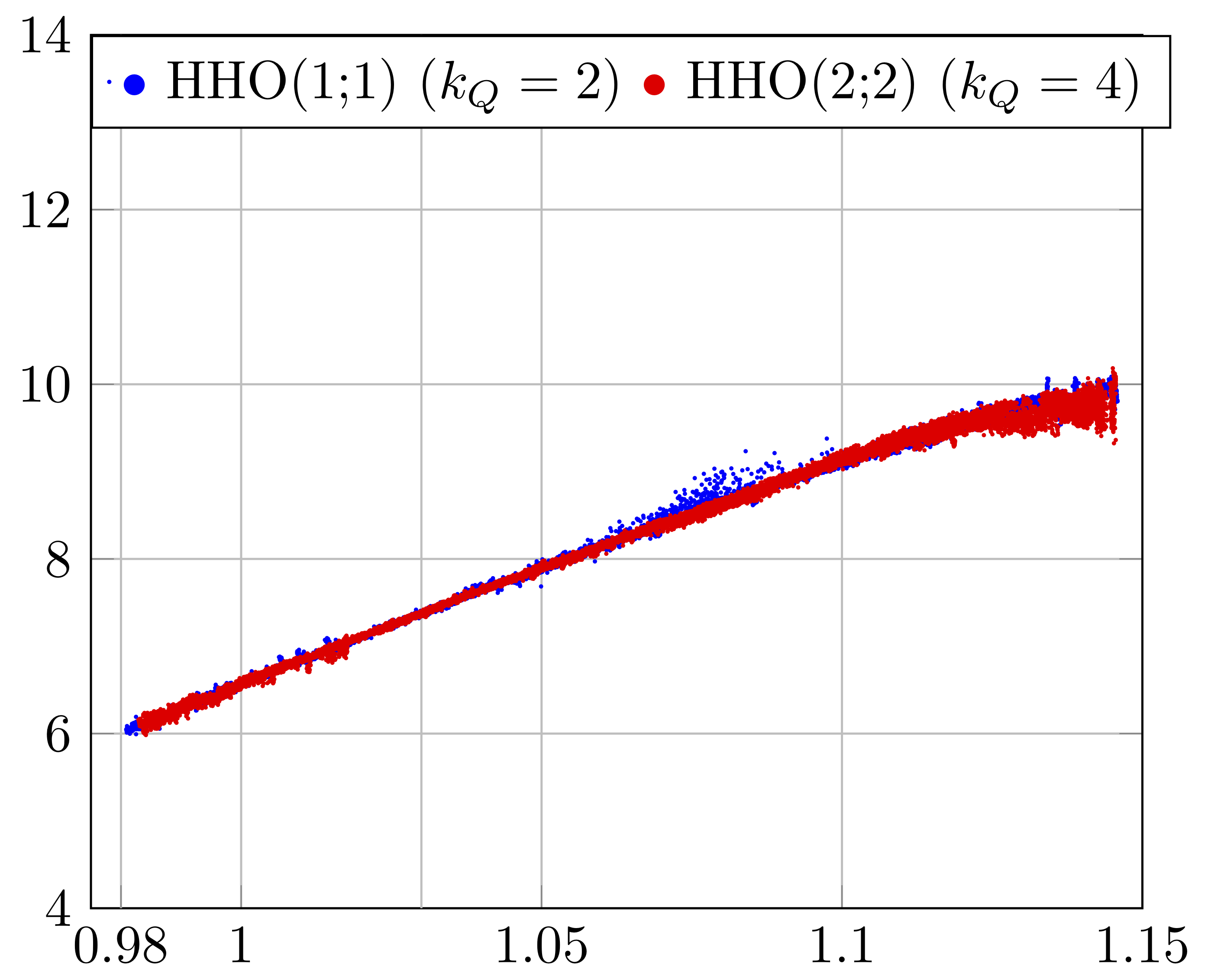} 
  }
    ~ 
   \subfloat[HHO with $k_Q = 2k+2$]{
        \centering 
        \includegraphics[scale=0.9]{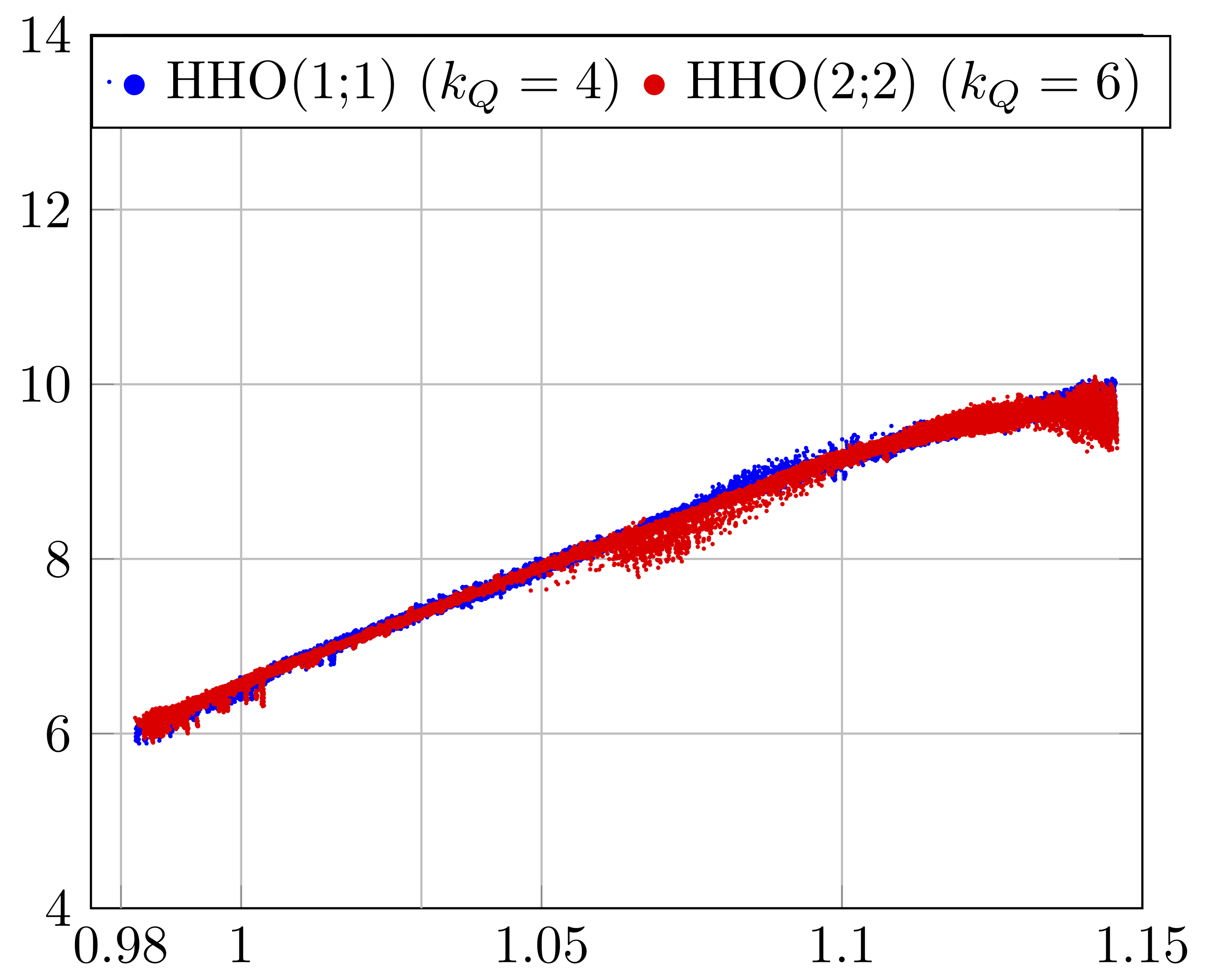} 
  }

    \subfloat[UPG]{
        \centering
	    \includegraphics[scale=0.9]{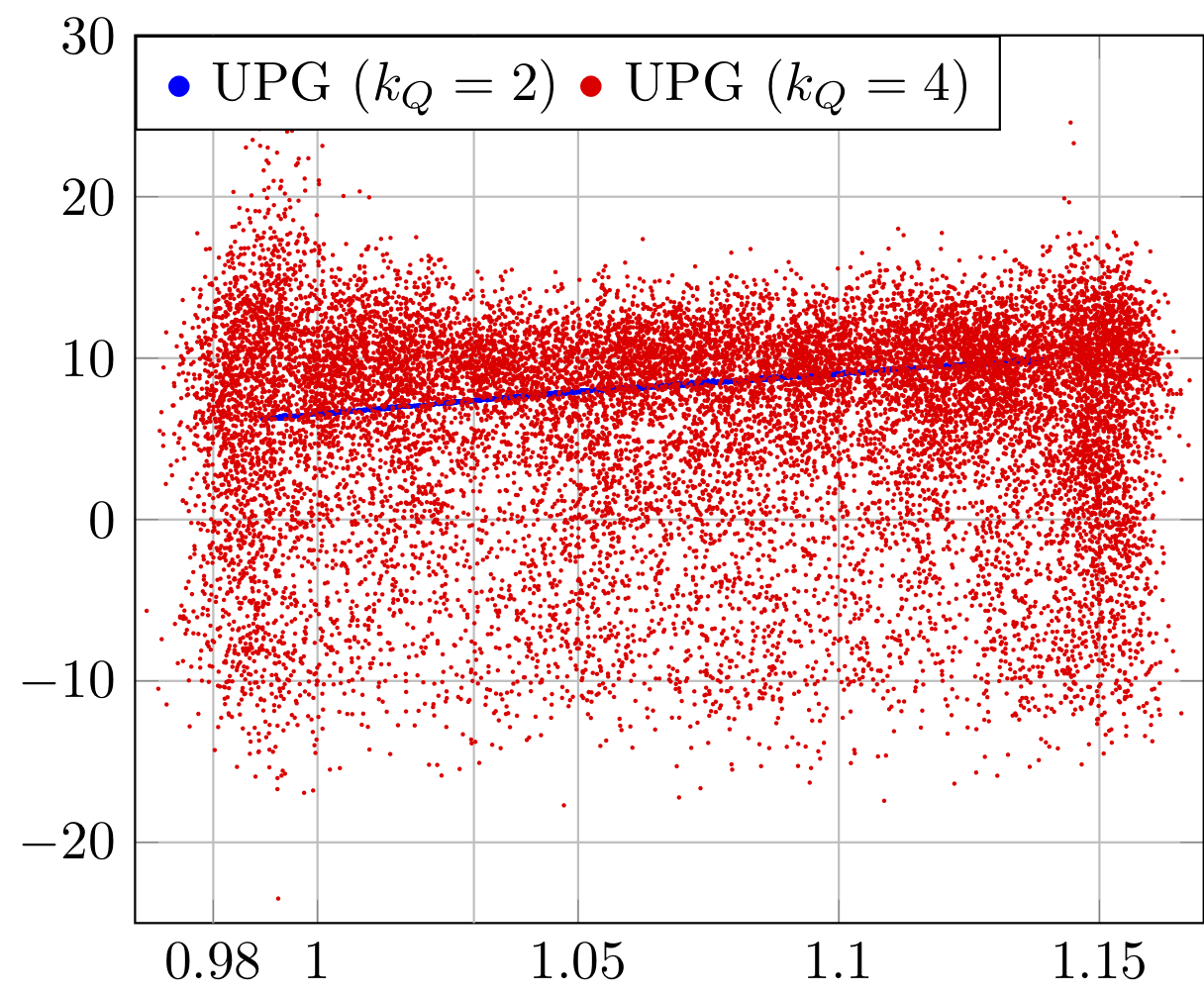}
    }  
    \caption{Compressible sphere ($\nu=0.3$) under internal pressure: trace of the Cauchy stress tensor $\stress$ (in $\MPa$)  vs. deformed radius $r$ (in $\mm$) for the different methods at all the quadrature points and for the limit load.}
    \label{fig::sphere_trace_comp}
\end{figure}

We next investigate the influence of the quadrature order $k_Q$ on the accuracy of the solution. The trace of the Cauchy stress tensor $\stress$ is compared for HHO(1;1), HHO(2;2), and UPG methods in Fig.~\ref{fig::sphere_trace_kq} at all the quadrature points on the final configuration for the limit load, and for a quadrature order $k_Q$ higher than the one employed in Fig.~\ref{fig::sphere_trace} (HHO(1;2) and HHO(2;3) give similar results and are not shown for brevity). We remark that when we increase the quadrature order, UPG locks for quasi-incompressible finite deformations, whereas HHO does not lock, and the results are (only) a bit more dispersed around the analytical solution. Moreover, HHO(2;2) is less sensitive than HHO(1;1) to the choice of the quadrature order $k_Q$. Note that this problem is not present for HHO methods with small deformations \cite{AbErPi2018a}. Furthermore, this sensitivity to the quadrature order seems to be absent for finite deformations when the elastic deformations are compressible (the plastic deformations are still incompressible). To illustrate this claim, we perform the same simulations as before but for a compressible material. The Poisson ratio is taken now as $\nu=0.3$ (recall that we used $\nu=0.499$ in the quasi-incompressible case) whereas the other material parameters are unchanged. Unfortunately, an analytical solution is no longer available in the compressible case. We compare again the trace of the Cauchy stress tensor $\stress$  for HHO(1;1), HHO(2;2), and UPG methods in Fig.~\ref{fig::sphere_trace_comp} at all the quadrature points on the final configuration and for different quadrature orders $k_Q$. We observe a quite marginal dependence on the quadrature order for HHO methods (as in the quasi-incompressible case); whereas the UPG method still locks if the order of the quadrature is increased. Moreover, in the compressible case, HHO(2;2) gives a more accurate solution than HHO(1;1). \\

\subsection{Summary of the above results}
The proposed HHO method has been tested successfully on four benchmarks in two and three dimensions. A first conclusion is that the proposed HHO method is robust for large elastoplastic deformations and is locking-free as mixed methods but without the need to introduce additional globally coupled unknowns. HHO(2;2) and HHO(2;3) give generally more accurate results both for the displacement and the Cauchy stress tensor than HHO(1;1), HHO(1;2), and UPG on a fixed mesh (cG methods lock). Moreover, contrary to the UPG method, HHO methods are not very sensitive to the choice of the quadrature order (particularly for $k=2$). Finally, HHO(1;1) appears to be more prone to the localization of the plastic deformations, contrary to the other variants HHO(1;2), HHO(2;2), and HHO(2;3).
\section{Further numerical investigations}\label{ss::num_inv}
In this section, we perform further numerical investigations to test other capacities of HHO methods such as the support of general meshes with possibly non-conforming interfaces, the possibility of considering the lowest-order case $k=0$, and the dependence on the stabilization parameter $\beta$.
\subsection{Polygonal meshes}\label{ss::poly_mesh}
\begin{figure}[htbp]
    \centering
    \subfloat[Reference solution with HHO(2;3) on a quadrangular mesh]{
        \centering
        \includegraphics[scale=0.35, trim= 55 0 70 0, clip=true]{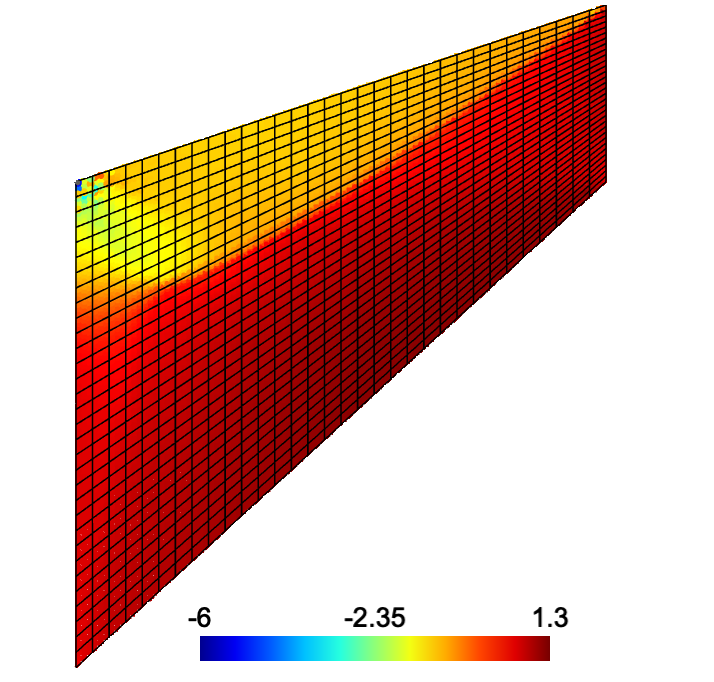}
  }
    ~ 
    \subfloat[HHO(1;1)]{
        \centering
	    \includegraphics[scale=0.375, trim= 53 -10 0 30, clip=true]{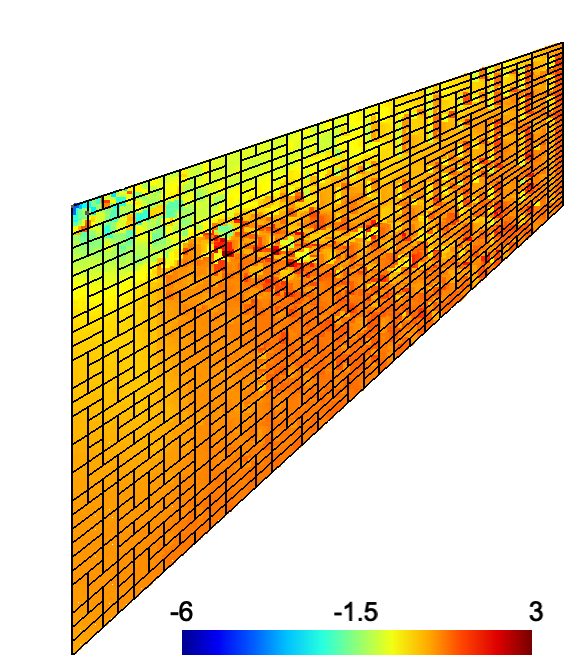}
    }
      ~ 
    \subfloat[HHO(1;2)]{
        \centering
	    \includegraphics[scale=0.35, trim= 60 0 70  0, clip=true]{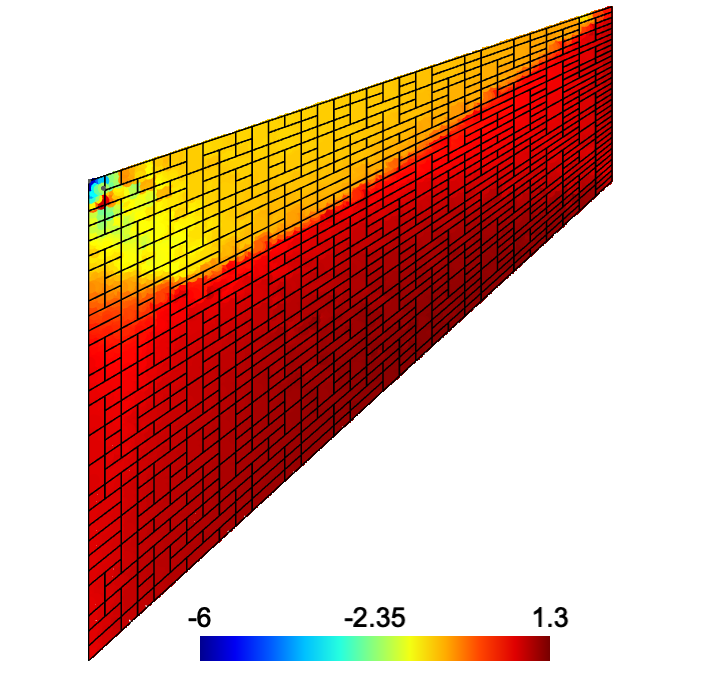}
    }
    
        \subfloat[HHO(2;2)]{
        \centering
	    \includegraphics[scale=0.35, trim= 60 0 70 0, clip=true]{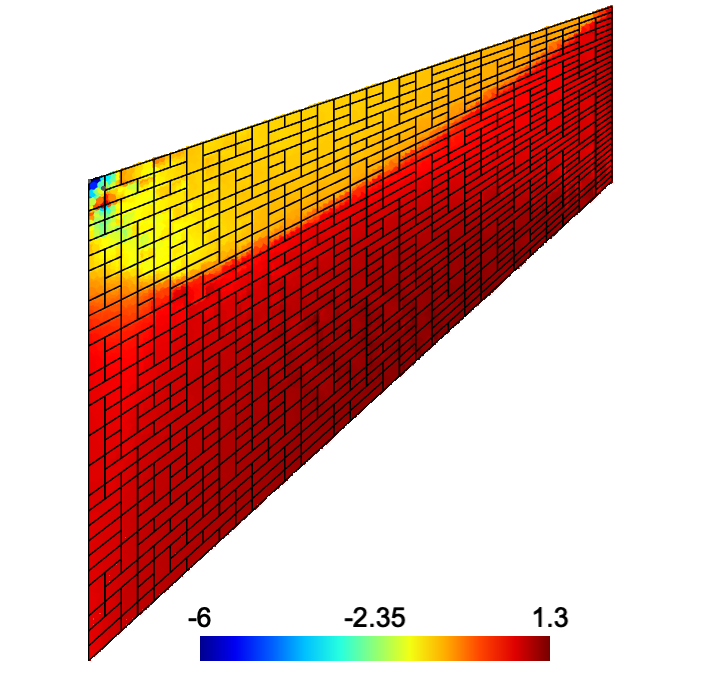}
    }
      ~ 
    \subfloat[HHO(2;3)]{
        \centering
	    \includegraphics[scale=0.35, trim= 60 0 70 0, clip=true]{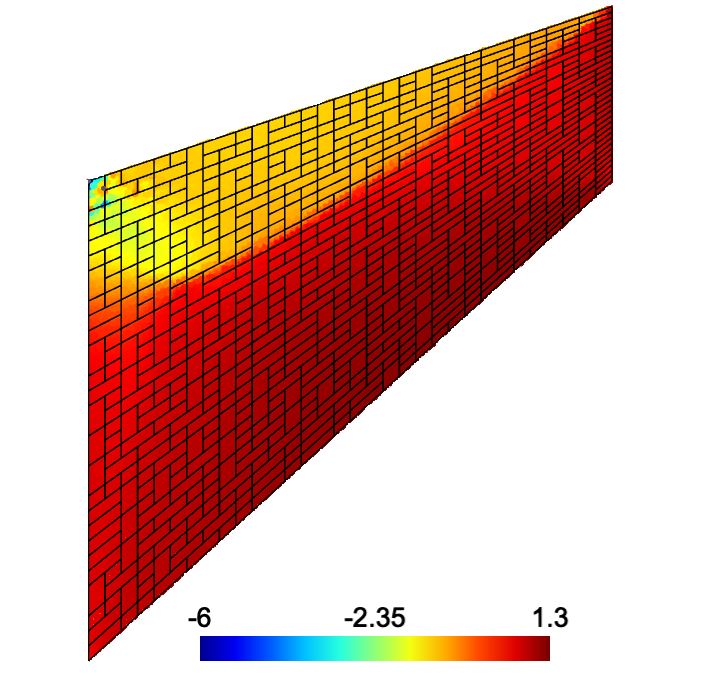}
    }
    \caption{Cook's membrane: trace of the Cauchy stress tensor $\stress$ (in $\GPa$) at the quadrature points on the reference configuration (a) Reference solution with HHO(2;3) on a $32 \times 32$ quadrangular mesh composed of 1024 cells. (b)-(e) Results for the different HHO variants on a mesh with hanging nodes composed of 719 polygonal cells.}
    \label{fig::cook_poly}
\end{figure}

\begin{figure}[htbp]
    \centering
    \subfloat[]{
        \centering
        \includegraphics[scale=1]{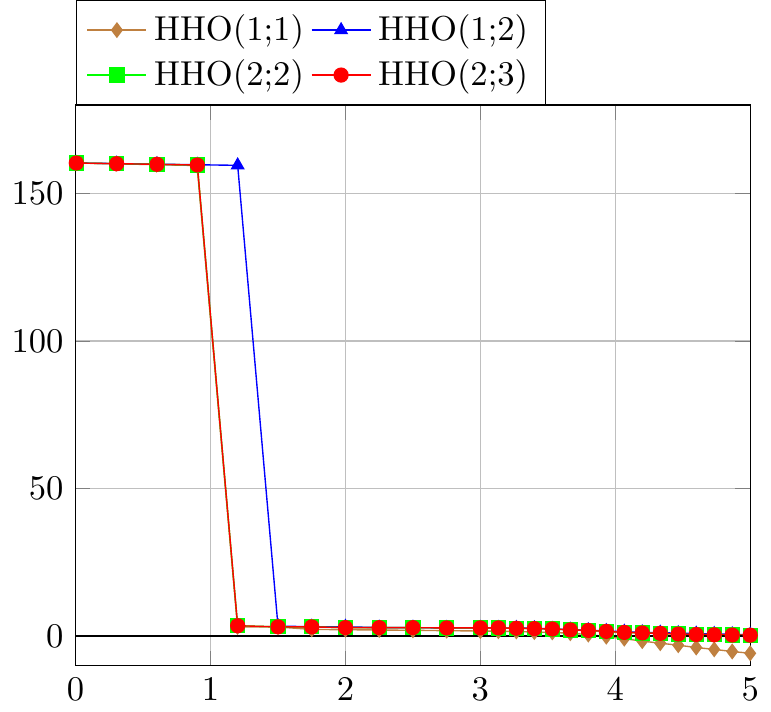}
  }
    ~ 
    \subfloat[]{
        \centering
	    \includegraphics[scale=1]{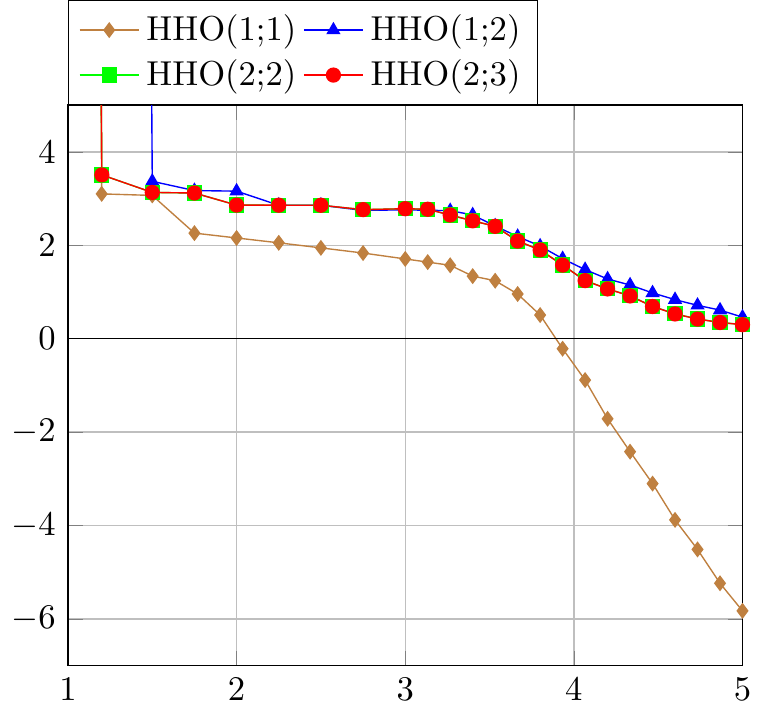}
    }
    
    \caption{Cook's membrane: evolution of the magnitude of the smallest eigenvalue $\theta_{\Th, Q}$  (in $\MPa$)  vs.  vertical load applied  $F_y$ (in $\kN$) for the different HHO methods on a non-conforming mesh with hanging nodes composed of 719 polygonal cells (a) during the complete loading (b) a zoom when the plastic evolution occurs.}
    \label{fig::cook_evol_vp}
\end{figure}

In the previous sections, the proposed HHO method has been tested on simplicial and hexahedral meshes so as to be able to compare it to the UPG method which only supports this type of meshes. Our goal is now to illustrate that the HHO method supports general meshes with possibly non-matching interfaces. For our test cases, the polygonal meshes are generated from quadrangular meshes by removing the common face for some pairs of neighbouring cells and then merging the two cells in question (about 30\% of the cells are merged) thereby producing non-matching interfaces materialized by hanging nodes for a significant portion of the mesh cells. We consider the Cook's membrane problem from Section~\ref{ss::cook} and we use a mesh composed of 719 polygonal cells including quads, pentagons (quads with one hanging node) and hexagons (quads with two hanging nodes). The trace of the Cauchy stress tensor $\stress$ is shown in Fig.~\ref{fig::cook_poly} at the quadrature points on the reference configuration; and we compare the results with a reference solution computed with HHO(2;3) on a $32\times 32$ quadrangular mesh. The results agree very well except for HHO(1;1) where the trace is not smooth due to the localization of the plastic deformations (as for the quadrangular mesh, see Section~\ref{ss::cook}).
A reason for the localization of the plastic deformations is the loss of coercivity of the consistent elastoplastic tangent modulus $\depmodulePK$.  The evolution of the magnitude of the smallest eigenvalue $\theta_{\Th, Q}$ of $\depmodulePK$ during the loading is plotted in Fig.~\ref{fig::cook_evol_vp}. We remark that the magnitude of $\theta_{\Th, Q}$ decreases quickly when the plastic evolution begins (around $F_y = 1~\kN$), then $\theta_{\Th, Q}$ continues to decrease more slowly and remains positive for all HHO methods except for HHO(1;1) where it decreases more quickly and becomes negative  (so that Theorem~\ref{th::coer_newton} is no longer valid).  As mentioned above, this loss of positive-definiteness of $\depmodulePK$ for HHO(1;1) can explain the presence of nonphysical plastic localization. Moreover,  the distribution of the smallest eigenvalue  $\theta^{\min}(\depmodulePK)$ of $\depmodulePK$ at the quadrature points at the end of the loading $F_y = 5~\kN $ for the different HHO methods is summarized in Table~\ref{tab::cook_vp}. Only HHO(1;1) has negative eigenvalues. We remark that the number of quadrature points where $\depmodulePK$ has negative eigenvalues for HHO(1;1) or close to 0 (<0.5) for the others HHO methods is small compared to the total number of quadrature points. This confirms that the loss of positive-definiteness and the presence of plastic localization are local and confined to few quadrature points. Note that for a total vertical load $F_y > 6.1~\kN $, $\theta_{\Th, Q}$ is negative for all HHO methods.

\begin{table}
\centering
\resizebox{\textwidth}{!}{\begin{tabular}{|c|c|c|c|c|c|c|c|c|c|c|}
\hline 
\multirow{2}{*}{Method} & $\theta_{\Th, Q}$ & \multicolumn{8}{c|}{Total number of eigenvalues by sub-interval over [-6; 170] (in $\MPa$)} &  \\ 
\cline{3-11}
 & (in $\MPa$) & [-6; -1] & (-1; -0.5] & (-0.5; 0] & (0; 0.5] & (0.5; 1] & (1; 5] & (5; 150] & (150; 170] & Total \\ 
\hline 
HHO(1;1) & -5.83 & 5 & 2 & 5 & 10 & 92 & 5435 & 1 & 546 & 6096 \\ 
\hline 
HHO(1;2) & 0.46 & 0 & 0 & 0  &  3 & 968 & 4735 & 0 & 390 & 6096 \\ 
\hline 
HHO(2;2) & 0.30 & 0 & 0 & 0 & 7 & 2073 & 9731 & 0 & 1005 &12816  \\ 
\hline 
HHO(2;3) & 0.29 & 0 & 0 & 0 & 8 & 2073 & 9729 & 0 & 1006 & 12816 \\ 
\hline 
\end{tabular} }
\caption{Cook's membrane: distribution of the smallest eigenvalue  $\theta^{\min}(\depmodulePK)$  (in $\MPa$)  of the consistent elastoplastic tangent modulus $\depmodulePK$ at  the quadrature points at the end of the loading $F_y = 5~\kN $ for the different HHO methods on a mesh with hanging nodes composed of 719 polygonal cells.}\label{tab::cook_vp}
\end{table}

\subsection{Lowest-order variant}
\begin{figure}
    \centering  
      \subfloat[Equivalent plastic strain $p$]{
        \centering
        \includegraphics[scale=0.33, trim= 550 0 500 0, clip=true]{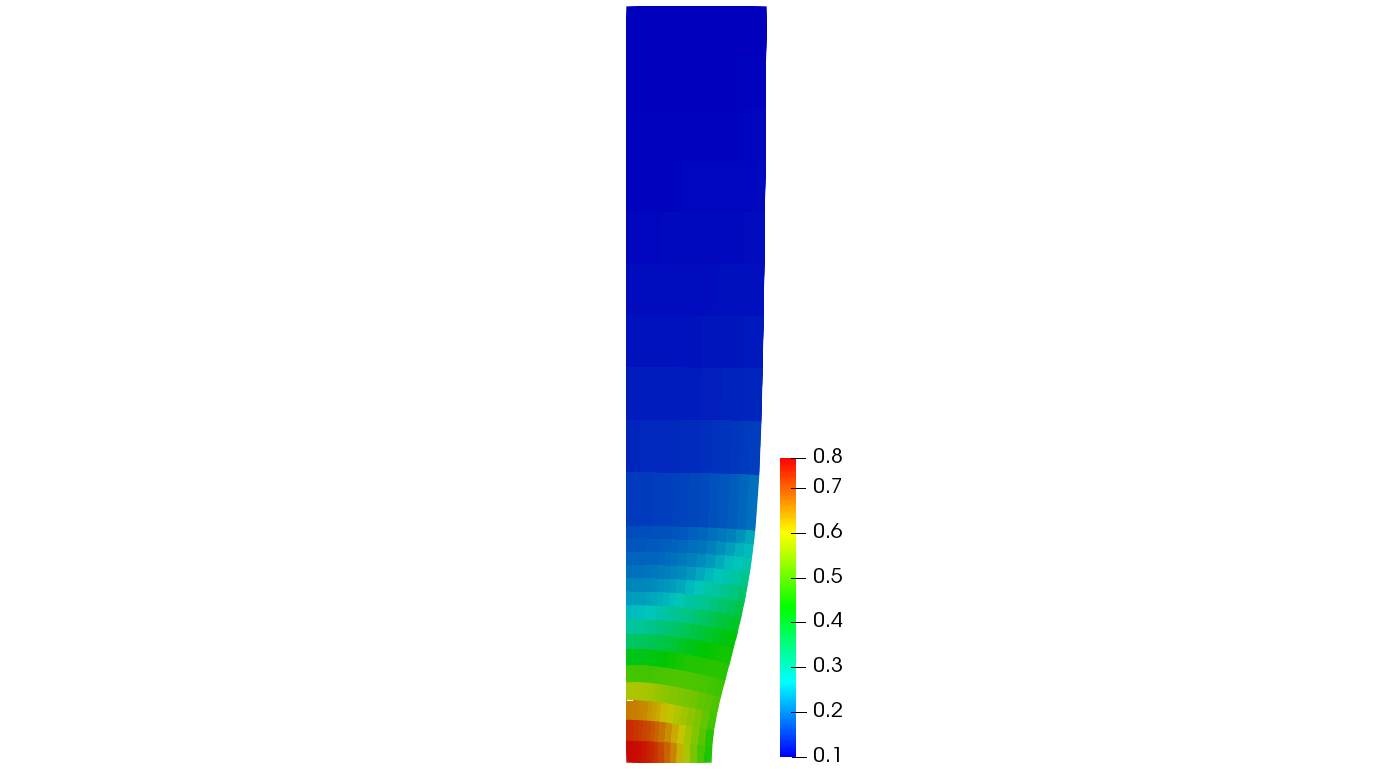}
  } 
 ~ 
    \subfloat[Trace of the Cauchy stress tensor $\stress$]{
        \centering
        \includegraphics[scale=0.33, trim= 550 0 425 0, clip=true]{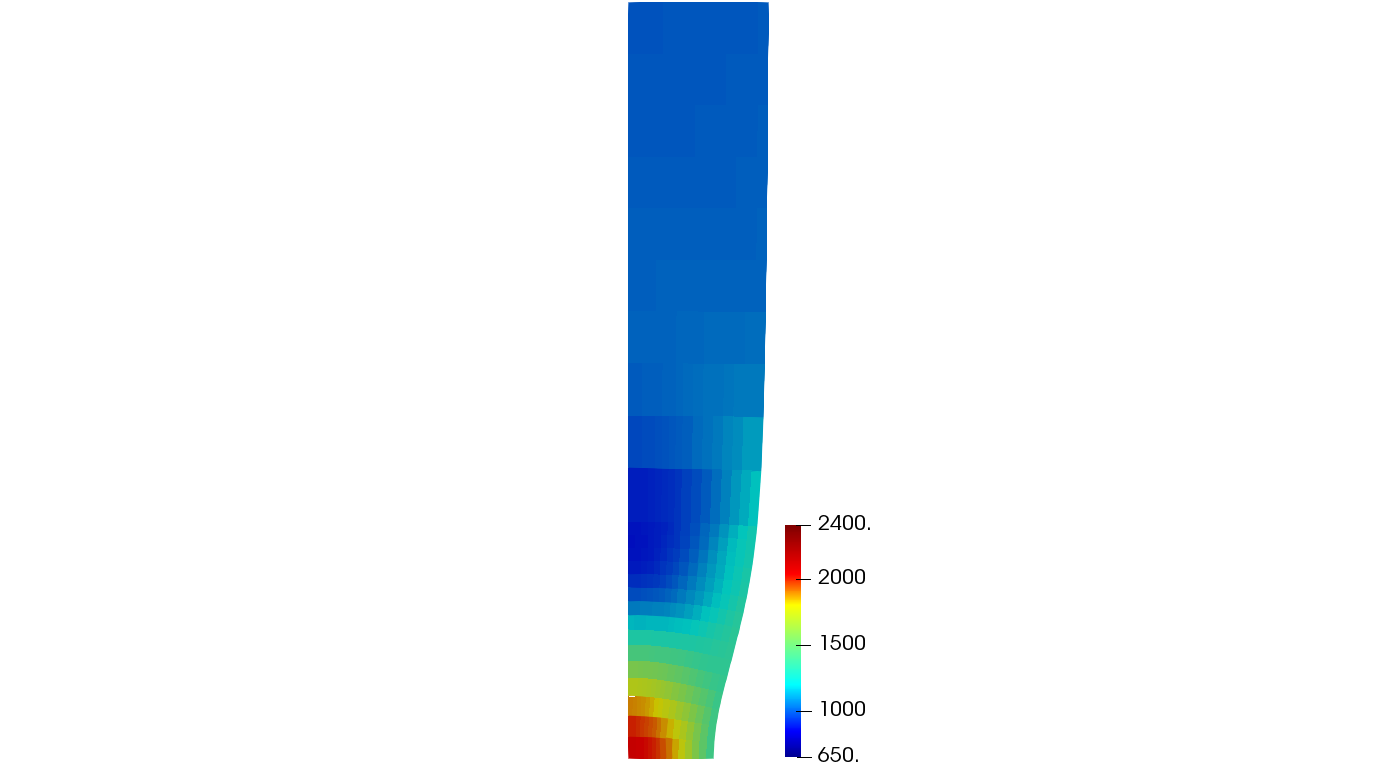}
  }
    \caption{Necking of a 2D rectangular bar with low-order variant HHO(0;1): (a) equivalent plastic strain $p$ and (b) trace of the Cauchy stress tensor $\stress$ (in $\MPa$) at the quadrature points on the final configuration.}
        \label{fig::low-order}
\end{figure}

The main reason to take $k \geq 1$ in the HHO method applied to the linear elasticity problem is that the rigid-body motions $\vecteur{RM}(T)$ are then a subset of $\vecteur{U}_{T}^{1,1}$. The lowest-order case $k=0$ and $l=1$ is interesting since there are only $d$ unknowns per face, i.e, two in 2D and three in 3D; and we could expect that the energy error, resp. the $L^2$-error, converges as $h|\vecteur{u}|_{\vecteur{H}^{2}(\Omega_0)}$, resp. as $h^2|\vecteur{u}|_{\vecteur{H}^{2}(\Omega_0)}$, for the linear elasticity problem. The difficulty with this lowest-order case is to deal with the rigid-body motions on the faces since unfortunately $\vecteur{RM}(T){}_{|\dT} \nsubseteq \Poly^{0}_{d-1}(\FT; \Rd)$. Therefore, at the theoretical level, it is not clear that Lemma~\ref{lemma_stability_stab} still holds true. Nevertheless, we observed numerically that for the linear elasticity problem, the energy error, resp. the $L^2$-error, converges as $h|\vecteur{u}|_{\vecteur{H}^{2}(\Omega_0)}$, resp. as $h^2|\vecteur{u}|_{\vecteur{H}^{2}(\Omega_0)}$, (the expected optimal rates) if all the cells have at least $2d$ faces, i.e, four faces in 2D and six faces in 3D. This observation seems to be confirmed for small elastoplasticity. However, the conclusions are less clear for finite elastoplasticity. The equivalent plastic strain $p$ and the trace of the Cauchy stress tensor $\stress$ are plotted in Fig.~\ref{fig::low-order} at the quadrature points on the final configuration for the necking of a rectangular bar (see Section~\ref{ss::bar}) approximated using the HHO(0;1) variant. We observe the absence of volumetric locking and that the results are close to those obtained for $k\geq 1$ (see Fig.~\ref{fig::necking_p} and Fig.~\ref{fig::necking_trace}). However, for the Cook's membrane problem (see Section~\ref{ss::cook}), the displacement is not correct (not shown for brevity). 
\begin{remark}
In\cite{Fu2015}, a theoretical study of an HDG method for the linear elasticity problem in the equal-order case $k=0$ is performed where the main difference with the present HHO method is the stabilization weight which is of the form $O(1)$ and no longer $O(h^{-1})$ as here. In this case, for the linear elasticity problem, the the energy error, the $L^2$-error, and the stress error converge as $h|\vecteur{u}|_{\vecteur{H}^{2}(\Omega_0)}$, $h|\vecteur{u}|_{\vecteur{H}^{2}(\Omega_0)}$,  as $h^{\frac12}|\vecteur{u}|_{\vecteur{H}^{2}(\Omega_0)}$, respectively, on general meshes. Moreover, recent numerical results still for the linear elasticity problem \cite{Sevilla2019} indicate that the stress error can converge as $h|\vecteur{u}|_{\vecteur{H}^{2}(\Omega_0)}$. Nevertheless, the $L^2$-error converges slower for this HDG variant than what we could expect for the lowest-order HHO method.
\end{remark}
\subsection{Influence of the stabilization parameter}\label{ss:stab}
\begin{figure}[htbp]
    \centering
    \subfloat[Number of total Newton's iterations vs. $\beta_0$ for the Cook's membrane problem]{
        \centering
        \includegraphics[scale=1]{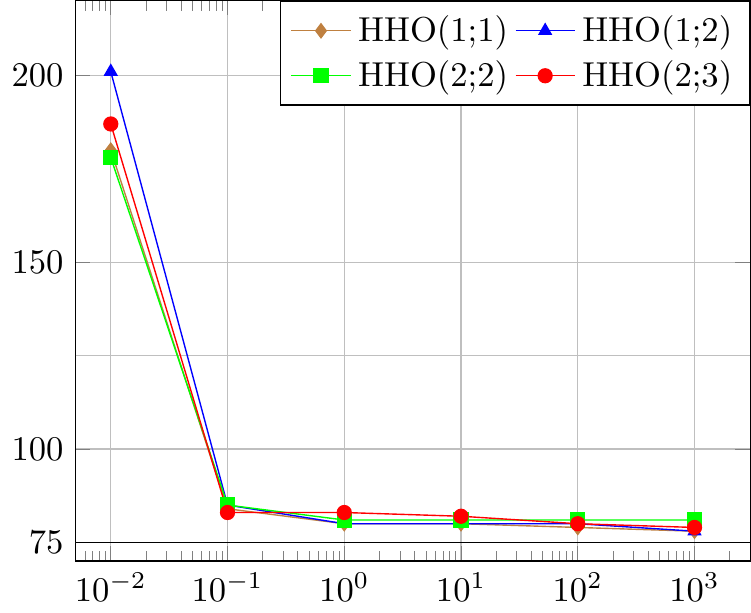}
        \label{fig:cook_beta}
  }
    ~ 
    \subfloat[Number of total Newton's iterations vs. $\beta_0$ for the quasi-incompressible sphere under internal pressure]{
        \centering
	    \includegraphics[scale=1]{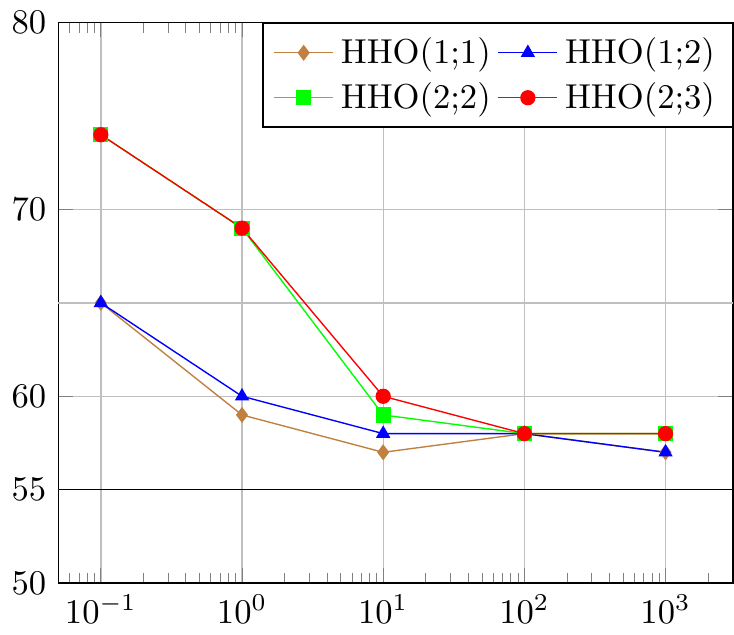}
	    \label{fig:sphere_beta}
    }
   \caption{Influence of the stabilization parameter: Total number of Newton's iterations vs. $\beta_0$ for (a) the Cook's membrane and (b) the quasi-incompressible sphere under internal pressure.}
\end{figure}

To evaluate the influence of the stabilization parameter $\beta_0$,  we compare the total number of Newton's iterations needed to solve the nonlinear problem \eqref{discrete_problem_ptv} versus the magnitude of the stabilization parameter $\beta_0$. The Newton's iterations are stopped under the relative criterion  $\norme[\Th]{R_h(\dvTh,\dvFh)} \leq 10^{-6} \norme[\Th]{F_{\textrm{int}}(\dvTh,\dvFh)}$ where $F_{\textrm{int}} $ are the internal forces. We perform this comparison on two of the previous benchmarks, the Cook's membrane problem (see Section~\ref{ss::cook}) and the quasi-incompressible sphere under internal pressure (see Section~\ref{ss::sphere}).
In Fig.~\ref{fig:cook_beta}, we report the total number of Newton's iterations for the Cook's membrane problem with strain-hardening plasticity. We use a $32 \times 32$ quadrangular mesh, and 15 load increments of equal size are considered. On the one hand, we remark that the different HHO variants need almost the same total number of Newton's iterations (around 78 compared to 75 for UPG) if $\beta_0 \geq 0.1$; On the other hand, if $\beta_0  < 0.01$, the Newton's method stops converging whatever the HHO variant and the number of load increments.
For the quasi-incompressible sphere under internal pressure, the pressure is applied in 15 increments of equal size. Recall that this experiment is particularly challenging since  we are considering here perfect plasticity for which the stability result from Theorem~\ref{th::coer_newton} is not applicable. In Fig.~\ref{fig:sphere_beta}, we plot the total number of Newton's iterations to perform the simulation. On the one hand, if $\beta_0 \geq 10$, all the HHO variants need almost to the same total number of Newton's iterations (around 57 compared to 55 for UPG). On the other hand, if $\beta_0 \leq 1$,  the HHO variants with $k =2$ need more Newton's iterations than the HHO variants with $k=1$. As previously, if $\beta_0 < 0.1$, the Newton's method stops converging.

 A first conclusion is therefore that the proposed HHO methods are stable for a large range of values of the stabilization parameter $\beta_0$. A second conclusion is that it seems reasonable to take $\beta_0 \in [1, 100]$ since the number of Newton's iterations is lower and close to the value for UPG, and the condition number does not increase too much. Note that for extremely large values of $\beta_0$, HHO methods reduce to $H^1$-conforming methods due to the matching of the face unknowns with the trace of the cell unknowns, and volumetric-locking can appear (not shown here for brevity).

\section{Conclusions}

We have devised and evaluated numerically a Hybrid High-Order (HHO) method to approximate finite elastoplastic deformations within a logarithmic strain framework. This framework allows one to re-use behavior laws developed originally for small deformations in the context of finite deformations.The HHO method exhibits a robust behavior for strain-hardening plasticity as well as for perfect plasticity, and produces accurate solutions with a moderate number of degrees of freedom for various benchmarks from the literature. In particular, as mixed methods, the HHO method avoids volumetric locking due to plastic incompressiblity, but with less unknowns than mixed methods for the same accuracy. Moreover, the HHO method supports general meshes with non-matching interfaces. 

This work can be pursued in several directions. One could use a non-local plasticity model, as for example a strain-gradient plasticity model, to take into account scale-dependent effects \cite{McBride2009} and possibly prevent unphysical localization of the plastic deformations. Furthermore, the extension of the present HHO method to contact and friction problems is the subject of ongoing work.

\ifCM
\else
\bibliographystyle{abbrvnat}
\fi
\bibliography{Bibliographie}
\end{document}